\documentclass[11pt]{article}
\usepackage{a4wide}

\usepackage{bm}

\usepackage{amsmath,amssymb}
\DeclareFontFamily{U}{mathx}{}
\DeclareFontShape{U}{mathx}{m}{n}{<-> mathx10}{}
\DeclareSymbolFont{mathx}{U}{mathx}{m}{n}
\DeclareMathAccent{\widehat}{0}{mathx}{"70}
\DeclareMathAccent{\widecheck}{0}{mathx}{"71}
\newcommand{\wc}[1]{\widecheck{#1}}

\usepackage{comment}

\usepackage[dvipdfmx]{graphicx}
\usepackage{graphicx}
\usepackage{color}

\usepackage{xcolor}
\usepackage{braket}
\usepackage{slashed}
\usepackage[bookmarks=true,bookmarksnumbered=true,setpagesize=false]{hyperref}
\usepackage{enumerate}

\DeclareMathOperator{\ephogehoge}{\epsilon}
\newcommand{\ep}[1]{\sideset{}{\scriptstyle{[#1]}}\ephogehoge}

\DeclareMathOperator{\tr}{tr}
\DeclareMathOperator{\Tr}{Tr}

\DeclareMathOperator{\diag}{diag}

\renewcommand{\t}{\text{t}}

\newcommand{\commutator}[2]{\left[#1,\,#2\right]}
\newcommand{\anticommutator}[2]{\left\{#1,\,#2\right\}}

\usepackage{pict2e}
\makeatletter
\DeclareRobustCommand{\intprod}{%
  \mathbin{\mathpalette\int@prod{(0.1,0)(0.85,0)(0.85,0.7)}}%
}
\DeclareRobustCommand{\intprodr}{%
  \mathbin{\mathpalette\int@prod{(0.1,0.7)(0.1,0)(0.85,0)}}}

\newcommand{\int@prod}[2]{%
  \begingroup
  \sbox\z@{$\m@th#1+$}%
  \setlength\unitlength{\wd\z@}%
  \begin{picture}(1,1)
  \roundcap
  \polyline#2
  \end{picture}%
  \endgroup
}
\makeatother

\newcommand{\al}[1]{\begin{align}#1\end{align}}
\newcommand{\als}[1]{\begin{align*}#1\end{align*}}
\newcommand{\ov}{\over}
\newcommand{\nn}{\nonumber\\}
\newcommand{\tx}{\text}

\newcommand{\paren}[1]{\left(#1\right)}
\newcommand{\pn}[1]{\left(#1\right)}
\newcommand{\sqbr}[1]{\left[#1\right]}
\newcommand{\ab}[1]{\left|#1\right|}

\newcommand{\fn}[1]{\!\left(#1\right)}
\newcommand{\fnl}[1]{\!\left[#1\right]}
\newcommand{\pa}[1]{\left(#1\right)\!{}}
\newcommand{\mt}[1]{\left[#1\right]\!{}}	

\newcommand{\Pn}[1]{\bigl(#1\bigr)}
\newcommand{\Sqbr}[1]{\bigl[#1\bigr]}

\newcommand{\Fn}[1]{\!\bigl(#1\bigr)}

\newcommand{\vect}[1]{\bm{#1}}
\newcommand{\bs}{\boldsymbol}

\newcommand{\df}{\text{d}}
\newcommand{\D}{\mathscr{D}}

\newcommand{\I}{I}

\newcommand{\mc}{\mathcal}

\usepackage{mathrsfs}
\newcommand{\ms}{\mathscr}

\newcommand{\mf}{\mathfrak}
\newcommand{\bmat}[1]{\begin{bmatrix}#1\end{bmatrix}}

\newcommand{\p}{\partial}

\newcommand{\ol}{\overline}
\newcommand{\pr}{\prime}

\newcommand{\wt}{\widetilde}
\newcommand{\MP}{M_\text{P}}

\newcommand{\ba}{\textbf{a}}
\newcommand{\bb}{\textbf{b}}
\newcommand{\bc}{\textbf{c}}
\newcommand{\bd}{\textbf{d}}

\definecolor{darkgreen}{rgb}{0,0.75,0}
\definecolor{darkred}{rgb}{0.75,0,0}
\definecolor{darkyellow}{rgb}{0.75,0.75,0}
\definecolor{darkcyan}{rgb}{0,0.75,0.75}
\definecolor{darkmagenta}{rgb}{0.75,0,0.75}

\newcommand{\diff}{\textsf{diff}}

\newcommand{\powe}[1]{\left\lfloor#1\right\rfloor} 

\newcommand{\sr}[2]{\stackrel{#1}{#2}}

\newcommand{\vep}{\varepsilon}

\newcommand{\tb}[1]{\textbf{#1}}
\newcommand{\ola}[1]{\overleftarrow{#1}}

\newcommand{\os}[2]{\overset{#1}{#2}{}}

\newcommand{\lra}{\longrightarrow}

\newbox{\ORCIDicon}
\sbox{\ORCIDicon}{\large\includegraphics[width=0.5em]{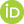}}

\begin{document}
\title{
Irreversible vierbein postulate:\\
Emergence of spacetime
from quantum phase transition
}\maketitle
\begin{center}

\author{
Yadikaer Maitiniyazi,$^{*\href{https://orcid.org/0009-0004-0826-1130}{\usebox{\ORCIDicon}}}$\footnote{E-mail: \tt ydqem22@mails.jlu.edu.cn}
\
Shinya Matsuzaki,$^{*\href{https://orcid.org/0000-0003-4531-0363}{\usebox{\ORCIDicon}}}$\footnote{E-mail: \tt synya@jlu.edu.cn}
\
Kin-ya Oda,$^{\dagger\href{https://orcid.org/0000-0003-3021-1661}{\usebox{\ORCIDicon}}}$\footnote{E-mail: \tt odakin@lab.twcu.ac.jp}
\ and
Masatoshi Yamada$^{*\href{https://orcid.org/0000-0002-1013-8631}{\usebox{\ORCIDicon}}}$\footnote{E-mail: \tt yamada@jlu.edu.cn}
}
\end{center}
\begin{center}\small\it
$^*$Center for Theoretical Physics and College of Physics, Jilin University, Changchun 130012, China\smallskip\\
$^{\dagger}$Department of Mathematics, Tokyo Woman’s Christian University, Tokyo 167-8585, Japan
\end{center}
\bigskip

\begin{abstract}\noindent
We formulate a model for quantum gravity based on the local Lorentz symmetry and general coordinate invariance. A key idea is the irreversible vierbein postulate that a tree-level action for the model at a certain energy scale does not contain an inverse vierbein. Under this postulate, only the spinor becomes a dynamical field, and no gravitational background field is introduced in the tree-level action. In this paper, after explaining the transformation rules of the local Lorentz and general-coordinate transformations in detail, a tree-level action is defined. We show that fermionic fluctuations can induce a nonvanishing gravitational background field.
\end{abstract}

\newpage
\tableofcontents

\setcounter{footnote}{0}

\newpage
\section{Introduction}
Quantum gravity---how to consistently quantize interacting spacetime fluctuations---is one of the most profound mysteries that has defied human challenges over the past century. On the other hand, it has been established experimentally that there indeed exist spacetime fluctuations that propagate over cosmic distances with the speed of light, namely, the gravitational waves, consistently described by Einstein's general relativity; see e.g.\ Refs.~\cite{Taylor:1982zz, LIGOScientific:2016aoc} for classic examples. Given the tremendous success of the Standard Model (SM) of particle physics based on quantum field theory (see, e.g.,\ Ref.~\cite{ParticleDataGroup:2022pth} for a review), it is natural to expect that the gravitational field governing the observed spacetime fluctuation must be quantized too. Whether it is really quantized or not will be experimentally explored within the forthcoming decades as a form of quantized free spacetime fluctuations, gravitons, on a curved classical background during inflation in observations of the B-mode polarization of the cosmic microwave background~\cite{POLARBEAR:2016wwl, CORE:2017oje, LiteBIRD:2023iei} and further in direct observation of the cosmic gravitational-wave background~\cite{Kawamura:2020pcg}; see also Ref.~\cite{Page:1981aj}.

Recent advances in cooling, control, and measurement of mechanical systems in the quantum regime, particularly using matter-wave and optomechanical systems, have set the stage for potential first observations of quantum gravitational effects, as predicted by various low-energy quantum-gravity models, though with certain challenges~\cite{Carney:2018ofe}. Concurrently, recent table-top experiments in quantum-gravity phenomenology reassess classical descriptions by focusing on gravitational effects from delocalized quantum sources, aiming to uncover interactions beyond the Newtonian potential and deepen our understanding of gravity's quantum nature~\cite{Chen:2024xvm}.

It is well known, however, that the quantization of the metric based on Einstein's general relativity is perturbatively nonrenormalizable, requiring an infinite number of counterterms and thus spoiling its predictability at the quantum level due to the infinite number of free parameters; see, e.g.,\ Refs.~\cite{tHooft:1974toh, Goroff:1985sz, Goroff:1985th}, and also Ref.~\cite{Percacci:2017fkn} for a review.
Furthermore, the truncation of the gravitational action up to the dimension-two Einstein-Hilbert term with the Ricci curvature scalar $\mathcal R$ yields a conformal mode that has a wrong-sign kinetic term, which makes Euclidean quantum gravity ill-defined for both directions of Wick rotations such that either the wrong-sign mode or the other fields become exponentially growing along the imaginary time direction; see, e.g.,\ Refs.~\cite{Gibbons:1978ac,tHooft:2011aa} and Appendix~A in Ref.~\cite{Lee:2023wdm} for a simple review.

It is known that a higher-derivative gravity involving the $\mathcal{R}^2$ and $\mathcal{R}_{\mu\nu}^2$ terms, in addition to the Einstein-Hilbert term, is perturbatively renormalizable. However, this leads to a loss of unitarity in the theory~\cite{Stelle:1976gc}, though this issue of nonunitarity is under an attempt to be circumvented by recent works in Refs.~\cite{Anselmi:2018kgz,Donoghue:2021cza} and references therein; see also Refs.~\cite{Kubo:2023lpz,Kubo:2024ysu} for further discussions.

If we allow the theory to discard the Lorentz symmetry, it could be perturbatively renormalizable~\cite{Horava:2009uw}, though the speed of light depends on particle species and hence we need additional fine-tunings; see also Refs.~\cite{EmirGumrukcuoglu:2017cfa,Gong:2018vbo,Barausse:2019yuk,Frusciante:2020gkx}. In any case, it appears that the realization of a renormalizable theory of gravity in perturbation theory is difficult if we retain all the essential symmetries and properties, in particular, both the Lorentz symmetry and unitarity.

The above perturbative nonrenormalizability argument is based on quantum theory around a free theory. In the Wilsonian viewpoint, the perturbative gravity is constructed around the Gaussian (trivial) fixed point. On the other hand, the notion of renormalizability in quantum field theory is generalized to the nonperturbative realm. This scenario is known as asymptotically safe gravity~\cite{Hawking:1979ig,Reuter:1996cp,Souma:1999at}; see also Refs.~\cite{Oda:2015sma,Niedermaier:2006wt,Niedermaier:2006ns,Percacci:2007sz,Codello:2008vh,Reuter:2012id,Percacci:2017fkn,Reuter:2019byg,Wetterich:2019qzx,Pawlowski:2020qer,Bonanno:2020bil,Reichert:2020mja,Eichhorn:2017egq,Eichhorn:2018yfc} for reviews. There is accumulating evidence that there exists a nontrivial (interacting) ultraviolet (UV) fixed point in gravitational systems by means of the functional renormalization group method. Quantum gravity is perturbatively nonrenormalizable but might be nonperturbatively renormalizable. This situation is similar to the $O(N)$ nonlinear sigma model. (This will be discussed in Section~\ref{sec: Degrees of freedom}.)

The statements mentioned above are based on an assumption that the metric field (spin-2 symmetric tensor field) is the fundamental degree of freedom. The view that Einstein's general relativity is a local Lorentz (LL) gauge theory~\cite{Utiyama:1956sy} is found almost at the same time as the (nonabelian) gauge theory itself~\cite{Yang:1954ek} and has been developed in Refs.~\cite{Kibble:1961ba,Sciama:1964wt}. To write down the LL symmetry, it is essential to rewrite the metric degrees of freedom by the vierbein (tetrad) ones. The vierbein is also indispensable to writing down a spinor field on a curved space, namely, the matter field in our Universe.\footnote{
One may consider replacing the vierbein degrees of freedom by promoting the gamma matrices $\gamma_\mu\!\left(x\right)=e^{\mathbf a}{}_\mu\!\left(x\right)\gamma_{\mathbf a}$ as dynamical variables~\cite{Gies:2013noa,Lippoldt:2016ayw}. The fluctuation of $\gamma_\mu\!\left(x\right)$ can be decomposed into that of metric and $SL(4,\mathbb C)$ transformation. If this $SL(4,\mathbb C)$ transformation is not anomalous, the corresponding degrees of freedom become redundant, unless there is a higher-dimensional operator that includes derivatives of $\gamma_\mu\fn{x}$ in the action. We do not delve into this issue in this paper, and choose to take the vierbein as the fundamental degrees of freedom.
}
In this sense, the vierbein degrees of freedom are more fundamental than the metric ones.
In this paper, we postulate that the dynamical degrees of freedom that describe spacetime fluctuation are the vierbein and the LL-gauge field.\footnote{
It is worth noting that supergravity also uses the vierbein and the LL-gauge field as the fundamental (bosonic) degrees of freedom; contrary to the simplest model presented here, supergravity induces torsion in general due to the presence of a (fermionic) gravitino; see, e.g.,\ Ref.~\cite{VanNieuwenhuizen:1981ae} for a review.
}
The simplest gravitational model with the vierbein and the LL-gauge fields is the Einstein-Cartan gravity; see e.g.\ Ref.~\cite{Hehl:1976kj} for a review on classical Einstein-Cartan gravity and Refs.~\cite{Daum:2010qt, Daum:2013fu} for its quantization.

In this paper, we consider a model for gravity and matter based on the LL-gauge symmetry as well as the invariance under the general-coordinate (GC) transformation [sometimes interchangeably called diffeomorphism (\diff)] at a certain energy scale~$\Lambda_\tx{G}$~\cite{Matsuzaki:2020qzf}. In particular, we postulate that its tree-level action admits the degenerate limit of the vierbein~\cite{Tseytlin:1981ks, Horowitz:1990qb}. This forbids inverse vierbeins in the action, and therefore we call it the ``irreversible vierbein postulate."

Under the irreversible vierbein postulate, only spinor fields can have kinetic terms, while the other fields become dynamical due to the quantum effects of spinor fields below $\Lambda_\tx{G}$. The main purpose of this work is to demonstrate possible generation of a spacetime background, i.e., the emergence of a nonvanishing background vierbein field, due to quantum fluctuations of the spinor field. This idea might also be viewed along the direction of pregeometry; see, e.g.,\ Refs.~\cite{Matsuzaki:2020qzf,Akama:1979tm,Akama:1994ehj,Akama:1991np,Akama:1981kq,Terazawa:1980vf,Akama:1981dk,Akama:1978pg,Wetterich:2021ywr,Wetterich:2021cyp,Wetterich:2021hru,Floreanini:1990cf,Volovik:2021wut}.

This paper is organized as follows: We start with a brief overview of degrees of freedom and symmetries in gravitational theories in Section~\ref{sec: Degrees of freedom}. In Section~\ref{sec: Local-Lorentz and general coordinate transformations}, we introduce our notation and explain transformation laws under the LL and GC transformations in detail. In particular, together with Appendices~\ref{LD, GC, and gauge transformations} and \ref{Lie derivative on spinor}, we intend to highlight differences between earlier works and ours. Then, we implement the degenerate limit on the action in Section~\ref{sec: action setting}, where we refer to Appendices~\ref{app: Degenerate limit of vierbein} and \ref{topological section} for detailed calculations. After briefly explaining transformation laws for the background fields in Section~\ref{sec: Background fields}, we demonstrate a generation of a nonvanishing flat background field of vierbein due to quantum effects of fermionic degrees of freedom in Section~\ref{sec: dynamical vierbein}. In Section~\ref{sec: Summary and discussion}, we summarize this work and discuss future prospects.

\section{Degrees of freedom of gravitational fields}
\label{sec: Degrees of freedom}
In this section, we first review the ordinary minimal Einstein gravity in the metric formalism and then in the vielbein formalism.

\subsection{Minimal Einstein gravity in metric formalism}
It is known that the Einstein-Hilbert action as the metric formalism
\al{
S_{\rm EH} =\int \df^Dx\sqrt{-g}\left[  -\Lambda_{\rm cc} + \frac{M_\tx{P}^2}{2} \mathcal R(g)\right]
\label{eq: action for EH gravity}
}
well describes the classical gravitational interactions in $D=4$. Here, $M_\tx{P}^2=1/(8\pi G_N)$ is the Planck mass squared or inverse Newtonian coupling constant and $\Lambda_{\rm cc}$ is the cosmological constant. The Ricci scalar curvature $\mathcal R(g)$ is given by the metric field $g_{\mu\nu}$ and its inverse $g^{\mu\nu}$.
The metric field is a symmetric tensor, so it classically has $D(D+1)/2$ degrees of freedom in $D$-dimensional spacetimes.

It is known that the quantum theory based on the action \eqref{eq: action for EH gravity} is nonrenormalizable in terms of the perturbative expansion of $G_N$~\cite{tHooft:1974toh}. 
The simplest explanation for the perturbative nonrenormalizability is the negative mass dimensionality of $G_N$.
This may however be somewhat naive. Indeed, the Einstein-Hilbert action in three-dimensional spacetime is renormalizable even though the Newton coupling has the negative mass dimensionality~\cite{Witten:1988hc}. This is because the Einstein-Hilbert action in three-dimensional spacetime becomes a topological theory and thus can be formulated as a Chern-Simons theory due to the peculiarity of the three-dimensional spacetime. In other words, this is because there are no propagating degrees of freedom of a graviton. The simple dimensional counting of the coupling constant cannot fully capture the property of renormalizability.

Another viewpoint why the perturbation theory for the Einstein-Hilbert action becomes nonrenormalizable is the existence of the inverse metric which is defined by
\al{
g_{\mu\lambda}g^{\lambda\nu} =\delta_\mu^\nu.
\label{eq: inverse metric}
}
This reversibility condition \eqref{eq: inverse metric} for the metric field induces an infinite number of interactions: When one considers the metric fluctuation field $h_{\mu\nu}$ around a background field $\bar g_{\mu\nu}$, namely $g_{\mu\nu}=\bar g_{\mu\nu} + h_{\mu\nu}$, the inverse metric is expanded so as to satisfy Eq.~\eqref{eq: inverse metric} and is given by    
\begin{align}
    g^{\mu\nu} = \bar g^{\mu\nu} - h^{\mu\nu} + h^\mu{}_\alpha h^{\alpha\nu} +\cdots.
    \label{eq: inverse metric expansion}
\end{align}
 This series continues infinitely around a certain background field. That is, once the inverse metric is defined by Eq.~\eqref{eq: inverse metric}, the Ricci scalar curvature in the action \eqref{eq: action for EH gravity} generally contains an infinite number of vertices of metric fluctuations, whereas all vertices have a common coupling constant $G_N$. In general, one cannot remove all the UV divergences arising from quantum loops including an arbitrary number of vertices by only a single coupling constant.

\subsection{Nonlinear and linear sigma models}
The above situation is quite similar to the $O(N)$ nonlinear sigma model which is a low-energy effective model of pions, $\pi^i$ ($i=1,\ldots, N-1$). Its action is given by
\al{
S_\text{NLS}= \frac{f_\pi^2}{2}\int \df^D x \left[-\p_\mu \pi^i \p^\mu \pi^i - \pn{\pi^i \p^2 \pi^i}^2+\cdots \right],
\label{eq: action of NLS model}
}
where $f_\pi$ is the pion decay constant. This theory can be obtained from the spontaneous symmetry breaking in the $O(N)$ linear sigma model whose action reads
\al{
S_\text{LS}=\int \df^D x \left[ -\frac{1}{2}\pn{\p_\mu \phi^i}^2 -\frac{m^2}{2}\pn{\phi^i \phi^i} -\frac{\lambda}{4}\pn{\phi^i \phi^i}^2  \right].
\label{eq: action of LS model}
}
Here, $\phi^i=\pn{\pi^j,\sigma}$ with $i=1,\ldots, N$ and $j=1,\ldots, N-1$.
For $m^2<0$, a nontrivial vacuum
\al{
\langle\phi^i\phi^i\rangle = \frac{2\ab{m^2}}{\lambda}
\label{LS constraint}
}
is realized and $O(N)$ symmetry is broken into $O(N-1)$. As a consequence, the $\sigma$ mode becomes massive and decouples from the low-energy dynamics, while $\pi^i$ are massless and remain as effective degrees of freedom in the low-energy regime. In this case, one has the constraint on fields~\eqref{LS constraint} with which integrating out the $\sigma$ mode in Eq.~\eqref{eq: action of LS model} [with the constraint $\sigma=\sqrt{f_\pi^2- (\pi^i)^2}$] results in the action of the nonlinear sigma model \eqref{eq: action of NLS model}. The decay constant just corresponds to the vacuum expectation value $f_\pi=\sqrt{2\ab{m^2}/\lambda}$.
To summarize, the nonlinear sigma model \eqref{eq: action of NLS model} is obtained from the expansion of the linear sigma model~\eqref{eq: action of LS model} around the vacuum~\eqref{LS constraint}.

An important fact is that for $D>2$, the nonlinear sigma model is perturbatively nonrenormalizable, while the linear sigma model is perturbatively renormalizable. The parameter~$f_\pi$, which arises from the consequence of the $O(N)$ symmetry breaking in the linear sigma model, is a free parameter in the nonlinear sigma model. In particular, the massless pions are realized only at the vacuum at $\langle \sigma \rangle=f_\pi$ in the linear sigma model as a consequence of the Nambu-Goldstone theorem. In this viewpoint, one has an inconsistency in the nonlinear sigma model between the massless pion condition and a free choice of $f_\pi$, and there is a range of validity for pion field fluctuations: $\pi^i\lesssim f_\pi$. This makes the system nonrenormalizable.\footnote{
Note here that the nonlinear sigma model in $D=3$ is an asymptotically safe theory; i.e., there exists a nontrivial UV fixed point at which a nonperturbatively renormalizable theory is constructed~\cite{Friedan:1980jf,Friedan:1980jm,Codello:2008qq}. Quantum gravity as metric theories could be an asymptotically safe theory as well~\cite{Hawking:1979ig,Reuter:1996cp,Souma:1999at}.
}

\subsection{Minimal Einstein gravity in vielbein formalism}
In the metric theory describing gravity, Eq.~\eqref{eq: inverse metric} may be regarded as the constraint analogous to Eq.~\eqref{LS constraint}.
Following the argument above, the inconsistency at high energies in the metric formalism may be between an expansion of the metric field around a background field, e.g.\ a flat background metric $\langle g_{\mu\nu}\rangle =\bar g_{\mu\nu}=\eta_{\mu\nu}$, and the existence of massless metric fields. We expect that there exists an appropriate high-energy theory of the metric theory and the generation of a vacuum $\langle g_{\mu\nu}\rangle =\bar g_{\mu\nu}\neq0$ may imply the appearance of the massless metric field. 
The reversibility condition \eqref{eq: inverse metric} for the metric field enforces a finite domain where quantum fields can fluctuate analogously to $\pi^i\lesssim f_\pi$ in the nonlinear sigma model. Beyond such a domain, we expect that new degrees of freedom appear and participate in the dynamics.

We notice at this point that the gravitational theory with the action \eqref{eq: action for EH gravity} is similar to the nonlinear sigma model \eqref{eq: action of LS model}. Hence, we intend to construct a gravitational theory with new additional degrees of freedom analogous to the meson $\sigma$ in the linear sigma model.

Having this viewpoint in mind, we are motivated to consider a theory for gravity with more degrees of freedom. Let us here deal with a formulation for the gravitational theory with vielbein $e^\ba{}_\mu\fn{x}$ and LL-gauge field $\omega^\ba{}_{\bb\mu}\fn{x}$, known as the Einstein-Cartan gravity based on $SO(1,d)$ LL symmetry. Its minimal form of action is given by
\al{
S_\text{EC} = \int \df^Dx\,\ab{e(x)} \left[ -\Lambda_{\rm cc} + {\MP^2\ov2}e_{\ba}{}^\mu\fn{x}e_{\bb}{}^\nu\fn{x}\os\omega{\mc F}^{\ba\bb}{}_{\mu\nu}\fn{x}\right],
\label{eq: action for EC gravity}
}
where $\ab{e(x)}=\det_{\ba,\mu}e^\ba{}_\mu(x)$ is the determinant of the vielbein, and $\os\omega{\mc F}^{\ba\bb}{}_{\mu\nu}\fn{x}$ is the field strength of LL-gauge field $\omega^\ba{}_{\bb\mu}\fn{x}$. Note here that LL indices $\ba,\bb,\dots$ are antisymmetric in $\omega_{\ba\bb\mu}\fn{x}$ and $\os\omega{\mc F}_{\ba\bb\mu\nu}\fn{x}$, namely, $\omega_{\ba\bb\mu}\fn{x}=-\omega_{\bb\ba\mu}\fn{x}$ and $\os\omega{\mc F}_{\ba\bb\mu\nu}\fn{x}=-\os\omega{\mc F}_{\bb\ba\mu\nu}\fn{x}$, due to the $SO(1,d)$ algebra; see Section~\ref{section on miscellaneous points} below for details.
It is worth stressing that the action \eqref{eq: action for EC gravity} does not contain the LL-gauge kinetic term $\os\omega{\mc F}_{\ba\bb\mu\nu}\fn{x}\os\omega{\mc F}^{\ba\bb}{}^{\mu\nu}\fn{x}$, whereas the existence of the term $e_\ba{}^\mu\fn{x}e_\bb{}^\nu\fn{x}\os\omega{\mc F}^{\ba\bb}{}_{\mu\nu}\fn{x}$ is peculiar in the Einstein-Cartan theory, as compared to an ordinary gauge theory that does not have such a term.

In this formulation, it seems that there are apparently $D(3D-1)/2$ independent classical degrees of freedom because $e^{\ba}{}_\mu\fn{x}$ and $\omega^\ba_{\bb\mu}\fn{x}$ have $D^2$ and $D(D-1)/2$ degrees of freedom, respectively. In the action \eqref{eq: action for EC gravity}, however, there are no apparent kinetic terms for the vielbein or the LL-gauge field, so these fields are auxiliary fields at this stage; i.e., they are not dynamical degrees of freedom yet.
Imposing the equation of motion on $\omega^\ba{}_{\bb\mu}\fn{x}$, i.e.\ $\delta S_\text{EC}/\delta \omega_\mu=0$, one obtains its solution to $\omega^\ba{}_{\bb\mu}\fn{x}$ as a function of the vielbein, namely, the Levi-Civita spin connection: $\omega^\ba{}_{\bb\mu}\fn{x}=\os e\Omega^\ba{}_{\bb\mu}\fn{x}$; see Eq.~\eqref{Levi-Civita spin connection} below for its explicit definition. By substituting this into $\omega^\ba{}_{\bb\mu}\fn{x}$ in the field strength, the LL-gauge field disappears from the action, and the action is written in terms of only vielbein with the kinetic term.
At this point, the second term on the right-hand side of Eq.~\eqref{eq: action for EC gravity} just turns into the Einstein-Hilbert term written in terms of the vielbein, and the number of dynamical degrees of freedom is $D(D+1)/2$ which is the same as that of the symmetric metric field, being the composite of vielbein fields $g_{\mu\nu}\fn{x}=\eta_{\ba\bb}e^\ba{}_\mu\fn{x} e^\bb{}_\nu\fn{x}$.

From the fact above, it turns out that a certain condition between $e^\ba{}_\mu\fn{x}$ and $\omega^\ba{}_{\bb\mu}\fn{x}$, such as the equation of motion for $\omega^\ba{}_{\bb\mu}\fn{x}$, reduces the original (tree-level) auxiliary degrees of freedom to the ``dynamical'' ones that have the kinetic term. A question now is whether such a condition can be generalized or not. 

In this paper, we advocate the irreversible vierbein postulate~\cite{Matsuzaki:2020qzf}: At a certain energy scale~$\Lambda_{\rm G}$, the action for gravity must admit the degenerate limit of the vielbein~\cite{Tseytlin:1981ks,Horowitz:1990qb,Floreanini:1991cw} in which an arbitrary set of eigenvalues of the vielbein goes to zero, and hence the inverse vielbein cannot be defined. In a sense, the action at $\Lambda_{\rm G}$ corresponds to the linear sigma model~\eqref{eq: action of LS model}. 
The irreversible vierbein postulate shares the same assumption that, in the language of the linear sigma model, we do not take into account inverse powers of $O(N)$ invariant $\phi^i\phi^i$ such as $(\phi^i\phi^i)^{-1}$ and $(\phi^i\phi^i)^{-2}$. Even though they do not spoil renormalizability in terms of power counting, they do prevent defining the symmetric phase $\langle \phi^i\phi^i\rangle=0$. In this sense, the irreversible vierbein postulate ensures a well-defined symmetric phase $\langle g_{\mu\nu}\rangle=0$. 

Indeed, solving the equation of motion for the auxiliary field $\omega^\ba{}_{\bb\mu}\fn{x}$ requires the inverse vielbein. Thus, at $\Lambda_{\rm G}$, we cannot impose the equation of motion. If one introduced the inverse vielbein {\it a priori}, it could kinematically reduce the degrees of freedom. By contrast, in our postulate, we claim that the reduction of degrees of freedom takes place dynamically below the scale $\Lambda_\tx{G}$. Hence, the inverse vielbein is defined thanks to quantum dynamics.

We consider a gravitational theory which is based on $SO(1,d) \times \tx{GC}$ in the degenerate limit which entails $\langle g_{\mu\nu}\rangle= 0$ at the tree level. Its dynamics realizes $\langle g_{\mu\nu}\rangle\neq 0$ and the massless metric field as a consequence of spontaneous symmetry breaking: $SO(1,d) \times \tx{GC} \to \tx{GC}$. In the following sections, we explain the transformation laws under the LL and GC transformations and introduce corresponding gauge fields in detail.

\section{Local-Lorentz and general coordinate transformations}
\label{sec: Local-Lorentz and general coordinate transformations}

In this section, we clarify how the fields transform under the LL and GC symmetries in details. 
In Sec.~\ref{Field content section}, we spell out the field content.
In Sec.~\ref{LL tf section} and \ref{GC tf section}, we present their transformation laws under the LL and GC symmetries, respectively.
In Sec.~\ref{section on miscellaneous points}, we show the field strength for the LL-gauge field and argue that we do not need an extra GC-gauge field or its field strength.
Through this section, we work in $d+1$ spacetime dimensions. Later, we will specify $d=3$ when constructing a concrete action.

This section is intended to be mainly a review; see, e.g.,\ Refs.~\cite{Hehl:1976kj,Hehl:1994ue} for further details. Nevertheless, to our best knowledge, the following are the first to be clearly stated in our paper comparing with the literature:
\begin{itemize}
\item The reduction condition of $GL(4)$ to GC in Eq.~\eqref{GC condition on M}
\item The fact that the antisymmetric part becomes unnecessary for GC as shown in Eq.~\eqref{needs only symmetric part}
\item The $GL(4)$ field strength being a differential operator as shown in Eq.~\eqref{GL field strength different from GC one} 
\item The distinction between our GC and what we call the LD transformations.
\end{itemize}

\subsection{Field content}\label{Field content section}
We introduce the fields and symmetries to clarify our notations and to construct an action.
Our starting assumption is that at a certain scale $\Lambda_\tx{G}$, the action enjoys the LL and GC symmetries. In particular, the gravitational sector consists of the vielbein (vierbein in four dimensional spacetime) and the LL-gauge field.

The gravity sector consists of the vielbein field $e^\ba{}_\mu\fn{x}$ and the LL-gauge field $\omega^\ba{}_{\bb\mu}\fn{x}$, where $\mu,\nu,\dots$ ($\ba,\bb,\dots$) run for the spacetime (tangent-space) indices $0,\dots,d$ ($\tb0,\dots,\bs d$).
Here and hereafter, we make the dependence on a specific coordinate system $x^\mu$ explicit on each chart, unless otherwise stated, since it is anyway necessary for any realistic calculation of a dynamical quantity; this will make a distinction between a variable and constant more apparent.
From the vielbein, we construct the metric field
\al{
g_{\mu\nu}\fn{x}
	&=	\eta_{\ba\bb}\,e^\ba{}_\mu\fn{x}e^\bb{}_\nu\fn{x},
}
where the tangent-space metric and its inverse are
\al{
\bmat{\eta_{\ba\bb}}_{\ba,\bb=\tb0,\dots,\bs d}
	=	\bmat{\eta^{\ba\bb}}_{\ba,\bb=\tb0,\dots,\bs d}
	&=	\diag\fn{-1,1,\dots,1},
	\label{tangent-space metric}
}
in which ``diag'' denotes the corresponding diagonal matrix.
We note that the metric field $g_{\mu\nu}\fn{x}$ can be constructed without using the inverse vielbein field $e_\ba{}^\mu\fn{x}$, whereas construction of an inverse metric field~$g^{\mu\nu}\fn{x}$ does require the inverse vielbein.

The matter sector of an effective field theory consists of scalar, spinor, and 1-form fields $\phi\fn{x}$, $\psi\fn{x}$ and $\mc A_\mu\fn{x}$ with spin-0, -1/2, and -1, respectively.\footnote{
The existence of a nearly massless spin-3/2 field, gravitino, implies nearly unbroken local supersymmetry, supergravity, which does not seem to be realized in our Universe at low energies. It may still be interesting to include it since our scale $\Lambda_\tx{G}$ is supposed to be much higher than the electroweak one; see Appendix~\ref{Rarita-Schwinger field}.
}
Precisely speaking, $\mc A_\mu\fn{x}$ are the components of the 1-form field $\mc A\fn{x}:=\mc A_\mu\fn{x}\df x^\mu$, but we sloppily call these components a 1-form field too.
Below, $\Psi\fn{x}$ will denote either $\phi\fn{x}$ or $\psi\fn{x}$ fields collectively.
Also, $\Phi\fn{x}$ will denote any one of the fields, including both the gravity and matter sectors.

Here, we take the 1-form field $\mc A_\mu\fn{x}$ rather than the corresponding vector field $\mc A^\mu\fn{x}:=g^{\mu\nu}\fn{x}\mc A_\nu\fn{x}$ as a fundamental degree of freedom
because the former rather than the latter primarily appears in a gauge covariant derivative\footnote{
For a compact gauge group $G$ with the corresponding Lie algebra $\mf g$, one usually writes
$\mc A_\mu\fn{x}=ig_GA_\mu\fn{x}=ig_GA^a_\mu\fn{x}T^a$ where $g_G$ is the gauge coupling and $T^a$ ($a=1,\dots,\dim\mf g$) are the Hermitian generators of the gauge symmetry. In particular, the kinetic terms of $\mc A_\mu\fn{x}$ and $A_\mu^a\fn{x}$ have opposite signs.
\label{sign of guage kinetic}
}
\al{
\os{\mc A}{\mc D}_\mu:=\p_\mu+\mc A_\mu\fn{x}.
	\label{gauge covariant derivative}
}
More explicitly, on a field $\Psi\fn{x}$ in the fundamental representation of a gauge group, 
\al{
\Psi\fn{x}\to U\fn{x}\Psi\fn{x},
\label{eq: gauge transformation for psi}
}
the covariant derivative $\os{\mc A}{\mc D}_\mu=\p_\mu+\mc A_\mu\fn{x}$ transforms covariantly,
\al{
\os{\mc A}{\mc D}_\mu\Psi\fn{x}
	&\to	U\fn{x}\os{\mc A}{\mc D}_\mu\Psi\fn{x},
		\label{covariance of D}
}
due to the gauge transformation of the gauge field $\mc A_\mu\fn{x}$:
\al{
\mc A_\mu\fn{x}\to U\fn{x}\mc A_\mu\fn{x}U^{-1}\fn{x}-\p_\mu U\fn{x}U^{-1}\fn{x},
	\label{ordinary gauge field transformed}
}
or for an infinitesimal transformation $U\fn{x}=\I+\vartheta\fn{x}$,
\al{
\mc A_\mu\fn{x}
	\to	\mc A_\mu\fn{x}
		+\commutator{\vartheta\fn{x}}{\mc A_\mu\fn{x}}-\p_\mu\vartheta\fn{x},
		\label{infinitesimal ordinary gauge}
}
where $\I$ is the identity matrix and the commutator is $\commutator AB:=AB-BA$.

In the following sections, for totally symmetric and antisymmetric tensors denoted here by $\mathcal S$ and $\mathcal A$, respectively, we use the notation
\al{
&\mathcal S_{\mu_1\cdots \mu_n} =\mathcal S_{(\mu_1\cdots \mu_n)},&
&\mathcal A_{\mu_1\cdots \mu_n} =\mathcal A_{[\mu_1\cdots \mu_n]}.
}
In particular, second-rank tensors are given by\footnote{
Note that in this notation, $\anticommutator{\mathcal S_\mu}{\mathcal S_\nu}=2\mathcal S_{(\mu}\mathcal S_{\nu)}$ and $\commutator{\mathcal A_\mu}{\mathcal A_\nu}=2\mathcal A_{[\mu}\mathcal A_{\nu]}$.
\label{normalization footnote}
}
\al{
\mathcal S_{\mu\nu} &= \mathcal S_{(\mu\nu)} =\frac{\mathcal S_{\mu\nu}+\mathcal S_{\nu\mu}}{2},&
\mathcal A_{\mu\nu} &= \mathcal A_{[\mu\nu]} =\frac{\mathcal A_{\mu\nu}-\mathcal A_{\nu\mu}}{2}.
\label{eq: symmetric and antisymmetric notation}
}

\subsection{Local Lorentz transformation}\label{LL tf section}
Here, we review the LL transformations on various fields to spell out our notation.
The LL transformation is a local rotation of the tangent-space basis; therefore, in particular, it does not act on a spacetime-scalar field.

\subsubsection{LL transformation on the gravity sector}
Under an LL transformation that satisfies the defining relation of the $SO(1,d)$ symmetry
\al{
\Lambda^\bc{}_\ba\fn{x}\eta_{\bc\bd}\,\Lambda^\bd{}_\bb\fn{x}
	&=	\eta_{\ba\bb},
	\label{LL defining relation with indices}
}
the gravitational fields transform as
\al{
e^\ba{}_\mu\fn{x}
	&\sr{\tx{LL}}\to	\Lambda^\ba{}_\bb\fn{x} e^\bb{}_\mu\fn{x},\\
\omega^\ba{}_{\bb\mu}\fn{x}
	&\sr{\tx{LL}}\to	\Lambda^\ba{}_\bc\fn{x}\omega^\bc{}_{\bd\mu}\fn{x}\pa{\Lambda^{-1}}^\bd{}_\bb\fn{x}
			-\p_\mu\Lambda^\ba{}_\bc\fn{x}\pa{\Lambda^{-1}}^\bc{}_\bb\fn{x},
			\label{omega under LL with indices}
}
where $\pa{\Lambda^{-1}}^\ba{}_\bb=\Lambda_\bb{}^\ba$, with their indices being lowered and raised by the tangent-space metric and its inverse~\eqref{tangent-space metric}.
Here and hereafter, a derivative such as $\p_\mu:={\p\ov\p x^\mu}$ acts only on its neighbor:
\al{
\p_\mu AB
	&:=	\pn{\p_\mu A}B,&
\p_\mu\pn{AB}C
	&:=	\Pn{\p_\mu\pn{AB}}C,
	\label{neighboring notation}
}
etc.

We will also use a matrix notation such as
\al{
\sqbr{\omega_\mu\fn{x}}{}^\ba{}_\bb
	&:=	\omega^\ba{}_{\bb\mu}\fn{x},
		\label{matrix notation}
}
leading to
\al{
\Lambda^\t\fn{x}\eta\,\Lambda\fn{x}
	&=	\eta,
	\label{LL defining relation}
}
and
\al{
e_\mu\fn{x}
	&\sr{\tx{LL}}\to	\Lambda\fn{x}e_\mu\fn{x},\\
\omega_\mu\fn{x}
	&\sr{\tx{LL}}\to	\Lambda\fn{x}\omega_\mu\fn{x}\Lambda^{-1}\fn{x}
			-\p_\mu\Lambda\fn{x}\Lambda^{-1}\fn{x},
			\label{omega under LL}
}
where the superscript ``t'' denotes the transpose.

For an infinitesimal transformation $\Lambda\fn{x}=\I+\theta\fn{x}$ in the matrix notation, or more explicitly $\Lambda^\ba{}_\bb\fn{x}=\delta^\ba_\bb+\theta^\ba{}_\bb\fn{x}$, we have
\al{
\omega_\mu\fn{x}
	&\sr{\tx{LL}}\to
		\omega_\mu\fn{x}
		+\commutator{\theta\fn{x}}{\omega_\mu\fn{x}}-\p_\mu\theta\fn{x},
		\label{infinitesimal LL in matrix}
}
or more explicitly
\al{
\omega^\ba{}_{\bb\mu}\fn{x}
	&\sr{\tx{LL}}\to
		\omega^\ba{}_{\bb\mu}\fn{x}
		+\theta^\ba{}_\bc\fn{x}\omega^\bc{}_{\bb\mu}\fn{x}
		-\omega^\ba{}_{\bc\mu}\fn{x}\theta^\bc{}_\bb\fn{x}
		-\p_\mu\theta^\ba{}_\bb\fn{x}.
		\label{infinitesimal LL more explicit}
}
Note that the defining relation for $SO(1,d)$ in Eq.~\eqref{LL defining relation with indices}, or \eqref{LL defining relation}, implies the antisymmetry $\theta_{\bb\ba}\fn{x}=-\theta_{\ba\bb}\fn{x}$.

To summarize, the vielbein transforms as a fundamental representation of the LL symmetry, while being a spacetime 1-form. Recall that the Higgs field transforms as a fundamental representation of gauge symmetry while being a spacetime scalar. On the other hand, the LL-gauge field transforms just as an $SO(1,d)$ gauge field under the LL symmetry.

It is the transformation~\eqref{omega under LL with indices}, or \eqref{omega under LL}, that makes the LL-covariant derivative on an LL-vector (spacetime-scalar) field $V^\ba\fn{x}$,
\al{
\os\omega{\mc D}_\mu V^\ba\fn{x}
	&:=	\p_\mu V^\ba\fn{x}+\omega^\ba{}_{\bb\mu}\fn{x}V^\bb\fn{x},
		\label{LL-covariant derivative}
}
to be covariant:
\al{
\os\omega{\mc D}_\mu V^\ba\fn{x}
	&\sr{\tx{LL}}\to	\Lambda^\ba{}_\bb\fn{x}\os\omega{\mc D}_\mu V^\bb\fn{x}.
		\label{covariance of LL-covariant D}
}
In the matrix notation, the above equations read
\al{
\os\omega{\mc D}_\mu V\fn{x}
	&:=	\Sqbr{\p_\mu+\omega_\mu\fn{x}}V\fn{x}
		\label{LL-covariant derivative in matrix notation}
}
and
\al{
\os\omega{\mc D}_\mu V\fn{x}
	&\sr{\tx{LL}}\to	\Lambda\fn{x}\os\omega{\mc D}_\mu V\fn{x}.
	\label{covariance of LL-covariant D in matrix notation}
}

\subsubsection{LL transformation on the matter sector}
Now we turn to the matter fields.
The bosonic matter fields transform as a scalar (namely, do not transform) under the LL symmetry:
\al{
\phi\fn{x}
	&\sr{\tx{LL}}\to	\phi\fn{x},\\
\mc A_\mu\fn{x}
	&\sr{\tx{LL}}\to	\mc A_\mu\fn{x}.
}
We here comment on the relation to the irreversible vierbein postulate which we will impose on the action.
The 1-form field $\mc A_\mu\fn{x}$ can be regarded as a composite field $\mc A_\mu\fn{x}=\mc A_\ba\fn{x}e^\ba{}_\mu\fn{x}$ made of the vielbein $e^\ba{}_\mu\fn{x}$ and an LL-vector\footnote{
For the LL symmetry, we call both the covariant vector $V_\ba$ and the contravariant vector $V^\ba$ the LL-covariant vectors.
}
spacetime scalar $\mc A_\ba\fn{x}$ that transforms as
\al{
\mc A_\ba\fn{x}
	&\sr{\tx{LL}}\to	\mc A_\bb\fn{x}\Lambda^\bb{}_\ba\fn{x}.
 \label{LL of ordinary gauge field}
}
Even when we regard $\mc A_\ba\fn{x}$ as a fundamental degree of freedom, we can always construct $\mc A_\mu\fn{x}$ without contradicting the irreversible vierbein postulate.
In contrast, if starting from the 1-form field $\mc A_\mu\fn{x}$, we need the inverse vielbein field $e_\ba{}^\mu\fn{x}$ to construct the LL-vector spacetime-scalar field $\mc A_\ba\fn{x}=e_\ba{}^\mu\fn{x}\mc A_\mu\fn{x}$.
That is, we cannot reconstruct $\mc A_\ba\fn{x}$ from $\mc A_\mu\fn{x}$ under the irreversible vierbein postulate at the scale $\Lambda_\tx{G}$.

The fermionic matter field, spinor, transforms nontrivially under the LL symmetry.
In the matrix notation, we may parametrize an LL transformation as
\al{
\Lambda\fn{x}
	&=	e^{\theta\pn{x}},
}
that is,
\al{
\Lambda^\ba{}_\bb\fn{x}
	&=	\delta^\ba_\bb+\theta^\ba{}_\bb\fn{x}+{1\ov2!}\theta^\ba{}_\bc\fn{x}\theta^\bc{}_\bb\fn{x}+\cdots.
}
Now the spinor field transforms as
\al{
\psi\fn{x}
	&\sr{\tx{LL}}\to	S\Fn{\Lambda\fn{x}}\psi\fn{x},
}
where we define
\al{
S\fn{e^{\theta\pn{x}}}
	&:=	e^{{1\ov2}\theta_{\ba\bb}\fn{x}\sigma^{\ba\bb}},
}
in which the LL generators on the spinor representation are
\al{
\sigma^{\ba\bb}
	&:=	{\commutator{\gamma^\ba}{\gamma^\bb}\ov4},
}
with $\gamma^\ba$ being the gamma matrices that obey the Clifford algebra:
\al{
\anticommutator{\gamma^\ba}{\gamma^\bb}
	&=	2\eta^{\ba\bb}\I.
}
Here the anticommutator is defined by $\anticommutator{A}{B}:=AB+BA$.

Among the matter fields, the spinor field is the only nontrivial representation under the LL symmetry. As stressed in the Introduction, the LL symmetry is necessary to define a spinor field at all on a curved spacetime.
Note also that in our treatment, the LL symmetry is no different from the ordinary gauge symmetry other than it is under a noncompact group $SO(1,d)$.

\subsection{General coordinate transformation}\label{GC tf section}
Next, we discuss the GC transformation $x^\mu\to x^{\pr\mu}\fn{x}$, where, throughout this paper, the prime symbol $'$ exclusively denotes a quantity after the GC transformation and not a derivative.
Accordingly, the bases for 1-form and spacetime vector transform as
\al{
\df x^\mu
	&\sr{\tx{GC}}\to	\df x^{\pr\mu}={\p x^{\pr\mu}\ov\p x^\nu}\df x^\nu,\\
\p_\mu
	&\sr{\tx{GC}}\to	\p'_\mu={\p x^\nu\ov\p x^{\pr\mu}}\p_\nu.
}
The GC transformation is generally identified with 
\diff, while it is often said that the transformation under \diff\ is given by the Lie-derivative (LD) transformation.
One might regard that these three transformations were equivalent.
Strictly speaking, however, GC/\diff\ and the LD transformation should be distinguished. Indeed, the GC transformation introduced in this section is not given by the LD. A detailed comparison between the GC transformation and {\diff} is given in Appendix~\ref{LD, GC, and gauge transformations}.
In the main body of this paper, we use the terminology ``GC" rather than \diff.

In a matrix notation
\al{
M^\mu{}_\nu\fn{x}
	&:=	{\p x^{\pr\mu}\ov\p x^\nu},&
\pa{M^{-1}}^\nu{}_\lambda\fn{x}
	&=	{\p x^\nu\ov\p x^{\pr\lambda}},
		\label{M defined}
}
and writing similarly to Eq.~\eqref{matrix notation} such as $\mt{M\fn{x}}^\mu{}_\nu:=M^\mu{}_\nu\fn{x}$,
the above GC transformations on the bases read
\al{
\df x^\mu
	&\sr{\tx{GC}}\to	\df x^{\pr\mu}=M^\mu{}_\nu\fn{x}\df x^\nu,\\
\p_\mu
	&\sr{\tx{GC}}\to	\p'_\mu=\mt{M^{-1}\fn{x}}^\nu{}_\mu\,\p_\nu.
}
It is important that the ``matrix'' $M$ satisfies the extra ${d\pn{d+1}^2\ov2}$ conditions
\al{
\p_{[\lambda}M^\mu{}_{\nu]}\fn{x}
	&=	0
		\label{GC condition on M}
}
for the GC transformation~\eqref{M defined}.
Conversely, it is also true that any function $M^\mu{}_\nu\fn{x}$ that satisfies the condition~\eqref{GC condition on M} can always be written (locally) in terms of $(d+1)$ functions $x^{\pr\mu}\fn{x}$ ($\mu=0,\dots,d$) as in Eq.~\eqref{M defined}.
The transformation by $M^\mu{}_\nu$ without the condition \eqref{GC condition on M} corresponds to the general linear (GL) transformation, i.e., $GL(d+1)$.

From $\p_\lambda (M^{-1}M)=0$, we obtain $\p_\lambda M^{-1}\,M=-M^{-1}\p_\lambda M$, or specifying indices, it is given by $\p_\lambda\pa{M^{-1}}^\mu{}_\rho\,M^\rho{}_\nu=-\pa{M^{-1}}^\mu{}_\rho\p_\lambda M^\rho{}_\nu$.
By antisymmetrizing $\lambda$ and $\nu$, we get
\al{
M^\rho{}_{[\nu}\p_{\lambda]}\pa{M^{-1}}^\mu{}_\rho=0.
	\label{GC condition 2}
}
Similarly, the derivative of the inverse function gives
\al{
0	&=	{\p^2x^\lambda\ov\p x^{\pr[\mu}\p x^{\pr\nu]}}
	=	{\p x^\alpha\ov\p x^{\pr[\mu|}}{\p\ov\p x^\alpha}{\p x^\lambda\ov\p x^{\pr|\nu]}}
	=	\p_\alpha\pn{M^{-1}}^\lambda{}_{[\nu}\pn{M^{-1}}^\alpha{}_{\mu]},
		\label{GC condition 3}
}
where vertical lines (between the antisymmetrization symbols in indices) denote that the indices between the vertical lines are not antisymmetrized. For instance, in Eq.~\eqref{GC condition 3}, the index $\alpha$ between the vertical lines is not antisymmetrized.
In actual computations such as will be done in Eq.~\eqref{Levi-Civita transformed}, it is more convenient to use the coordinate notation on the right-hand sides in Eq.~\eqref{M defined} rather than to use these relations~\eqref{M defined}, \eqref{GC condition 2}, \eqref{GC condition 3}, etc.\ in the matrix notation on the left-hand sides in Eq.~\eqref{M defined}.
The matrix notation is of use for more conceptual understanding.

\subsubsection{GC transformation on fields}
Under the GC transformation, the gravitational fields transform as
\al{
e^\ba{}_\mu\fn{x}
	&\sr{\tx{GC}}\to	e^{\pr\ba}{}_\mu\fn{x'}=e^\ba{}_\nu\fn{x}{\p x^\nu\ov\p x^{\pr\mu}},
 \label{eq: GC for vielbein}
 \\
\omega^\ba{}_{\bb\mu}\fn{x}
	&\sr{\tx{GC}}\to	\omega^{\pr\ba}{}_{\bb\mu}\fn{x'}=\omega^\ba{}_{\bb\nu}\fn{x}{\p x^\nu\ov\p x^{\pr\mu}},
}
and the matter fields as
\al{
\phi\fn{x}
	&\sr{\tx{GC}}\to	\phi'\fn{x'}=\phi\fn{x},
  \label{eq: GC for scalar}
 \\
\psi\fn{x}
	&\sr{\tx{GC}}\to	\psi'\fn{x'}=\psi\fn{x},
 \label{eq: GC for spinor}\\
\mc A_\mu\fn{x}
	&\sr{\tx{GC}}\to	\mc A'_\mu\fn{x'}=\mc A_\nu\fn{x}{\p x^\nu\ov\p x^{\pr\mu}}.
  \label{eq: GC for gauge field}
}
The transformed scalar field $\phi'\fn{x'}$ is defined to satisfy $\phi'\Fn{x'\fn{x}}=\phi\fn{x}$ such that the pullback of the function $\phi'\fn{x'}$ by the function $x'\fn{x}$ becomes $\phi\fn{x}$.
Equivalently, the pullback of the function $\phi\fn{x}$ by the inverse function $x\fn{x'}$ is $\phi'\fn{x'}$, namely, $\phi\Fn{x\fn{x'}}=\phi'\fn{x'}$.
Here, the spinor field also transforms the same as the scalar field (namely as the pullback of the function) under the GC transformation; see Appendix~\ref{Lie derivative on spinor} for another point of view and our opinion on it. Note that the spinor field transforms as scalar under the GC transformation because it does not have any spacetime index, while the ``LD transformation'' gives different transformation laws from Eq.~\eqref{eq: GC for spinor}. They are discussed in Appendix~\ref{Lie derivative on spinor}.

In the matrix notation, with Eq.~\eqref{M defined}, the GC transformation is, on the gravitational fields,
\al{
e^\ba{}_\mu\fn{x}
	&\sr{\tx{GC}}\to	e^{\pr\ba}{}_\mu\fn{x'}=e^\ba{}_\nu\fn{x}\mt{M^{-1}\fn{x}}^\nu{}_\mu,\\
\omega^\ba{}_{\bb\mu}\fn{x}
	&\sr{\tx{GC}}\to	\omega^{\pr\ba}{}_{\bb\mu}\fn{x'}=\omega^\ba{}_{\bb\nu}\fn{x}\mt{M^{-1}\fn{x}}^\nu{}_\mu,
}
and, on the matter fields,
\al{
\phi\fn{x}
	&\sr{\tx{GC}}\to	\phi'\fn{x'}=\phi\fn{x},\\
\psi\fn{x}
	&\sr{\tx{GC}}\to	\psi'\fn{x'}=\psi\fn{x},\\
\mc A_\mu\fn{x}
	&\sr{\tx{GC}}\to	\mc A'_\mu\fn{x'}=\mc A_\nu\fn{x}\mt{M^{-1}\fn{x}}^\nu{}_\mu.
}
In the matrix notation, a spacetime vector $V^\mu$ transforms like a fundamental representation under the GC transformation: $V^\mu\to M^\mu{}_\nu V^\nu$.

\subsubsection{GC-gauge field}\label{GC gauge field section}
Now we want to define a GC-covariant derivative.
To this end, let us first suppose that there exists a GC-gauge field that transforms as
\al{
\Upsilon_\mu\fn{x}
	&\sr{\tx{GC}}\to	\Upsilon'_\mu\fn{x'}
	=	\mt{M\fn{x}\Upsilon_\nu\fn{x}M^{-1}\fn{x}-\p_\nu M\fn{x}M^{-1}\fn{x}}\mt{M^{-1}\fn{x}}^\nu{}_\mu,
		\label{transformation of GC gauge field in matrix}
}
where we employ the matrix notation $\mt{\Upsilon_\mu\fn{x}}^\alpha{}_\beta:=\Upsilon^\alpha{}_{\beta\mu}\fn{x}$ similarly to Eq.~\eqref{matrix notation}. More explicitly, the transformation~\eqref{transformation of GC gauge field in matrix} means
\al{
\Upsilon^\alpha{}_{\beta\mu}\fn{x}
	&\sr{\tx{GC}}\to	\Upsilon^{\pr\alpha}{}_{\beta\mu}\fn{x'}
	=	\pn{M^\alpha{}_\gamma\fn{x}\Upsilon^\gamma{}_{\delta\nu}\fn{x}\pa{M^{-1}}^\delta{}_\beta\fn{x}-\p_\nu M^\alpha{}_\gamma\fn{x}\pa{M^{-1}}^\gamma{}_\beta\fn{x}}\pa{M^{-1}}^\nu{}_\mu\fn{x}.
		\label{transformation of GC gauge field more explicit}
}
Here, we stress that the difference from the gauge transformations of the ordinary and LL-gauge fields in Eqs.~\eqref{ordinary gauge field transformed} and \eqref{omega under LL with indices} [or \eqref{omega under LL}], respectively, is the last $M^{-1}$ factor that rotates the spacetime index too.

Then, one can construct a GC-covariant derivative on a spacetime-vector field $V^\mu\fn{x}$ and a 1-form field $W_\mu\fn{x}$: In the matrix notation, we write
\al{
\mt{\os\Upsilon\nabla_\mu V\fn{x}}^\alpha
	&:=	\mt{\Pn{\p_\mu+\Upsilon_\mu\fn{x}}V\fn{x}}^\alpha,
		\label{GC-covariant derivative on vector in matrix notation}\\
\mt{\os\Upsilon\nabla_\mu W\fn{x}}_\alpha
	&:=	\mt{ W\fn{x}\pn{\ola{\p_\mu}-\Upsilon_\mu\fn{x}}}_\alpha,
}
where the left derivative reads $A\ola{\p_\mu}:=\p_\mu A$, with the neighboring notation $AB\ola{\p_\mu}:=A\pn{\p_\mu B}$, $A\pn{BC}\ola{\p_\mu}:=A\Pn{\p_\mu\pn{BC}}$, etc.,\ similar to Eq.~\eqref{neighboring notation}. More explicitly, they are expressed as
\al{
\os\Upsilon\nabla_\mu V^\alpha\fn{x}
	&:=	\p_\mu V^\alpha\fn{x}+\Upsilon^\alpha{}_{\beta\mu}\fn{x}V^\beta\fn{x},
		\label{GC-covariant derivative on vector}\\
\os\Upsilon\nabla_\mu W_\alpha\fn{x}
	&:=	\p_\mu W_\alpha\fn{x}-W_\beta\fn{x}\Upsilon^\beta{}_{\alpha\mu}\fn{x}.
}
It is straightforward to check their covariance under the GC transformation: In the matrix notation,
\al{
\os\Upsilon\nabla_\mu V\fn{x}
	\sr{\tx{GC}}\to	\os\Upsilon\nabla'_\mu V'\fn{x'}
	&=	\pn{\p_\mu'+\Upsilon'_\mu\fn{x'}}V'\fn{x'}\nn
	&=	\sqbr{\pn{\p_\nu+M\Upsilon_\nu M^{-1}-\p_\nu MM^{-1}}\Pn{MV}}\pa{M^{-1}}^\nu{}_\mu\nn
	&=	\sqbr{M\os\Upsilon\nabla_\nu V}\pa{M^{-1}}^\nu{}_\mu,\\
\os\Upsilon\nabla_\mu W\fn{x}
	\sr{\tx{GC}}\to	\os\Upsilon\nabla'_\mu W'\fn{x'}
	&=	W'\fn{x'}\pn{\ola{\p_\mu'}-\Upsilon'_\mu\fn{x'}}\nn
	&=	\sqbr{\pn{WM^{-1}}\pn{\ola{\p_\nu}-M\Upsilon_\nu M^{-1}+\p_\nu MM^{-1}}}\pa{M^{-1}}^\nu{}_\mu\nn
	&=	\sqbr{\os\Upsilon\nabla_\nu W\,M^{-1}}\pa{M^{-1}}^\nu{}_\mu,
}
where we have suppressed the dependence on $x$ on the right-hand side and have used the identity $\p_\mu\pn{MM^{-1}}=\p_\mu M\,M^{-1}+M\p_\mu M^{-1}=0$. Recall that we are employing the neighboring notation for derivatives as given in Eq.~\eqref{neighboring notation}. More explicitly, the above transformations read
\al{
\os\Upsilon\nabla_\mu V^\alpha\fn{x}
	&\sr{\tx{GC}}\to	M^\alpha{}_\beta\fn{x}\,\os\Upsilon\nabla_\nu V^\beta\fn{x}\,\pa{M^{-1}}^\nu{}_\mu\fn{x},\\
\os\Upsilon\nabla_\mu W_\alpha\fn{x}
	&\sr{\tx{GC}}\to	\os\Upsilon\nabla_\nu W_\beta\fn{x}\,\pa{M^{-1}}^\beta{}_\alpha\fn{x}\,\pa{M^{-1}}^\nu{}_\mu\fn{x}.
}

We may separate the GC-gauge field $\Upsilon_\mu$ into symmetric and antisymmetric parts:
\al{
\Upsilon^\alpha{}_{\beta\mu}\fn{x}
	&=	\Upsilon^\alpha{}_{(\beta\mu)}\fn{x}+\Upsilon^\alpha{}_{[\beta\mu]}\fn{x},
		\label{separation of Gamma}
}
where the parentheses and square brackets for the indices are defined in Eq.~\eqref{eq: symmetric and antisymmetric notation}.
Note that we have mixed the indices $\beta$ and $\mu$ that correspond to an internal gauge index and a spacetime index, respectively, for the case of the ordinary/LL-gauge field.
The symmetric and antisymmetric parts in the first and second terms of Eq.~\eqref{separation of Gamma} have ${\pn{d+1}^2\pn{d+2}\ov2}$ and ${d\pn{d+1}^2\ov2}$ degrees of freedom, respectively.
The number of degrees of freedom of the antisymmetric part $\Upsilon^\alpha{}_{[\beta\mu]}$ is the same as that of the GC conditions~\eqref{GC condition on M}. This fact suggests that it is redundant for the GC symmetry.

Let us see that this is indeed the case.
Under the GC transformation~\eqref{transformation of GC gauge field more explicit}, the antisymmetric part of the GC-gauge field transforms homogeneously:
\al{
\Upsilon^\alpha{}_{[\beta\mu]}\fn{x}
	\sr{\tx{GC}}\to	\Upsilon^{\pr\alpha}{}_{[\beta\mu]}\fn{x'}
	&=	
			M^\alpha{}_\gamma\Upsilon^\gamma{}_{\delta\nu}
				\pa{M^{-1}}^\delta{}_{[\beta}
				\pa{M^{-1}}^\nu{}_{\mu]}
			-\p_\nu M^\alpha{}_\gamma
				\pa{M^{-1}}^\gamma{}_{[\beta}
				\pa{M^{-1}}^\nu{}_{\mu]}\nn
	&=		M^\alpha{}_\gamma\Upsilon^\gamma{}_{[\delta\nu]}
				\pa{M^{-1}}^{[\delta}{}_\beta
				\pa{M^{-1}}^{\nu]}{}_\mu
			-\p_{[\nu} M^\alpha{}_{\gamma]}
				\pa{M^{-1}}^{[\gamma}{}_\beta
				\pa{M^{-1}}^{\nu]}{}_\mu\nn
	&=		M^\alpha{}_\gamma\Upsilon^\gamma{}_{[\delta\nu]}
				\pa{M^{-1}}^\delta{}_\beta
				\pa{M^{-1}}^\nu{}_\mu,
				\label{needs only symmetric part}
}
where we have omitted the dependence on $x$ on the right-hand side for simplicity and have used the GC condition~\eqref{GC condition on M} in the last step.
That is, the GC covariance of the GC-covariant derivative is maintained even if we do not include the antisymmetric part $\Upsilon^\alpha{}_{[\beta\mu]}\fn{x}$.
(Though it means that we do not need the antisymmetric part at all in order to realize the GC covariance of the GC covariant derivative, this argument itself does not prohibit having the antisymmetric part.)

\subsubsection{Levi-Civita (spin) connection}\label{Levi-Civita section}
Conventionally, the Levi-Civita connection $\os g\Gamma$ has been used as the GC-gauge field:\footnote{
This is the case in supergravity too~\cite{VanNieuwenhuizen:1981ae}: ``One has four choices: $\omega$ or $\omega\fn{e}$ for Lorentz connection, and $\Gamma$ or $\Gamma\fn{g}$ for the other connection. The choice appropriate for local supersymmetry is $\omega$ and $\Gamma\fn{g}$. Any other choice would do as well, but one would need extra complicated terms in the action.''
\label{supergravity footnote}
}
\al{
\os g\Gamma{}^\alpha{}_{\beta\mu}\fn{x}
	&:=	{g^{\alpha\gamma}\fn{x}\ov2}\Pn{-\p_\gamma g_{\beta\mu}\fn{x}+\p_\beta g_{\mu\gamma}\fn{x}+\p_\mu g_{\gamma\beta}\fn{x}},
 \label{Levi-Civita given}
}
which is the solution to the metricity condition on $\Upsilon$:
\al{
\os \Upsilon\nabla_\alpha g_{\beta\mu}\fn{x}=0.
}
By construction, it has only the symmetric part
$\os g\Gamma{}^\alpha{}_{\beta\mu}\fn{x}=\os g\Gamma{}^\alpha{}_{(\beta\mu)}\fn{x}$; recall the discussion in the paragraphs containing Eqs.~\eqref{separation of Gamma} and \eqref{needs only symmetric part}.
The transformation of the Levi-Civita connection can be found, as in any textbook of general relativity, e.g. Ref.~\cite{Wald:1984rg}, to be the same as Eq.~\eqref{transformation of GC gauge field in matrix}, or \eqref{transformation of GC gauge field more explicit}:
\al{
\os g\Gamma{}^\alpha{}_{\beta\mu}\fn{x}
	&\sr{\tx{GC}}\to
		\mt{M\fn{x}\os g\Gamma_\nu\fn{x}M^{-1}\fn{x}}^\alpha{}_\beta\mt{M^{-1}\fn{x}}^\nu{}_\mu
		+{\p x^{\pr\alpha}\ov\p x^\gamma}{\p^2x^\gamma\ov\p x^{\pr\mu}\p x^{\pr\beta}}\nn
	&\quad=
		\mt{M\fn{x}\os g\Gamma_\nu\fn{x}M^{-1}\fn{x}}^\alpha{}_\beta\mt{M^{-1}\fn{x}}^\nu{}_\mu
		-{\p x^\nu\ov\p x^{\pr\mu}}{\p^2x^{\pr\alpha}\ov\p x^\nu\p x^\gamma}{\p x^\gamma\ov\p x^{\pr\beta}}\nn
	&\quad=
		\mt{M\fn{x}\os g\Gamma_\nu\fn{x} M^{-1}\fn{x}}^\alpha{}_\beta\mt{M^{-1}\fn{x}}^\nu{}_\mu
		-\pa{M^{-1}}^\nu{}_\mu\fn{x}
			\p_{\nu} M^\alpha{}_{\gamma}\fn{x}
			\pa{M^{-1}}^\gamma{}_\beta\fn{x}\nn
	&\quad=
		\mt{M\fn{x}\os g\Gamma_\nu\fn{x} M^{-1}\fn{x}-\p_\nu M\fn{x}M^{-1}\fn{x}}^\alpha{}_\beta
		\pa{M^{-1}}^\nu{}_\mu\fn{x}.
		\label{Levi-Civita transformed}
}

We note that the Levi-Civita connection requires an inverse metric $g^{\mu\nu}$ and hence an inverse vielbein $e_\ba{}^\mu$.
Therefore, it cannot be used under the irreversible vierbein postulate imposed on our action at the scale~$\Lambda_\tx{G}$.
That is, the GC-gauge field is absent at $\Lambda_\tx{G}$ since we do not further introduce it as extra degrees of freedom; see also footnote~\ref{supergravity footnote}. We will come back to this point below.

Once the Levi-Civita connection $\os g\Gamma_\mu$ is introduced (in our scenario, it is induced by quantum fluctuations below the scale $\Lambda_\tx{G}$), then another LL-gauge field can also be induced, namely the Levi-Civita spin connection $\os e\Omega_\mu$:
\al{
\os e\Omega{}^\ba{}_{\bb\mu}\fn{x}
	&:=	e^\ba{}_\lambda\fn{x}\os g\nabla_\mu e_\bb{}^\lambda\fn{x}
	:=	e^\ba{}_\lambda\fn{x}\pn{\p_\mu e_\bb{}^\lambda\fn{x}+\os g\Gamma{}^\lambda{}_{\sigma\mu}\fn{x}e_\bb{}^\sigma\fn{x}}.
	\label{Levi-Civita spin connection}
}
It is straightforward to check that the Levi-Civita spin connection $\os e\Omega_\mu\fn{x}$ transforms in the same way as the LL-gauge field $\omega_\mu\fn{x}$ under the LL and GC transformations.

\subsection{Field strengths, Riemann tensor, and \texorpdfstring{$GL(d+1)$}{$GL(d+1)$}}\label{section on miscellaneous points}
Now we come back to considering the general gravitational gauge fields $\omega$ and $\Upsilon$.

The field strengths for the ordinary and LL-gauge fields are given in the matrix notation, respectively, as\footnote{
For a compact gauge group such as that introduced in footnote~\ref{sign of guage kinetic}, it is more common to use $F_{\mu\nu}\fn{x}=\p_\mu A_\nu\fn{x}-\p_\nu A_\mu\fn{x}+ig_G\commutator{A_\mu\fn{x}}{A_\nu\fn{x}}$ that follows from $\os{\mc A}{\mc F}_{\mu\nu}\fn{x}=ig_GF_{\mu\nu}^a\fn{x}T^a$.
}
\al{
\os{\mc A}{\mc F}_{\mu\nu}\fn{x}
	&:=	\p_\mu\mc A_\nu\fn{x}-\p_\nu\mc A_\mu\fn{x}
		+\commutator{\mc A_\mu\fn{x}}{\mc A_\nu\fn{x}},
			\label{F for A}\\
\os\omega{\mc F}_{\mu\nu}\fn{x}
	&:=	\p_\mu\omega_\nu\fn{x}-\p_\nu\omega_\mu\fn{x}
		+\commutator{\omega_\mu\fn{x}}{\omega_\nu\fn{x}}.
			\label{F for omega}
}
More explicitly, the LL field strength reads
\al{
\os\omega{\mc F}{}^\ba{}_{\bb\mu\nu}\fn{x}
	&=	\p_\mu\omega^\ba{}_{\bb\nu}\fn{x}-\p_\nu\omega^\ba{}_{\bb\mu}\fn{x}
		+\omega^\ba{}_{\bc\mu}\fn{x}\omega^\bc{}_{\bb\nu}\fn{x}
		-\omega^\ba{}_{\bc\nu}\fn{x}\omega^\bc{}_{\bb\mu}\fn{x}.
}

We can rewrite the field strength as a commutator of the covariant derivatives~\eqref{gauge covariant derivative} and \eqref{LL-covariant derivative}, or \eqref{LL-covariant derivative in matrix notation}, on the fundamental representation:
\al{
\os{\mc A}{\mc F}_{\mu\nu}\fn{x}
	&=	\commutator{\os{\mc A}{\mc D}_\mu}{\os{\mc A}{\mc D}_\nu},\\
\os\omega{\mc F}_{\mu\nu}\fn{x}
	&=	\commutator{\os\omega{\mc D}_\mu}{\os\omega{\mc D}_\nu}.
}
It is important that the field strengths reduce to functions~\eqref{F for A} and \eqref{F for omega} when deriving their covariance
\al{
\os{\mc A}{\mc F}_{\mu\nu}\fn{x}
	&\to	U\fn{x}\os{\mc A}{\mc F}_{\mu\nu}\fn{x}U^{-1}\fn{x},\\
\os{\omega}{\mc F}_{\mu\nu}\fn{x}
	&\sr{\tx{LL}}\to	\Lambda\fn{x}\os{\omega}{\mc F}_{\mu\nu}\fn{x}\Lambda^{-1}\fn{x},
}
from the covariance of the covariant derivatives~\eqref{covariance of D} and \eqref{covariance of LL-covariant D in matrix notation}, respectively, as follows:
\al{
\os{\mc A}{\mc F}_{\mu\nu}\fn{x}\Psi\fn{x}
	=	\commutator{\os{\mc A}{\mc D}_\mu}{\os{\mc A}{\mc D}_\nu}\Psi\fn{x}
	\to	U\fn{x}\commutator{\os{\mc A}{\mc D}_\mu}{\os{\mc A}{\mc D}_\nu}\Psi\fn{x}
	&=	U\fn{x}\commutator{\os{\mc A}{\mc D}_\mu}{\os{\mc A}{\mc D}_\nu}\pn{U^{-1}\fn{x}U\fn{x}\Psi\fn{x}}\nn
	&=	\pn{U\fn{x}\os{\mc A}{\mc F}_{\mu\nu}\fn{x}U^{-1}\fn{x}}U\fn{x}\Psi\fn{x},\nn
\os\omega{\mc F}_{\mu\nu}\fn{x}V\fn{x}
	=	\commutator{\os\omega{\mc D}_\mu}{\os\omega{\mc D}_\nu}V\fn{x}
	\sr{\tx{LL}}\to	\Lambda\fn{x}\commutator{\os\omega{\mc D}_\mu}{\os\omega{\mc D}_\nu}V\fn{x}
	&=	\Lambda\fn{x}\commutator{\os\omega{\mc D}_\mu}{\os\omega{\mc D}_\nu}\pn{\Lambda^{-1}\fn{x}\Lambda\fn{x}V\fn{x}}\nn
	&=	\pn{\Lambda\fn{x}\os{\mc\omega}{\mc F}_{\mu\nu}\fn{x}\Lambda^{-1}\fn{x}}\Lambda\fn{x}\Psi\fn{x}.
}

Now let us take the commutator of the GC-covariant derivative~\eqref{GC-covariant derivative on vector in matrix notation}, or \eqref{GC-covariant derivative on vector}, on a spacetime-vector field $V\fn{x}=V^\mu\fn{x}\p_\mu$ that transforms as a fundamental representation under GC transformation: In the matrix notation,
\al{
\os\Upsilon{\mc F}_{\mu\nu}\fn{x}V\fn{x}
	&:=	\commutator{\os\Upsilon\nabla_\mu}{\os\Upsilon\nabla_\nu}V\fn{x}\nn
	&=	\pn{
		\p_\mu\Upsilon_\nu\fn{x}-\p_\nu\Upsilon_\mu\fn{x}+\commutator{\Upsilon_\mu\fn{x}}{\Upsilon_\nu\fn{x}}
		+2\,\I\,\Upsilon^\rho{}_{[\nu\mu]}\fn{x}\os\Upsilon\nabla_\rho
		}V\fn{x},
  \label{GL field strength different from GC one}
}
or more explicitly,
\al{
\os\Upsilon{\mc F}{}^\alpha{}_{\beta\mu\nu}\fn{x}V^\beta\fn{x}
	&=	\Big(\p_\mu\Upsilon^\alpha{}_{\beta\nu}\fn{x}
		-\p_\nu\Upsilon^\alpha{}_{\beta\mu}\fn{x}
		+\Upsilon^\alpha{}_{\gamma\mu}\fn{x}\Upsilon^\gamma{}_{\beta\nu}\fn{x}
		-\Upsilon^\alpha{}_{\gamma\nu}\fn{x}\Upsilon^\gamma{}_{\beta\mu}\fn{x}\nn
	&\qquad
		+2\delta^\alpha_\beta\Upsilon^\rho{}_{[\nu\mu]}\fn{x}\os\Upsilon\nabla_\rho
		\Big)V^\beta\fn{x}.
}
The last term is peculiar to the GC-field strength:
The antisymmetric part $\Upsilon^\rho{}_{[\nu\mu]}\fn{x}$ is not only in vain in covariantizing the GC-covariant derivative, but also, it is an obstacle to making the GC-field strength a function rather than a differential operator.
This fact disfavors an introduction of the extra GC-gauge field $\Upsilon_\mu$ with the antisymmetric part. See also the discussion around Eq.~\eqref{needs only symmetric part} for the redundancy of the antisymmetric part.

Let us comment on the relation to the Levi-Civita (spin) connection.
It is noteworthy that if the GC-gauge field is identified with the Levi-Civita connection~\eqref{Levi-Civita given},
\al{
\Upsilon^\alpha{}_{\beta\mu}\equiv\os g\Gamma{}^\alpha{}_{\beta\mu},
    \label{GC gauge field as Levi-Civita}
}
the GC-field strength becomes the Riemann tensor itself:
\al{
\os g{\mc F}{}^\alpha{}_{\beta\mu\nu}\fn{x}
	&=	\mc R^\alpha{}_{\beta\mu\nu}\fn{x},
 \label{eq: Riemann tensor}
}
where $\os g{\mc F}_{\mu\nu}\fn{x}:=\os{\os g\Gamma}{\mc F}_{\mu\nu}\fn{x}$. This can be shown as follows:
\al{
\os g{\mc F}_{\mu\nu}\fn{x}V\fn{x}
	:=	\commutator{\os g\nabla_\mu}{\os g\nabla_\nu}V\fn{x}
	&=	2\pn{\p_{[\mu}+\os g\Gamma{}_{[\mu}\fn{x}}\pn{\p_{\nu]}+\os g\Gamma{}_{\nu]}\fn{x}}V\fn{x}\nn
	&=	\pn{\p_\mu\os g\Gamma{}_\nu\fn{x}-\p_\nu\os g\Gamma{}_\mu\fn{x}+\os g\Gamma{}_\mu\fn{x}\os g\Gamma{}_\nu\fn{x}-\os g\Gamma{}_\nu\fn{x}\os g\Gamma{}_\mu\fn{x}}V\fn{x},
}
or more explicitly,
\al{
\mt{\os g{\mc F}_{\mu\nu}\fn{x}}^\alpha{}_\beta
	&=	\mt{\p_\mu\os g\Gamma{}_\nu\fn{x}-\p_\nu\os g\Gamma{}_\mu\fn{x}+\os g\Gamma{}_\mu\fn{x}\os g\Gamma{}_\nu\fn{x}-\os g\Gamma{}_\nu\fn{x}\os g\Gamma{}_\mu\fn{x}}^\alpha{}_\beta,\nn
\os g{\mc F}{}^\alpha{}_{\beta\mu\nu}\fn{x}
	&=	\p_\mu\os g\Gamma{}^\alpha{}_{\beta\nu}\fn{x}
		-\p_\nu\os g\Gamma{}^\alpha{}_{\beta\mu}\fn{x}
		+\os g\Gamma{}^\alpha{}_{\gamma\mu}\fn{x}\os g\Gamma{}^\gamma{}_{\beta\nu}\fn{x}
		-\os g\Gamma{}^\alpha{}_{\gamma\nu}\fn{x}\os g\Gamma{}^\gamma{}_{\beta\mu}\fn{x};
}
the right-hand side is nothing but the Riemann tensor.
Under the assumption~\eqref{GC gauge field as Levi-Civita}, the antisymmetric part of the GC-gauge field $\Upsilon^{\alpha}{}_{\beta\mu}$ does not take part and play any role.

In the same manner, we can define the Riemann tensor from the LL-field strength with the Levi-Civita spin connection, i.e., $\Upsilon^\alpha{}_{\beta\mu}\equiv\os e\Omega^\alpha{}_{\beta\mu}$, together with a vielbein and its inverse:
\al{
e_\ba{}^\alpha\fn{x}e^\bb{}_\beta\fn{x}\os e{\mc F}{}^\ba{}_{\bb\mu\nu}\fn{x}
	&=	\mc R^\alpha{}_{\beta\mu\nu}\fn{x},\nn
\os e{\mc F}{}^\ba{}_{\bb\mu\nu}\fn{x}
	&=	e^\ba{}_\alpha\fn{x}e_\bb{}^\beta\fn{x}\mc R^\alpha{}_{\beta\mu\nu}\fn{x},
}
where $\os e{\mc F}{}^\ba{}_{\bb\mu\nu}\fn{x}:=\os{\os e\Omega}{\mc F}{}^\ba{}_{\bb\mu\nu}\fn{x}$, namely, in the matrix notation,
\al{
\os e{\mc F}_{\mu\nu}\fn{x}
	:=	\commutator{\os e{\mc D}_\mu}{\os e{\mc D}_\nu}
	&=	2\pn{\p_{[\mu}+\os e\Omega{}_{[\mu}\fn{x}}\pn{\p_{\nu]}+\os e\Omega{}_{\nu]}\fn{x}}\nn
	&=	\p_\mu\os e\Omega_\nu\fn{x}-\p_\nu\os e\Omega_\mu\fn{x}
		+\os e\Omega_\mu\fn{x}\os e\Omega_\nu\fn{x}
		-\os e\Omega_\nu\fn{x}\os e\Omega_\mu\fn{x},
}
in which $\os e{\mc D}_\mu:=\p_\mu+\os e\Omega_\mu$ [see Eq.~\eqref{Levi-Civita spin connection}], or more explicitly,
\al{
\mt{\os e{\mc F}_{\mu\nu}\fn{x}}^\ba{}_\bb
	&=	\mt{
		\p_\mu\os e\Omega_\nu\fn{x}
		-\p_\nu\os e\Omega_\mu\fn{x}
		+\os e\Omega_\mu\fn{x}\os e\Omega_\nu\fn{x}
		-\os e\Omega_\nu\fn{x}\os e\Omega_\mu\fn{x}
		}^\ba{}_\bb,\nn
\os e{\mc F}^\ba{}_{\bb\mu\nu}\fn{x}
	&=	\p_\mu\os e\Omega^\ba{}_{\bb\nu}\fn{x}
		-\p_\nu\os e\Omega^\ba{}_{\bb\mu}\fn{x}
		+\os e\Omega^\ba{}_{\bc\mu}\fn{x}\os e\Omega^\bc{}_{\bb\nu}\fn{x}
		-\os e\Omega^\ba{}_{\bc\nu}\fn{x}\os e\Omega^\bc{}_{\bb\mu}\fn{x}.
}

Under the irreversible-vielbein postulate, we expect that physically the following scenario takes place: In the action at $\Lambda_{\rm G}$, there is no specific background of the vielbein. The quantum dynamics induces a nontrivial background vielbein $\bar e^\ba{}_\mu\fn{x}$ that has its inverse $\bar e_\ba{}^\mu\fn{x}$ everywhere, namely, a nondegenerate $\bar e^\ba{}_\mu\fn{x}$, on which the expectation value of the LL-gauge field should become the Levi-Civita spin connection:
\al{
\bigl\langle\omega_\mu\fn{x}\bigr\rangle_{\bar e\pn{x}}
	&\sr{!}=	\bar\Omega_\mu\fn{x},
		\label{LL-gauge field vev}
}
where
\al{
\bar\Omega_\mu\fn{x}
	&:=	\left.\os e\Omega_\mu\fn{x}\right|_{e\pn{x}\to\bar e\pn{x}}.
 \label{eq: background LC spin connection}
}
This way, our formulation would reproduce the conventional covariant derivative acting on the spinor field in the metric formulation.
It suffices therefore that only the LL-field strength exists to construct the Riemann tensor at lower energies below $\Lambda_\tx{G}$ if the physical expectation~\eqref{LL-gauge field vev} is met. We do not need to prepare the GC-field strength as a source for the Riemann tensor from the beginning at $\Lambda_\tx{G}$.

We comment on the application order of transformations, namely, the GC transformation after a gauge transformation, or the other way around. The GC transformation acts on all spacetime indices, so that gauge transformations are affected by it. In other words, one may think that elements of their transformations do not commute. Fortunately, our definition of the GC transformations \eqref{eq: GC for vielbein}--\eqref{eq: GC for gauge field} commute with any gauge transformation, while \diff\ defined by the LD transformation does not. This means that the former is given by a direct product ``$\tx{GC} \times \tx{gauge}.$" while the latter is by the semidirect product ``$\tx{GC} \ltimes \tx{gauge}$." These facts are discussed in Appendix~\ref{app: (Semi-)direct product}.

Finally, let us comment on the $GL(d+1)$ theory. If we do not impose the GC conditions~\eqref{GC condition on M}--\eqref{GC condition 3}, etc.,\ and we regard $M^\mu{}_\nu$ as a general $\pn{d+1}\times\pn{d+1}$ matrix, then the theory becomes a $GL(d+1)$ gauge theory; see, e.g.,\ Refs.~\cite{Floreanini:1989hq,Fukaya:2016zqw}.
This theory might be of interest in itself, but we do not go in this direction and do not introduce the extra GC-gauge field~$\Upsilon_\mu$ at $\Lambda_\tx{G}$ because of the abovementioned points: (i) the nonnecessity of its antisymmetric part for the covariance of the GC-covariant derivative, (ii)~the GC-field strength becoming a differential operator rather than a function due to the antisymmetric part, and (iii)~the nonnecessity as a source for constructing the Riemann tensor.

\subsection{Summary on covariant derivatives}
For general LL and GC-gauge fields $\omega$ and $\Upsilon$, respectively, we summarize our notation for the covariant derivatives:
\al{
&\tx{LL only:}&
\os\omega{\mc D}_\mu e^\ba{}_\nu\fn{x}
    &=  \p_\mu e^\ba{}_\nu\fn{x}+\omega^\ba{}_{\bb\mu}\fn{x}e^\bb{}_\nu\fn{x},\\
&\tx{GC only:}&
\os\Upsilon\nabla_\mu e^\ba{}_\nu\fn{x}
    &=  \p_\mu e^\ba{}_\nu\fn{x}-e^\ba{}_\lambda\fn{x}\Upsilon^\lambda{}_{\nu\mu}\fn{x},\\
&\tx{LL and GC:}&
\os{\omega,\Upsilon}\D_\mu e^\ba{}_\nu\fn{x}
    &:=  \p_\mu e^\ba{}_\nu\fn{x}+\omega^\ba{}_{\bb\mu}\fn{x}e^\bb{}_\nu\fn{x}-e^\ba{}_\lambda\fn{x}\Upsilon^\lambda{}_{\nu\mu}\fn{x}\nn
&&
    &=  \os\omega{\mc D}_\mu e^\ba{}_\nu\fn{x}
            -e^\ba{}_\lambda\fn{x}\Upsilon^\lambda{}_{\nu\mu}\fn{x}\nn
&&
    &=  \os\Upsilon\nabla_\mu e^\ba{}_\nu\fn{x}
            +\omega^\ba{}_{\bb\mu}\fn{x}e^\bb{}_\nu\fn{x}.
}

\section{Action under irreversible vierbein postulate}
\label{sec: action setting}
In this section, we construct a ``tree" action given at a certain scale $\Lambda_\tx{G}$ based on the LL and GC symmetries, i.e. ${\rm GC} \times SO(1,3)$.   
As discussed in the Introduction, a central assumption at $\Lambda_\tx{G}$ is the irreversible vierbein postulate that forbids the action at $\Lambda_\tx{G}$ to contain an inverse of the vielbein.

In Sec.~\ref{sec: Levi-Civita tensor}, we start with the introduction of the Levi-Civita tensor which is independent of the vielbein and inverse vielbein. Then, in Sec.~\ref{sec: Irreversible vierbein and action}, the definition of the irreversible vierbein postulate is explained and the action respecting this postulate is shown.

This section fully explains, for the first time, the idea briefly sketched in the preceding Letter~\cite{Matsuzaki:2020qzf} to this paper.

\subsection{Levi-Civita tensor}
\label{sec: Levi-Civita tensor}
To write down the action, we spell out our notation on the totally antisymmetric tensor, etc.
We first introduce the Levi-Civita symbol:
\al{
\ep{\mu_0\dots\mu_d}
	&=	\begin{cases}
		1	&	\tx{when $\pn{\mu_0,\dots,\mu_d}$ is even permutation of $\pn{0,\dots,d}$,}\\
		-1	&	\tx{when $\pn{\mu_0,\dots,\mu_d}$ is odd permutation of $\pn{0,\dots,d}$,}\\
		0	&	\tx{otherwise,}
		\end{cases}
}
and similarly,
\al{
\ep{\bs{a_0}\dots\bs{a_d}}
	&=	\begin{cases}
		1	&	\tx{when $\pn{\bs{a_0},\dots,\bs{a_d}}$ is even permutation of $\pn{\bs0,\dots,\bs d}$,}\\
		-1	&	\tx{when $\pn{\bs{a_0},\dots,\bs{a_d}}$ is odd permutation of $\pn{\bs0,\dots,\bs d}$,}\\
		0	&	\tx{otherwise.}
		\end{cases}
}
We write the determinant of the vielbein and metric as
\al{
\ab{e\fn{x}}
	:=	\det_{\ba,\mu}e^\ba{}_\mu\fn{x}
	&=	\ep{\bs{a_0}\dots\bs{a_d}}e^{\bs{a_0}}{}_0\fn{x}\cdots e^{\bs{a_d}}{}_d\fn{x}\nn
	&=	\ep{\mu_0\dots\mu_d}e^{\bs0}{}_{\mu_0}\fn{x}\cdots e^{\bs d}{}_{\mu_d}\fn{x}\nn
	&=	{1\ov\pn{d+1}!}\ep{\bs{a_0}\dots\bs{a_d}}\ep{\mu_0\dots\mu_d}e^{\bs{a_0}}{}_{\mu_0}\cdots e^{\bs{a_d}}{}_{\mu_d},\\
\ab{g\fn{x}}
	:=	\det_{\mu,\nu}g_{\mu\nu}\fn{x}
	&=	\ep{\mu_0\dots\mu_d}g_{\mu_00}\fn{x}\cdots g_{\mu_dd}\fn{x}\nn
	&=	\ep{\nu_0\dots\nu_d}g_{0\nu_0}\fn{x}\cdots g_{d\nu_d}\fn{x}\nn
	&=	{1\ov\pn{d+1}!}\ep{\mu_0\dots\mu_d}\ep{\nu_0\dots\nu_d}g_{\mu_0\nu_0}\fn{x}\cdots g_{\mu_d\nu_d}\fn{x},
}
where the summation over repeated indices is understood for the Levi-Civita symbol as well.
It follows that
\al{
\ab{e\fn{x}}
	&=	\sqrt{-\ab{g\fn{x}}},
}
where $\ab{g\fn{x}}$ is always negative due to the Lorentzian signature.

From the Levi-Civita symbol, we define the Levi-Civita tensor for the LL transformation,
\al{
\epsilon_{\bs{a_0}\dots\bs{a_d}}
	&:=	\ep{\bs{a_0}\dots\bs{a_d}},\\
\epsilon^{\bs{a_0}\dots\bs{a_d}}
	&:=	\eta^{\bs{a_0}\bs{b_0}}\cdots\eta^{\bs{a_d}\bs{b_d}}\,\epsilon_{\bs{b_0}\dots\bs{b_d}}
	=	-\ep{\bs{a_0}\dots\bs{a_d}},
}
and for the GC transformation,
\al{
\vep_{\mu_0\dots\mu_d}\fn{x}
	&:=	\ab{e\fn{x}}\ep{\mu_0\dots\mu_d},\\
\vep^{\mu_0\dots\mu_d}\fn{x}
	&:=	g^{\mu_0\nu_0}\fn{x}\cdots g^{\mu_d\nu_d}\fn{x}\vep_{\nu_0\dots\nu_d}\fn{x}
	=	-{\ep{\mu_0\dots\mu_d}\ov\ab{e\fn{x}}}.
}
Note that the Lorentzian signature leads to
\al{
{1\ov\pn{d+1}!}\epsilon_{\bs{a_0}\dots\bs{a_d}}\epsilon^{\bs{a_0}\dots\bs{a_d}}
	&=	-1,\\
{1\ov\pn{d+1}!}\vep_{\mu_0\dots\mu_d}\fn{x}\vep^{\mu_0\dots\mu_d}\fn{x}
	&=	-1,
}
which follows from the $p=d$ case of more general identities: for $0\leq p\leq d$,
\al{
{1\ov \pn{p+1}!}\ep{\mu_0\dots\mu_p\,\mu_{p+1}\dots\mu_d}\ep{\mu_0\dots\mu_p\,\mu_{p+1}'\dots\mu_d'}
	&=	\pn{d+1-p}!\,\delta_{[\mu_{p+1}}^{\mu_{p+1}'}\cdots\delta_{\mu_d]}^{\mu_d'},
		\label{identity for Levi-Civita symbol}
}
etc.\footnote{
See footnote~\ref{normalization footnote} for the normalization.
}
Using the Levi-Civita tensor, we can write down the volume element in terms of the local coordinate system of each chart:
\al{
\star1
	&=	{1\ov\pn{d+1}!}\vep_{\mu_0\dots\mu_d}\fn{x}\df x^{\mu_0}\wedge\cdots\wedge\df x^{\mu_d}
	=	\ab{e\fn{x}}\df x^0\wedge\cdots\wedge\df x^d
	=	\ab{e\fn{x}}\df^{d+1}x,
}
where $\star$ is the Hodge dual, which is defined for a $p$-form $\alpha\fn{x}={1\ov p!}\alpha_{\mu_0\dots\mu_{p-1}}\fn{x}\df x^{\mu_0}\wedge\cdots\wedge\df x^{\mu_{p-1}}$ by\footnote{
Or else, one may first define
\als{
\star\fn{\df x^{\mu_0}\wedge\cdots\df x^{\mu_{p-1}}}
	&:=	{1\ov\pn{d+1-p}!}
		\vep^{\mu_0\dots\mu_{p-1}}{}_{\nu_p\dots\nu_d}\fn{x}
		\df x^{\nu_p}\wedge\cdots\wedge\df x^{\nu_d}
}
so that
\als{
\star\alpha\fn{x}
	&:=	{1\ov p!}\alpha_{\mu_0\dots\mu_{p-1}}\fn{x}\,{\star\fn{\df x^{\mu_0}\wedge\cdots\df x^{\mu_{p-1}}}}.
}
}
\al{
\star\alpha\fn{x}
	&:=	{1\ov\pn{d+1-p}!}\pn{\star\alpha}_{\mu_p\dots\mu_d}\fn{x}\df x^{\mu_p}\wedge\cdots\wedge\df x^{\mu_d},
}
with
\al{
\pn{\star\alpha}_{\mu_p\dots\mu_d}\fn{x}
	&:=	{1\ov p!}\alpha_{\nu_0\dots\nu_{p-1}}\fn{x}
		\vep^{\nu_0\dots\nu_{p-1}}{}_{\mu_p\dots\mu_d}\fn{x}.
}
Note that from the vielbein, the LL-gauge field, and its field strength, we can construct GC-scalar 1- and 2-form fields such that
\al{
e^\ba\fn{x}
	&:=	e^\ba{}_\mu\fn{x}\df x^\mu,\\
\omega^\ba{}_\bb\fn{x}
	&:=	\omega^\ba{}_{\bb\mu}\fn{x}\df x^\mu,\\
\os\omega{\mc F}^\ba{}_\bb\fn{x}
	&:=	{1\ov2}\os\omega{\mc F}^\ba{}_{\bb\mu\nu}\fn{x}\df x^\mu\wedge\df x^\nu.
}

\subsection{Irreversible vierbein and action}
\label{sec: Irreversible vierbein and action}

Let us now construct an action invariant under $\tx{GC}\times SO(1,3)$ symmetry. 
Hereafter, we work in the $d=3$ spatial dimensions, assuming that it is already settled down to be so at $\Lambda_\tx{G}$.
We call the vielbein for the $d+1=4$ spacetime dimensions the vierbein, accordingly.

The formulation of the irreversible vierbein postulate starts  by imposing regularity under the limit of any zero eigenvalues $\lambda_a\to 0$ of the vierbein among four eigenvalues in four-dimensional spacetime. 
Obviously, in such a case, the inverse vierbein cannot be defined. In other words, the inverse vierbein contains divergences. Then, the irreversible vierbein postulate at $\Lambda_{\rm G}$ states that the action at $\Lambda_{\rm G}$ does not diverge even for the (not necessarily simultaneous) zero eigenvalue limit of vierbein. We call this kind of limit the degenerate limit~\cite{Tseytlin:1981ks,Horowitz:1990qb,Floreanini:1991cw}.

There are some cases where, even when a term is apparently written down using the inverse vierbein, the irreversible vierbein postulate does not forbid such a term in the degenerate limit. 
An important observation for this is that some inverse vierbeins can be absorbed into the determinant $\ab{e\fn{x}}$ from the volume element. More specifically, from the identities~\eqref{identity for Levi-Civita symbol} and
\al{
\ab{e\fn{x}}\ep{\mu\nu\rho\sigma}
	&=	\ep{\ba\bb\bc\bd}e^\ba{}_\mu\fn{x}e^\bb{}_\nu\fn{x}e^\bc{}_\rho\fn{x}e^\bd{}_\sigma\fn{x},
	\label{eq: determinant 1}
}
we obtain
\al{
\ab{e\fn{x}}e_\ba{}^\mu\fn{x}
	&=	{1\ov3!}\ep{\ba\bb\bc\bd}\ep{\mu\nu\rho\sigma}e^\bb{}_\nu\fn{x}e^\bc{}_\rho\fn{x}e^\bd{}_\sigma\fn{x},
 \label{eq: determinant e}
 \\
\ab{e\fn{x}}e_{[\ba}{}^\mu\fn{x}e_{\bb]}{}^\nu\fn{x}
	&=	{1\ov2}\ep{\ba\bb\bc\bd}\ep{\mu\nu\rho\sigma}e^\bc{}_\rho\fn{x}e^\bd{}_\sigma\fn{x},
	\label{eq: determinant ee}
	\\
\ab{e\fn{x}}e_{[\ba}{}^\mu\fn{x}e_\bb{}^\nu\fn{x}e_{\bc]}{}^\rho\fn{x}
	&=	\ep{\ba\bb\bc\bd}\ep{\mu\nu\rho\sigma}e^\bd{}_\sigma\fn{x},
	\label{eq: determinant eee}
	\\
\ab{e\fn{x}}e_{[\ba}{}^\mu\fn{x}e_\bb{}^\nu\fn{x}e_\bc{}^\rho\fn{x}e_{\bd]}{}^\sigma\fn{x}
	&=	\ep{\ba\bb\bc\bd}\ep{\mu\nu\rho\sigma}.
	\label{eq: determinant eeee}
}
Though the left-hand sides of these equations appear to have the inverse vierbeins, the right-hand sides do not. Only these combinations of the inverse vierbeins can be used to write down the action at $\Lambda_\tx{G}$ without contradicting the irreversible vierbein postulate.

Let us now write down the action explicitly.
First, it turns out that the kinetic terms for the GC-scalar and vector fields contain the inverse vierbeins, all of which cannot be simultaneously absorbed into the volume element via Eqs.~\eqref{eq: determinant e}--\eqref{eq: determinant eeee}. They contain the inverse metric $g^{\mu\nu}$ that is symmetric for its indices (see Appendix~\ref{app: Degenerate limit of vierbein} for more an explicit discussion):\footnote{
See footnote~\ref{sign of guage kinetic} for the sign of the gauge kinetic term.
}
\al{
S_\tx{boson}
	&=	\int\df^{4}x\ab{e\fn{x}}\sqbr{
			-{1\ov2}g^{\mu\nu}\fn{x}\p_\mu\phi\fn{x}\p_\nu\phi\fn{x}
			+{1\ov2g_G^2}g^{\mu\rho}\fn{x}g^{\nu\sigma}\fn{x}\tr\pn{\os{\mc A}{\mc F}_{\mu\nu}\fn{x}\os{\mc A}{\mc F}_{\rho\sigma}\fn{x}}
			}.
}
Therefore, these terms are forbidden, and thus the GC-scalar and vector fields are not dynamical at $\Lambda_\tx{G}$. The kinetic terms for vierbein $e^\ba{}_\mu$ and LL-gauge fields $\omega^{\ba}{}_{\bb\mu}$ cannot be introduced due to the same reason.

On the other hand, the spinor kinetic term is consistent with the irreversible vierbein postulate because it contains only a single inverse vierbein, and thus we can use Eq.~\eqref{eq: determinant e}:
\footnote{\label{higher derivative terms}
One may regard the action~\eqref{spinor kinetic} as $\propto\int\ep{\ba\bb\bc\bd}e^\ba\fn{x}\wedge e^\bb\fn{x}\wedge e^\bc\fn{x}\wedge \Delta^\bd\fn{x}$, where $\Delta^\ba{}_\mu\fn{x}:=\ol\psi\fn{x}\gamma^\ba\os\omega{\mc D}_\mu\psi\fn{x}$.
This is nothing but a replacement from the action~\eqref{cc term}, being $\propto\int\ep{\ba\bb\bc\bd}e^\ba\fn{x}\wedge e^\bb\fn{x}\wedge e^\bc\fn{x}\wedge e^\bd\fn{x}$, of a single vierbein 1-form: $e^\bd{}\fn{x}\to \Delta^\bd\fn{x}$. In principle, one may replace any vierbein $e^\ba{}_\mu\fn{x}$ by $\Delta^\ba{}_\mu\fn{x}$ without contradicting the irreversible vierbein postulate. When one replaces all of the four vierbeins to the fermion bilinear in the cosmological constant term~\eqref{cc term}, one obtains the action for the spinor gravity $\propto\int\ep{\ba\bb\bc\bd}\Delta^\ba\fn{x}\wedge\Delta^\bb\fn{x}\wedge\Delta^\bc\fn{x}\wedge\Delta^\bd\fn{x}$~\cite{Wetterich:2003wr,Volovik:2021wut}. In this paper, we restrict ourselves to the lowest-derivative terms up to single $\Delta^\ba{}_\mu\fn{x}$. Further generalizations will be presented in a separate publication.
}
\al{
S_\tx{spinor}
	&=	\int\df^{4}x\ab{e\fn{x}}\sqbr{-\ol\psi\fn{x}e_\ba{}^\mu\fn{x}\gamma^\ba\pn{\p_\mu+{1\ov2}\omega_{\bb\bc\mu}\fn{x}\sigma^{\bb\bc}}\psi\fn{x}}\nn
	&=	\int\df^{4}x\sqbr{-{1\ov3!}\ep{\ba\bb\bc\bd}\ep{\mu\nu\rho\sigma}e^\bb{}_\nu\fn{x}e^\bc{}_\rho\fn{x}e^\bd{}_\sigma\fn{x}\ol\psi\fn{x}\gamma^\ba\pn{\p_\mu+{1\ov2}\omega_{\ba'\bb'\mu}\fn{x}\sigma^{\ba'\bb'}}\psi\fn{x}}.
		\label{spinor kinetic}
}

Unlike the ordinary gauge theory, the LL symmetry has a fundamental representation that is a spacetime vector, the vierbein. It allows one to write down an LL-invariant term constructed from a single field strength:
\al{
S_\tx{LL}
	&=	\int\df^4x\ab{e\fn{x}}\sqbr{
			{\MP^2\ov2}e_{[\ba}{}^\mu\fn{x}e_{\bb]}{}^\nu\fn{x}\os\omega{\mc F}^{\ba\bb}{}_{\mu\nu}\fn{x}
			}\nn
	&=	\int\df^4x
			\sqbr{
			{\MP^2\ov4}\ep{\ba\bb\bc\bd}\ep{\mu\nu\rho\sigma}e^\bc{}_\rho\fn{x}e^\bd{}_\sigma\fn{x}
			\os\omega{\mc F}^{\ba\bb}{}_{\mu\nu}\fn{x}
			}.
			\label{single LL term}
}
This term has become compatible with the irreversible vierbein postulate thanks to Eq.~\eqref{eq: determinant ee}.
Note that as mentioned in the paragraph containing Eq.~\eqref{eq: background LC spin connection}, it is not necessary to introduce the term with the field strength corresponding to the GC transformation $\os \Upsilon{\mc F}$.

Finally, one can also write down the cosmological constant term:
\al{
S_\tx{cc}
	&=	\int\df^4x\ab{e\fn{x}}\Lambda_\tx{cc}.
		\label{cc term}
}

Barring the topological terms (see discussion below) as well as the higher-derivative terms (see footnote~\ref{higher derivative terms}), the terms~\eqref{spinor kinetic}--\eqref{cc term} are the only combinations that are consistent with the irreversible vierbein postulate:
\al{
S_{\Lambda_\tx{G}}
	&=	\int\df^4x\ab{e\fn{x}}\Bigg[
			-Z_\psi\,\ol\psi\fn{x}e_\ba{}^\mu\fn{x}\gamma^\ba\pn{\p_\mu+{1\ov2}\omega_{\bb\bc\mu}\fn{x}\sigma^{\bb\bc}}\psi\fn{x}\nn
	&\phantom{=	\int\df^4x\ab{e\fn{x}}\Bigg[}
			+X_\omega {\MP^2\ov2}e_{[\ba}{}^\mu\fn{x}e_{\bb]}{}^\nu\fn{x}\os\omega{\mc F}^{\ba\bb}{}_{\mu\nu}\fn{x}
			-V
			\Bigg]
	,\label{starting action}
}
where $Z_\psi$, $X_\omega$, and $V$ are arbitrary functions of spacetime scalars at $x$ constructed by matter fields, say, $\phi\fn{x}$,\, $\ol\psi\fn{x}\psi\fn{x}$,\, etc. For example, $V$ includes the mass term for spinor $M_\psi\ol\psi\fn{x}\psi\fn{x}$ and the cosmological constant $\Lambda_\tx{cc}$ as well as the ordinary scalar potential.
Note that the kinetic term for the Rarita-Schwinger field, which is a spin-$3/2$ field, is also compatible with the degenerate limit (see Appendix~\ref{Rarita-Schwinger field}). In this work, we do not take this into account.

Various combinations of fields can be contracted with the LL metric $\eta_{\ba\bb}$ and the totally antisymmetric LL tensor $\ep{\ba\bb\bc\bd}$ to yield topological terms of the action, which are summarized in Appendix~\ref{topological section}.
In general, one may multiply, on these ``topological'' terms, arbitrary functions of GC scalars such as $Z_\psi$, $X_\omega$, and $V$ in Eq.~\eqref{starting action} so that they become dynamical (nontopological). Since all such interactions are higher dimensional, we neglect them in this paper. 
The inclusion of these terms might be of interest in itself, which we leave for future study.

Finally, we stress again the reason why we impose the irreversible vierbein postulate at the tree level. A typical criticism may be as follows: The existence of inverse vierbein in the tree action is harmless since terms with inverse vierbeins behave as $e^{-\mathcal O(e^{-1})}\to 0$ for $e\to 0$ within the path integral. On the other hand, the Standard Model of particle physics assumes that the action does not have inverse power of the fields such as $1/H^\dagger H$, where $H$ is the Higgs field, even though such terms can be perfectly consistent with all the gauge and spacetime symmetries. The absence of inverse power is particularly noteworthy in the effective field theory picture because such negative-power terms are more relevant than the normal ones toward the IR direction. We can interpret this as the requirement of the existence of the weak-field limit $H\to0$ for the action such that the symmetric phase $\langle H\rangle=0$ is well defined.\footnote{
The weak-field limit $H\to0$ here is different from that in the gravitational literature (see, e.g.,\ Ref.~\cite{Horowitz:1980fj}) in the sense that the latter means the limit of zero fluctuation around a background, that is, $\delta g_{\mu\nu}\to0$ for $g_{\mu\nu}=\bar g_{\mu\nu}+\delta g_{\mu\nu}$. For the former, the whole Higgs $H=\bar H+\delta H$ goes to zero rather than $H\to\bar H$. The existence of limit $H\to\bar H\neq0$ cannot forbid the negative power such as $1/H^\dagger H$.
}

The irreversible vierbein postulate introduces a well-defined symmetric phase in our quantum-gravity framework. Conventional quantum field theories, like the Standard Model, also assume the existence of the symmetric phase. Our framework facilitates the exploration of quantum spacetime dynamics where the spacetime metric \( g_{\mu\nu} \) approaches zero. By excluding inverse vierbeins, which are undefined in this degenerate limit, the postulate offers a novel approach to studying quantum gravitational phenomena under extreme conditions. This method extends beyond mathematical convenience, providing a framework that enhances our understanding of spacetime in scenarios such as near singularities or strong gravitational fields, and may offer valuable insights into phenomena like the early Universe and black hole interiors.

\section{Local-Lorentz and general coordinate transformations for Background fields}
\label{sec: Background fields}

One of the main purposes in this paper is to demonstrate the generation of nontrivial background fields due to quantum effects in four-dimensional spacetimes. This will be done in Section~\ref{sec: dynamical vierbein}. In this section, assuming that background vielbein and LL-gauge fields are induced in arbitrary spacetime dimensions, we discuss their transformation laws and covariance. Those are important for understanding of the low-energy effective theory from the action~\eqref{starting action}.
After these summary reviews, in Sec.~\ref{sec: Global background Lorentz invariance after spontaneous symmetry breaking}, we fully explain the idea briefly sketched in the preceding Letter~\cite{Matsuzaki:2020qzf}.

\subsection{Invertible background vielbein and Levi-Civita (spin) connection}
To begin with, we introduce a certain gravitational background $\bar\Phi=\pn{\bar e,\bar\omega}$, while we do not assume a classical background for the matter fields for simplicity: $\bar\Psi=(\bar\phi,\bar\psi,\bar{A}_\mu)=0$

Here, an important assumption is that the background vielbein field $\bar e^\ba{}_\mu\fn{x}$ is invertible that allows the inverse background vielbein $\bar e_\ba{}^\mu\fn{x}$ and the inverse background metric
\al{
\bar g^{\mu\nu}\fn{x}
	&:=	\eta^{\ba\bb}\,\bar e_\ba{}^\mu\fn{x}\bar e_\bb{}^\nu\fn{x},
		\label{background metric inverse defined}
}
where the background vielbein is defined to satisfy
\al{
\bar e_\ba{}^\mu\fn{x}\bar e^\ba{}_\nu\fn{x}
    &=  \delta^\mu_\nu,&
\bar e_\ba{}^\mu\fn{x}\bar e^\bb{}_\mu\fn{x}
    &=  \delta_\ba^\bb,&
\bar g^{\mu\nu}\fn{x}\bar g_{\nu\rho}\fn{x}
    &=  \delta^\mu_\rho.
\label{eq: inverse veirbein and metric}
}
In general, the full vielbein and metric fields
\al{
e^\ba{}_\mu\fn{x} = \bar e^\ba{}_\mu\fn{x} + \mathfrak{e}^\ba{}_\mu\fn{x},
    \label{vierbein expanded}\\
g_{\mu\nu}\fn{x}=\bar g_{\mu\nu}\fn{x}+\mf g_{\mu\nu}\fn{x},
}
will also become invertible, where $\mf e$ and $\mf g$ represent their quantum fluctuations, respectively.
However, we hereafter raise and lower the spacetime indices by the background vielbein and/or metric. In particular, we can now give
\al{
A_\ba\fn{x}
    &:=  \bar e_\ba{}^\mu\fn{x}A_\mu\fn{x};
}
recall the argument below Eq.~\eqref{LL of ordinary gauge field}.

The invertible background vielbein also allows us to write down the background Levi-Civita (spin) connection:
\al{
\bar\Gamma^\mu{}_{\rho\sigma}
	&:=	{\bar g^{\mu\nu}\ov2}\paren{-\p_\nu\bar g_{\rho\sigma}+\p_\rho\bar g_{\sigma\nu}+\p_\sigma\bar g_{\nu\rho}},
		\label{Levi-Civita defined}\\
\bar\Omega^\ba{}_{\bb\mu}
	&:=	\bar e^\ba{}_\lambda\bar\nabla_\mu\bar e_\bb{}^\lambda
	=	\bar e^\ba{}_\lambda\p_\mu \bar e_\bb{}^\lambda+\bar\Gamma^\ba{}_{\bb\mu},
	\label{eq: Omegaabmu}
}
where
\al{
\bar\nabla_\mu\bar e_\bb{}^\lambda
    &:= \p_\mu\bar e_\bb{}^\lambda+\bar\Gamma^\lambda{}_{\sigma\mu}\bar e_\bb{}^\sigma,\\
\bar\Gamma^\ba{}_{\bb\mu}
    &:= \bar e^\ba{}_\lambda \bar\Gamma^\lambda{}_{\sigma\mu}\bar e_\bb{}^\sigma,
}
etc.
Note that $\bar\Gamma$ and $\bar\Omega$ are solely made of the vielbein and its inverse. In the previous language,
\al{
\bar\Gamma^\mu{}_{\rho\sigma}\fn{x}
    &=  \os{\bar g}\Gamma^\mu{}_{\rho\sigma}\fn{x},&
\bar\Omega^\ba{}_{\bb\mu}\fn{x}
    &=  \os{\bar e}\Omega^\ba{}_{\bb\mu}\fn{x}.
}
The background spin connection $\bar\Omega$ transforms the same as the background LL-gauge field $\bar\omega$ under the background LL and GC transformations.

\subsection{Background covariance}

For the given background field of the gravitational fields $\bar\Phi$, we define the following LL-background-covariant (only) derivatives denoted by $\bar{\mc D}_\mu$ for matter fields
\al{
\bar{\mc D}_\mu\phi\fn{x}
    &:=  \p_\mu\phi\fn{x},\\
\bar{\mc D}_\mu\psi\fn{x}
    &:=  \p_\mu\psi\fn{x}+{\bar\omega_{\ba\bb\mu}\fn{x}\ov2}\Sigma^{\ba\bb}\psi\fn{x},\\
\bar{\mc D}_\mu A^\ba\fn{x}
    &:=  \p_\mu A^\ba\fn{x}+\bar\omega^\ba{}_{\bb\mu}\fn{x}A^\bb\fn{x},
}
and for gravitational fields
\al{
\bar{\mc D}_\mu e^\ba{}_\nu\fn{x}
    &:=  \p_\mu e^\ba{}_\nu\fn{x}+\bar\omega^\ba{}_{\bb\mu}\fn{x}e^\bb{}_\nu\fn{x},\\
\bar{\mc D}_\mu\omega^\ba{}_{\bb\nu}\fn{x}
    &:=  \p_\mu\omega^\ba{}_{\bb\nu}\fn{x}
        +\bar\omega^\ba{}_{\bc\mu}\fn{x}\omega^\bc{}_{\bb\nu}\fn{x}
        -\omega^\ba{}_{\bc\nu}\fn{x}\bar\omega^\bc{}_{\bb\mu}\fn{x}.
}
We also define the LL-and-GC-background-covariant derivative $\bar{\ms D}_\mu$. It acts the same as $\bar{\mc D}_\mu$ on the matter fields without the spacetime indices $\Psi$,
\al{
\bar\D_\mu\Psi\fn{x}
    &:=  \bar{\mc D}_\mu\Psi\fn{x},
}
whereas on the gravitational fields,\footnote{
For a given background $\bar e$ and $\bar\omega$, we may always construct a background GC-gauge field
\als{
\bar\gamma^\lambda{}_{\nu\mu}\fn{x}
    &:= \bar\omega^\lambda{}_{\nu\mu}\fn{x}
        -\bar e^\bc{}_\nu\fn{x}\p_\mu\bar e_\bc{}^\lambda\fn{x}
}
that is {\it defined} to satisfy the metricity:
\als{
\os{\bar\omega,\bar\gamma}\D_\mu\bar e^\ba{}_\nu\fn{x}
    &=  0.
}
We can explicitly check that $\bar\gamma^\lambda{}_{\nu\mu}$ is invariant under the background LL transformation.
}
\als{
\bar\D_\mu e^\ba{}_\nu\fn{x}
    &=  \bar{\mc D}_\mu e^\ba{}_\nu\fn{x}
        -e^\ba{}_\lambda\fn{x}\bar\Upsilon^\lambda{}_{\nu\mu}\fn{x}\nn
    &=  \p_\mu e^\ba{}_\nu\fn{x}+\bar\omega^\ba{}_{\bb\mu}\fn{x}e^\bb{}_\nu\fn{x}
        -e^\ba{}_\lambda\fn{x}\bar\Upsilon^\lambda{}_{\nu\mu}\fn{x},\\
\bar\D_\mu\omega^\ba{}_{\bb\nu}\fn{x}
    &=  \bar{\mc D}_\mu\omega^\ba{}_{\bb\nu}\fn{x}
        -\omega^\ba{}_{\bb\lambda}\fn{x}\bar\Upsilon^\lambda{}_{\nu\mu}\fn{x}
    \nn
    &=  \p_\mu\omega^\ba{}_{\bb\nu}\fn{x}
        +\bar\omega^\ba{}_{\bc\mu}\fn{x}\omega^\bc{}_{\bb\nu}\fn{x}
        -\omega^\ba{}_{\bc\nu}\fn{x}\bar\omega^\bc{}_{\bb\mu}\fn{x}
        -\omega^\ba{}_{\bb\lambda}\fn{x}\bar\Upsilon^\lambda{}_{\nu\mu}\fn{x}.
}

So far, $\bar\omega$ is pretty much unconstrained.\footnote{
We might end up with the following expression:
\als{
\bar\D_\mu\bar\omega^\ba{}_{\bb\nu}
    &=  \pn{\p_\mu\bar\omega^\ba{}_{\bb\bc}}\bar e^\bc{}_\nu
        +\bar\omega^\ba{}_{\bc\mu}\bar\omega^\bc{}_{\bb\nu}
        -\bar\omega^\ba{}_{\bc\nu}\bar\omega^\bc{}_{\bb\mu};
}
see, e.g., Ref.~\cite{Lippoldt:2016ayw}.
Using $\bar\D\bar e=0$, we may rewrite
\als{
\bar\D_\mu\bar\omega^\ba{}_{\bb\bc}
    &=  \p_\mu\bar\omega^\ba{}_{\bb\bc}
        +\bar\omega^\ba{}_{\bd\mu}\bar\omega^\bd{}_{\bb\bc}
        -\bar\omega^\ba{}_{\bd\bc}\bar\omega^\bd{}_{\bb\mu}.
}
In the language of differential forms,
\als{
\bar\D
    &:= \df x^\mu\bar\D_\mu,&
\bar\omega^\ba{}_\bb\fn{x}
    &:=\bar\omega^\ba{}_{\bb\nu}\fn{x}\df x^\nu,&
\df &:= \df x^\mu\p_\mu,&
\bar e^\bc\fn{x}
    &:= \bar e^\bc{}_\nu\fn{x}\df x^\nu\,.
}
This can be written as
\als{
\bar\D\bar\omega
    &=  \df\bar\omega\wedge\bar e
        +2\bar\omega\wedge\bar\omega.
}
}
In this paper, we assume that the background vielbein should obey the metricity
\al{
\bar{\ms D}_\mu\bar e^\ba{}_\nu\fn{x}=0,
}
the background LL-gauge field $\bar\omega^\ba{}_{\bb\mu}\fn{x}$ be Eq.~\eqref{LL-gauge field vev}, and the background GC connection be the Levi-Civita one
\al{
\bar\Gamma^\lambda{}_{\nu\mu}\fn{x}=\left.\os g\Gamma^\lambda{}_{\nu\mu}\fn{x}\right|_{g=\bar g};
}
recall Eq.~\eqref{Levi-Civita given}.

\subsection{Global background Lorentz invariance after spontaneous symmetry breaking}
\label{sec: Global background Lorentz invariance after spontaneous symmetry breaking}

For a general background $\bar e^\ba{}_\mu$, there remains a local $SO(1,3)\times \text{GC}$ background symmetry:
\al{
\bar\phi\fn{x}
    &\sr{SO(1,3)\times\tx{GC}}\lra    \bar\phi'\fn{x'},\\
\bar\psi\fn{x}
    &\sr{SO(1,3)\times\tx{GC}}\lra    \bar\psi'(x')=S(\Lambda(x'))\bar\psi\fn{x'},\\
\bar{\mc A}_\mu\fn{x}
	&\sr{SO(1,3)\times\tx{GC}}\lra 	\bar{\mc A}'_\mu\fn{x'}=\bar{\mc A}_\nu\fn{x}{\p x^\nu\ov\p x^{\pr\mu}},
}
and
\al{
\bar e^\ba{}_\mu\fn{x}
    &\sr{SO(1,3)\times\tx{GC}}\lra \bar e^{\pr\ba}{}_\mu\fn{x'}=\Lambda^\ba{}_\bb\fn{x} \bar e^\bb{}_\nu\fn{x} \pa{M^{-1}}^\nu{}_\mu\fn{x},
    \label{eq: background trans for vierbein}
    \\
\bar\omega^\ba{}_{\bb\mu}\fn{x}
    &\sr{SO(1,3)\times\tx{GC}}\lra    \bar\omega^\ba{}_{\bb\mu}\fn{x'}
    =       \sqbr{\Lambda\fn{x}\bar\omega_\nu\fn{x}\Lambda^{-1}\fn{x}
	        - \Pn{\p_\nu\Lambda\fn{x}}\Lambda^{-1}\fn{x}}\!{}^\ba{}_\bb
	        \pa{M^{-1}}^\nu{}_\mu\fn{x}.
}
The Lorentz transformation under the global $SO(1,3)$ is nothing but an accidental symmetry arising only when we take the flat background $\bar e^\ba{}_\mu=\delta^\ba_\mu$ for which $\bar\omega^{\ba}{}_{\bb \mu}=0$. That is, the gobal $SO(1,3)$ is the ``diagonal subgroup'': From Eq.~\eqref{eq: background trans for vierbein}, one has\footnote{
See the discussion around Eq.~\eqref{eq: global SO1d transformation} in Appendix~\ref{Lie derivative on spinor} for the reduction of $\pa{M^{-1}}^\nu{}_\mu\fn{x}\to\pa{\Lambda^{-1}}^\nu{}_\mu$.
}
\al{
\delta^\ba_\mu
    &\to    \Lambda^\ba{}_\bb \delta^\bb_\nu\pa{\Lambda^{-1}}^\nu{}_\mu=\delta^\ba_\mu,\\
x^\mu
    &\to \Lambda^\mu{}_\nu x^\nu,\quad\tx{(this is mere a reparametrization)}
}
so to say 
\al{
SO(1,3)\times \tx{GC}\to SO(1,3)_\tx{diag}.
}
Hence, under this transformation in the diagonal subgroup, the kinetic term of spinors transforms as usual, even though we assign them only a trivial representation under GC.

\section{Dynamical generation of flat spacetime from spinor loop} 
\label{sec: dynamical vierbein}

In this section, we derive the effective potential for a vierbein background field, assuming it to be a flat spacetime background, and then demonstrate that indeed a nonvanishing flat vierbein background is induced by quantum effects of the fermion field.

As emphasized in the preceding Letter~\cite{Matsuzaki:2020qzf}, both the vierbein and LL-gauge fields are auxiliary at $\Lambda_\tx{G}$, and both of them are shown to acquire the kinetic terms below $\Lambda_\tx{G}$. This situation is in accordance with the emergence of the hidden local symmetry applied in QCD; se,e e.g.,\ Ref.~\cite{Bando:1987br} for a review. Now we show the effective potential for the vierbein as a (linearly realized) Higgs field~\cite{Percacci:1990wy,Floreanini:1991cw,Floreanini:1993na,Percacci:2009ij,Volovik:2021wut}.

\subsection{Effective action for conformal mode}
\label{sec:Spinor-induced effective potential of vierbein}

We study now the dynamical symmetry breaking of the LL symmetry in four-dimensional spacetime. A central object for the observation of such a symmetry breaking is the effective potential for the vierbein field. To obtain it, we assume a background field for vierbein $\bar e^\ba{}_\mu$ and investigate the effective potential. The degenerate limit would enforce such a minimum to be located at $\bar e^\ba{}_\mu=0$. What we want to see in this section is whether quantum effects generate a nontrivial expectation value of $\bar e^\ba{}_\mu$ or not.
In this section, we consider the quantum effects of the spinor fields at the one-loop level on the effective potential for the vierbein, while we deal with the vierbein and the LL-gauge fields as classical fields.
The one-loop approximation might be justified by a large number of spinor degrees of freedom in the SM that is 90 without including the right-handed neutrinos, whereas we do not exclude the possibility of large effects from other sectors; we will come back to this point later.

We start with separating the vierbein field into the background and fluctuation as in Eq.~\eqref{vierbein expanded}.
For simplicity, we concentrate on a constant background field in this paper.
At the tree level of our action~\eqref{starting action}, the potential of the (constant) vierbein field is simply given by
\al{
V_\tx{tree}\fn{\bar e} = \Lambda_\tx{cc} \ab{\bar e}.
\label{eq: tree potential for e}
}
Here, we suppose that the kinetic term of the vierbein, which will be generated dynamically below $\Lambda_\tx{G}$, will take a ``correct'' (negative) sign so that the action is given by $S_{\rm eff}=\int \df^4x [-(\p_\mu \bar e^\ba{}_\nu)^2 -V_{\rm eff}(\bar e)]$. This assumption will be discussed at the end of this section.
For a negative cosmological constant $\Lambda_\tx{cc}<0$, we have an unbounded potential, and $|\bar e|=0$ is an unstable extremum, while for $\Lambda_\tx{cc}>0$, the potential is bounded and has a minimum at $\bar e=0$.
The former induces the spacetime background already at the tree level.
How about the latter?

We next consider quantum corrections to the effective potential. At the level of the action \eqref{starting action} at $\Lambda_\tx{G}$, only the spinor fields are dynamical and give the leading effects on the effective potential. 
Here, we compute the effective potential for an assumed background vierbein field value $\bar e^\ba{}_\mu\fn{x}$, taking into account the spinor one-loop correction. In order to evaluate the spinor action concretely, let us here assume a flat spacetime background, i.e., we parametrize
\al{
\bar e^\ba{}_\mu
	=	C\delta^\ba_\mu,
\label{eq: flat parametrization}
}
where $C$ is a dimensionless constant.
This parametrization is a special case in that all eigenvalues $\lambda_a$ of the  vierbein take the same value, namely, $\bar e^\ba{}_\mu= \diag(\lambda_0,\lambda_1,\lambda_2,\lambda_3)=\diag(C,C,C,C)$.
For such a flat background, the equation of motion may entail $\bar\omega_{\ba\bb\mu}=0$.

Hence, the point $C=0$ corresponds to the degenerate limit for the eigenvalues, and thus is identified with a ``symmetric phase'' of $\tx{GC}\times SO(1,3)$ in analogy to the ordinary Higgs mechanism~\cite{Percacci:1990wy}.
Needless to say, the point $C=0$ has no background spacetime at all and is intractable.
Our strategy to handle this ``symmetric'' point to compare with the broken phase\footnote{
This phase is sometimes called the Einstein phase in the literature~\cite{Tseytlin:1981ks}.
}
$C\neq0$ is first computing the effective potential for $C>0$ and then examining the limit $C\searrow0$.

Let us derive the effective potential for $C$ arising from the spinor loop.
The assumption~\eqref{eq: flat parametrization} for the background field gives $|\bar e|=C^4$ and $\bar e_\ba{}^\mu=C^{-1}\delta_\ba^\mu$.
In order to obtain the one-loop effects from a spinor under the background~\eqref{eq: flat parametrization}, it suffices to take its quadratic terms:
\al{
S_\tx{kin}
	&=	\int\df^4x\sqbr{
			-Z_\psi C^3\ol\psi\fn{x}\gamma^\mu \p_\mu\psi\fn{x}
			-C^4M_\psi\ol\psi\fn{x}\psi\fn{x}
			},
}
where we have included the spinor mass term coming from $V$ and have written $\gamma^\mu:=\delta^\mu_\ba\gamma^\ba$.\footnote{
Once we identify $\psi$ as one of the SM fermions, each $M_\psi$ should be regarded as the one from the electroweak symmetry breaking, with $M_\psi\ll\Lambda_\tx{G}$.
Here we treat it as a massive Dirac spinor for simplicity since the generalization is straightforward.
}
Now we redefine the spinor field as $\psi\fn{x}\to Z_\psi^{-1/2}C^{-3/2}\psi\fn{x}$:
\al{
S_\tx{kin}
	&=	\int\df^4x\sqbr{
			-\ol\psi\fn{x}\gamma^\mu \p_\mu\psi\fn{x}
			-Cm_\psi\ol\psi\fn{x}\psi\fn{x}
			},
}
where $m_\psi:=M_\psi/Z_\psi$.

Integrating out the spinor field yields the effective action
\al{
\Gamma_\text{eff}\fn{C}
	= - \int \df^4x\, C^4\Lambda_\tx{cc}
- i\Tr\log\fnl{
-\gamma^\mu\p_\mu -m_\psi C},
\label{eq: effective action of e}
}
where Tr acts on all internal degrees of freedom involved in the spinor field, e.g. eigenvalues of the covariant derivative, spinor space, and so on.
The second term in Eq.~\eqref{eq: effective action of e} may give a nontrivial form of the effective potential.

One can perform the Fourier transformation to obtain
\al{
\Gamma_\text{eff}\fn{C}&=- \int \df^4x\,C^4\sqbr{
		{\Lambda_\tx{cc}}
		+i\int\frac{\df^4p}{(2\pi)^4} \log\fnl{-i\gamma^\mu p_\mu -m_\psi C}
		},
}
where we have taken the spacetime volume from the momentum-space delta function $\delta^4\fn{0}$ in the trace as $\pn{2\pi}^4\delta^4\fn{0}=C^4\int\df^4x$, which can be naturally understood by first performing the heat-kernel expansion and then taking the limit~\eqref{eq: flat parametrization}.
Thus, the effective potential for $C$ under the flat background~\eqref{eq: flat parametrization} is given by
\al{
V_\text{eff}(C)= -\frac{\Gamma_\text{eff}\fn{C}}{\int\df^4x} 
={\Lambda_\tx{cc}} C^4 - C^4\int\frac{\df^4p_\tx{E}}{(2\pi)^4}\log\fnl{-i\gamma^\mu p_{\tx E\mu} -m_\psi C},
\label{eq: effective potential for C}
}
where we have performed the Wick rotation such that $p^0=ip_{0\tx E}$ and $p^i=p_{i\tx E}$. If we neglect the second term corresponding to the quantum correction from the spinor field, the effective potential would be simply $V_\text{eff}(C)={\Lambda_\tx{cc}} C^4$ and hence, for $\Lambda_\tx{cc}>0$, the effective potential would have a vacuum $C=0$, implying $\bar e^\ba{}_\mu=\langle e^\ba{}_\mu\rangle =0$.

Let us now perform the loop momentum integral in Eq.~\eqref{eq: effective potential for C}. To this end, we need regularizations.
Here, we attempt to employ the momentum-cutoff and dimensional regularization.
The use of the momentum cutoff such that $0<p_\tx{E}<\Lambda_\tx{G}$ and $C^2m_f^2\ll \Lambda_\tx{G}^2$ gives
\al{
V_\text{eff}(C) &={\Lambda_\tx{cc}} C^4 + \frac{C^4}{2(4\pi)^2}\int_0^{\Lambda_\tx{G}^2} \df(p_\tx{E}^2)(p_\tx{E}^2)\left[ - \frac{1}{2}\log\fn{ p_\tx{E}^2 + (m_fC)^2}\right] \nn
&={\Lambda_\tx{cc}} C^4
 + \frac{m_f^4}{2(4\pi)^2}C^4 \Big(  \log \Lambda_\tx{G}^2 - \log (m_f^2C^2) \Big)\nn
&\qquad
 + \frac{C^4}{2(4\pi)^2}\Bigg[ 
-\frac{\Lambda_\tx{G}^4}{8}  +\frac{\Lambda_\tx{G}^4}{4} \log \left(\Lambda_\tx{G}^2 \right)
 \Bigg]
+\frac{\Lambda_\tx{G}^2}{8(4\pi)^4}m_f^2C^2,
\label{eq: momentum cutoff potential}
}
while by dimensional regularization, we obtain
\al{
V_\text{eff}(C) &={\Lambda_\tx{cc}} C^4 - C^4\int\frac{\df^{4-\epsilon} p_\tx{E}}{(2\pi)^{4-\epsilon}}\left[
- \frac{1}{2}\log\fn{ p_\tx{E}^2 + (m_fC)^2} \right]
\nn
&= {\Lambda_\tx{cc}} C^4 + \frac{m_f^4}{2(4\pi)^2} C^4\left( \frac{2}{\bar\epsilon} - \log (m_f^2 C^2) \right) ,
\label{eq: dimensional regularization potential}
}
where $2/\bar\epsilon= 2/\epsilon - \gamma_E - \log 4\pi$ with $\epsilon=4-d$. 

The momentum regularization case \eqref{eq: momentum cutoff potential} is more complicated than the dimensional regularization case \eqref{eq: dimensional regularization potential}. To understand Eq.~\eqref{eq: momentum cutoff potential}, let us consider the chiral limit ($m_f\to 0$) for which
\al{
V_\text{eff}(C) &={\Lambda_\tx{cc}} C^4 + \frac{C^4}{2(4\pi)^2}\Bigg[ 
-\frac{\Lambda_\tx{G}^4}{8}  +\frac{\Lambda_\tx{G}^4}{4} \log \left(\Lambda_\tx{G}^2\right)
 \Bigg].
\label{eq: momentum cutoff potential without mass}
}
The quartically divergent terms $\sim \Lambda_\tx{G}^4$ can be subtracted by the additive renormalization for the cosmological constant $\Lambda_{\rm cc}$. More specifically, we prepare counterterms for $\Lambda_{\rm cc}$ such that $\delta\Lambda_{\rm cc} + \Lambda_{\rm cc,R}\delta_{\Lambda_{\rm cc}}$ where $\delta\Lambda_{\rm cc}$ additively subtracts terms proportional to $\Lambda_\tx{G}^4$, while the counterterm $\delta_{\Lambda_{\rm cc}}$ multiplicatively subtracts divergent terms. Therefore, by employing an appropriate renormalization condition, we would obtain
\al{
\delta\Lambda_{\rm cc}  =-\frac{1}{2(4\pi)^2}\Bigg[ 
-\frac{\Lambda_\tx{G}^4}{8}  +\frac{\Lambda_\tx{G}^4}{4} \log \left(\Lambda_\tx{G}^2\right)
 \Bigg].
}
This counterterm does not contribute to the running of the cosmological constant.
Indeed, such a prescription is analogous to the mass-independent scheme in scalar theories: For the scalar mass term, we give $\delta m^2 + m_R^2\delta_{m^2}$ and subtract quadratic divergences $\sim \Lambda^2$ by $\delta m^2$, while logarithmic divergences are removed by $m_R^2\delta_{m^2}$ and the running effects of scalar mass originate from $m_R^2\delta_{m^2}$, but not from quadratic divergences. The cancellation between the $\Lambda_\tx{G}^4$ terms and $\delta \Lambda_\tx{cc}$ is nothing but the cosmological constant problem~\cite{Weinberg:1988cp}. We do not intend to solve this problem in this work.

Next, we consider the last term on the right-hand side of Eq.~\eqref{eq: momentum cutoff potential}. It seems that this term cannot be subtracted because no counterterm proportional to $C^2$ exists in the original action~\eqref{starting action}. However, this is not the case. The momentum-cutoff regularization explicitly breaks the symmetries ${\rm GC}\times SO(1,3)$. In such a case, one has to estimate symmetry-breaking effects from the corresponding Ward-Takahashi identity and add counterterms to the action. Therefore, the last term on the right-hand side of Eq.~\eqref{eq: momentum cutoff potential} should be also removed from the effective action. To summarize, the effective potential under the momentum-cutoff regularization reads
\al{
V_\text{eff}(C) &={\Lambda_\tx{cc}} C^4
 + \frac{m_f^4}{2(4\pi)^2}C^4 \Big(  \log \Lambda_\tx{G}^2 - \log (m_f^2C^2) \Big)\,.
 \label{eq: potential with cutoff reg}
}
This is compatible with the result from dimensional regularization \eqref{eq: dimensional regularization potential} together with the identification $\log\Lambda_\tx{G}\leftrightarrow 2/\bar\epsilon$.

\begin{figure}
\centering
\includegraphics[width=8cm]{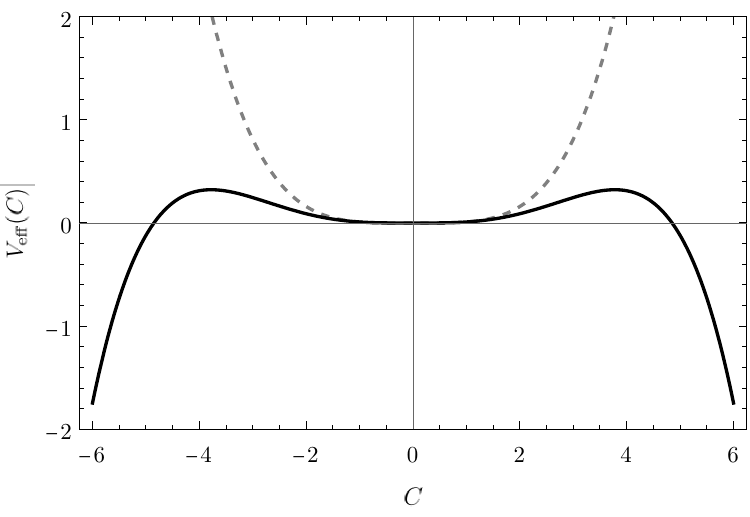}
\caption{
Plots of the potential as a function of $C$. The dashed line shows the tree-level potential \eqref{eq: tree potential for e}, $V_{\rm tree}(C)=\Lambda_{\rm cc}C^4$, while the effective potential \eqref{eq: renormalized effective potential for C} is depicted by the solid line. We set $\Lambda_\tx{G}=m_f=1$ and $\Lambda_\tx{cc}=0.01$ and assume a correct-sign kinetic term for $C$.
}
\label{fig:effective_potential_C}
\end{figure}

After removing the power divergences, which do not affect the running of theory parameters, the effective potential for $C$ becomes
\al{
V_\text{eff}(C) &={\Lambda_\tx{cc}} C^4
 - \frac{m_f^4}{2(4\pi)^2}C^4 \log \fn{\frac{m_f^2C^2}{\Lambda_\tx{G}^2}},
\label{eq: renormalized effective potential for C}
}
in the sense of bare perturbation theory around the scale $\Lambda_\tx{G}$.
In Fig.~\ref{fig:effective_potential_C}, we plot the effective potential~\eqref{eq: renormalized effective potential for C} as a function of $C$. Here, we set $\Lambda_\tx{G}=m_f=1$ and $\Lambda_{\rm cc}=0.01$ for displaying the effective potential. It is expected that the effective potential could yield a nonzero expectation value of $C$, i.e., $\langle e^\ba{}_\mu \rangle\neq 0$ due to the quantum tunneling effects.

\begin{figure}
\centering
\includegraphics[width=8cm]{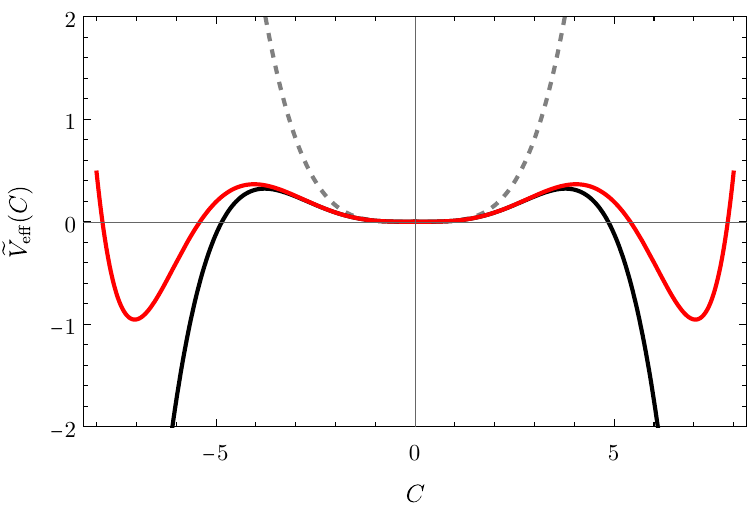}
\caption{
Plots of the sample potential (red line)~\eqref{sample potential} as a function of $C$ with $\lambda_8=0.02$ and the same values of other parameters as in Fig.~\ref{fig:effective_potential_C}. The black solid and gray dashed lines are the same as Fig.~\ref{fig:effective_potential_C}.
}
\label{fig:effective_potential_corrected}
\end{figure}

However, at this level of the approximation, the effective potential becomes a ``runaway" form. Thus, a nontrivial stable vacuum cannot be determined. So far, there are mainly two interpretations within the current model:
\begin{enumerate}[(i)]
\item Inclusion of higher-order effects such as loop effects of the vierbein and the LL-gauge field stabilize the effective potential.
\label{stabilized case}
\item The runaway behavior of $C$ implies the cosmological evolution of the scale factor~\cite{Giddings:1991qi}.
\label{runaway case}
\end{enumerate}
First we comment on the possibility~\eqref{stabilized case}.
In the tree-level action~\eqref{starting action}, there are no kinetic terms for the vierbein and the LL-gauge field, so these fields do not contribute to the effective potential at the leading order. In possibility~\eqref{stabilized case}, after the kinetic terms are induced by the fermionic quantum effects, the effective potential receives loop effects from the vierbein and the LL-gauge field and could be stabilized. 
As a demonstration, we plot in Fig.~\ref{fig:effective_potential_corrected} the potential with a possible higher-order correction $\propto\ab{e}^2$,
\al{
\wt V_\text{eff}(C) &={\Lambda_\tx{cc}} C^4
 - \frac{m_f^4}{2(4\pi)^2}C^4 \log \fn{\frac{m_f^2C^2}{\Lambda_\tx{G}^2}}
   +{\lambda_8\ov(4\pi)^4}C^8,
   \label{sample potential}
}
with a sample value $\lambda_8=0.02$. For other parameters, we use the same value as Fig.~\ref{fig:effective_potential_C}, i.e., $\Lambda_\tx{G}=m_f=1$ and $\Lambda_{\rm cc}=0.01$. In this case, there is a stable global vacuum at $\langle C\rangle =7.04$.
However, if $\lambda_8$ is large, the origin becomes the global minimum, and we do not get the emergence of spacetime.

Second, we comment on possibility~\eqref{runaway case}.
The field $C$ can be actually regarded as a conformal factor because the parametrization \eqref{eq: flat parametrization} with $C=e^\sigma$ gives the Weyl rescaling
\al{
\bar g_{\mu\nu} = \eta_{\ba\bb}\bar e^\ba{}_\mu \bar e^\bb{}_\nu = e^{2\sigma} \eta_{\mu\nu}.
}
Here, $\sigma$ is called the dilaton, the conformal field, or the scalaron, depending on the situation, and is associated with the scale symmetry. Indeed, in the potential~\eqref{eq: renormalized effective potential for C}, powers of $C$ reflect the canonical scaling of dimensionful parameters such as $\Lambda_{\rm cc}$ and $m_f$. Another possible interpretation is therefore that $C$ is regarded as a renormalization scale. The change of $C$ may give the running of renormalized couplings. The evolution of $C$ may be reasonable for realizing the expanding universe in cosmology.\footnote{
Obtaining a runaway potential for the vierbein has been discussed in Ref.~\cite{Giddings:1991qi} in a different mechanism: A background metric is assumed to be a cylinder of topology ${\mathbb R} \times {\mathcal M}^3$ with an arbitrary three manifold $\mathcal M^{3}$, i.e., $\df s^2= C^2(t)(-\df t^2 +\df {\vect x}^2)$ with time-dependent factor $C(t)$ and the effective potential for $C^2(t)$ is derived. In this scenario, a ``wrong-sign" kinetic term of $C(t)$ is crucial to obtain an unbounded potential for $C(t)$.
}
In this sense, the runaway potential is not excluded from possible scenarios. Moreover, the runaway potential may be also related to the wave function of the Universe~\cite{Hartle:1983ai}.
In any case, an important fact is that the solution $\langle C\rangle=0$ is an unstable vacuum, and then a nonvanishing vacuum is realized.

Finally, we comment on the kinetic term of the conformal mode $C$ in the vierbein. In the argument above, we have assumed a ``correct-sign" kinetic term of the conformal mode and then have obtained a runaway effective potential. Conversely, if the conformal mode has a wrong-sign kinetic term, the effective potential is unstabilized by the cosmological constant term and is bounded by the fermion loop effect. Consequently, we obtain a stable vacuum at
\al{
\langle C\rangle=\frac{\Lambda_\tx{G}}{m_f}\exp\left({\frac{16\pi^2\Lambda_{\rm cc}}{m_f^4}}\right).
}
The sign of the kinetic term for the conformal mode highly depends on interactions between gravitational fields and matter fields. Indeed, depending on the interaction between the scalar curvature $\mathcal R=\mathcal R^{\alpha\beta}{}_{\alpha\beta}$ [see Eq.~\eqref{eq: Riemann tensor} for the definition of the Riemann tensor] and scalars, the sign of the kinetic term for the conformal mode varies; see, e.g.,\ Appendix~A in Ref.~\cite{Lee:2023wdm}. In our model, it depends on whether or not the Planck mass in Eq.~\eqref{single LL term} is regarded as a function of scalars. Therefore, we do not specify the sign of the kinetic term for the conformal mode. Nonetheless, we stress that in any case, background vierbein has a nontrivial expectation value thanks to the fermion loop effect.

\section{Summary and discussion}
\label{sec: Summary and discussion}
In this paper, we have proposed a model for quantum gravity based on the LL-gauge and GC symmetries. In Section~\ref{sec: Degrees of freedom}, we have explained our viewpoint on constructing a quantum-gravity model. We have summarized transformation laws under these symmetries and the corresponding covariance carefully in Section~\ref{sec: Local-Lorentz and general coordinate transformations}.
Our main claim is to impose the irreversible vierbein postulate on the tree-level action at the scale $\Lambda_\tx{G}$ such that the action does not contain an inverse vierbein in invariant operators under the LL-gauge and GC symmetries. This postulate also prohibits, at $\Lambda_\tx{G}$, the background field of the gravitational fields, especially the vierbein.
It has been shown in Section~\ref{sec: action setting} that with matter fields, only three types of terms are admitted among operators up to dimension four in the tree-level action~\eqref{starting action}: the cosmological constant, the linear term in the field strength of the LL-gauge field, and the kinetic term for the spinor, possibly with their couplings being GC- and LL-invariant functions of matter fields. This means that only the spinor can behave as the dynamical quantum field at the lowest level.
Transformation laws for background fields have been summarized in Section~\ref{sec: Background fields}.
In Section~\ref{sec: dynamical vierbein}, we have argued the generation of a nonvanishing background vierbein field. There, supposing that a flat background vierbein field is induced, $\bar e^\ba{}_\mu=C\delta^\ba_\mu$, we have discussed the effective action for $C$. 
Depending on the signs of the kinetic term of $C$ and of the cosmological constant, the symmetric vacuum $\langle C\rangle =0$, which is consistent with the irreversible vierbein postulate, can be a stable minimum of the potential at the tree level, and then the fermionic fluctuations can make it unstable so that a nonvanishing value of $\langle C\rangle$ is realized. This implies the generation of the spacetime background.

With the nontrivial background vierbein generated by the quantum dynamics, we can discuss an effective theory. In the low-energy regime, gravitational interactions are well described by metric theories in which only 2 degrees of freedom within 10 degrees of freedom of symmetric tensor are physical, while our model~\eqref{starting action} has $4^2+6\times4=40$ degrees of freedom at the tree level since the vierbein and the LL-gauge fields have $16$ and $6\times4=24$ degrees of freedom, respectively. This discrepancy can be understood by the Higgs mechanism: The vierbein field plays the role of the Higgs field in generating the background spacetime as its vacuum expectation value. The LL-gauge field eats 6 degrees of freedom in the vierbein field and becomes massive~\cite{Matsuzaki:2020qzf}. The remaining 10 degrees of freedom of the vierbein field are the same as those of the symmetric tensor field. At the quantum level, only 2 degrees of freedom remain due to the subtraction of $8$ degrees of freedom by the gauge fixing and the ghost field. This picture provides a similar analogy to the nonlinear sigma model from the linear sigma model as discussed in Section~\ref{sec: Degrees of freedom}. Hence, in high energy, there are heavy modes for describing the gravitational interactions which cannot be observed in low-energy experiments; see Refs.~\cite{Ohanian:1969xhl,Nakanishi:1979fg,Floreanini:1991gi,Floreanini:1991cw,Floreanini:1994ypa,Floreanini:1995ie,Percacci:1990wy,Percacci:2009ij} for a similar mechanism in the $GL(4)$ case.

In this paper, we have not explored renormalizability of the theory because the existence of its further UV theories is unclear. String theory might be a high-energy theory for our model. Another possible candidate within the realm of quantum field theory is spinor gravity~\cite{Terazawa:1977xa,Hebecker:2003iw,Wetterich:2003wr,Volovik:2023pcm}, in which vierbein and LL-gauge fields are composites of spinor fields. This can be studied by the renormalization group with the compositeness condition~\cite{Akama:1998qy}.

\subsection*{Acknowledgement}
We thank Stefan Lippoldt, Hikaru Uchida, Daisuke Ida, Kunio Kaneta, Koichi Hamaguchi, Taichiro Kugo, and Christof Wetterich for their valuable discussions and comments.
The work of S.\,M.\ is supported in part by the National Science Foundation of China (NSFC) under Grants No.~11747308, No.~11975108, and No.~12047569 and the Seeds Funding of Jilin University. 
The work of K.\,O.\ is in part supported by JSPS Kakenhi Grant No.~19H01899.
The work of M.\,Y.\ is supported by the NSFC under Grant No.~12205116 and the Seeds Funding of Jilin University.
\appendix
\section*{Appendix}

\section{Lie derivative, general coordinate and gauge transformations}\label{LD, GC, and gauge transformations}

In this appendix, we summarize the transformation laws under the GC transformation and the Lie derivative. In particular, the difference between them is clarified in Appendices~\ref{app: Infinitesimal general coordinate transformation} and \ref{Lie derivative section}.

\subsection{Infinitesimal general-coordinate transformation}
\label{app: Infinitesimal general coordinate transformation}
For an infinitesimal GC transformation
\al{
x^{\pr\mu}\fn{x}=x^\mu+\xi^\mu\fn{x},
	\label{infinitesimal GC}
}
we have
\al{
M^\mu{}_\nu\fn{x}
	&=	\delta^\mu_\nu+\Theta^\mu{}_\nu\fn{x},
 \label{app: eq: Mmunu}
 \\
\pa{M^{-1}}^\mu{}_\nu\fn{x}
	&=	\delta^\mu_\nu-\Theta^\mu{}_\nu\fn{x},
}
where
\al{
\Theta^\mu{}_\nu\fn{x}
	&:=	\p_\nu\xi^\mu\fn{x}.
		\label{Theta equals p xi}
}
This implies
\al{
\p_{[\rho}\Theta^\mu{}_{\nu]}\fn{x}
	&=	0.\label{diff condition}
}
Conversely, a function $\Theta^\mu{}_\nu\fn{x}$ that satisfies the condition~\eqref{diff condition} can always be written (locally) as Eq.~\eqref{Theta equals p xi}.
The condition~\eqref{diff condition} is the infinitesimal version of the GC condition~\eqref{GC condition on M}.

For the infinitesimal transformation~\eqref{infinitesimal GC}, the variation of the bases becomes
\al{
\delta_\tx{GC}\,\df x^\mu
	&=	\Theta^\mu{}_\nu\fn{x}\df x^\nu
	=	\df x^\nu\,\p_\nu\xi^\mu\fn{x},
		\label{GC on dx}\\
\delta_\tx{GC}\,\p_\mu
	&=	-\Theta^\nu{}_\mu\fn{x}\p_\nu
	=	-\p_\mu\xi^\nu\fn{x}\p_\nu;
		\label{GC on p}
}
of the gravitational fields
\al{
\delta_\tx{GC}\,e^\ba{}_\mu\fn{x}
	&=	-e^\ba{}_\nu\fn{x}\Theta^\nu{}_\mu\fn{x}
	=	-\p_\mu\xi^\nu\fn{x}e^\ba{}_\nu\fn{x},
		\label{inifinitesimal GC on e}\\
\delta_\tx{GC}\,\omega^\ba{}_{\bb\mu}\fn{x}
	&=	-\omega^\ba{}_{\bb\nu}\fn{x}\Theta^\nu{}_\mu\fn{x}
	=	-\p_\mu\xi^\nu\fn{x}\omega^\ba{}_{\bb\nu}\fn{x};
}
and of the matter fields
\al{
\delta_\tx{GC}\,\phi\fn{x}
	&=	0,\\
\delta_\tx{GC}\,\psi\fn{x}
	&=	0,\\
\delta_\tx{GC}\,\mc A_\mu\fn{x}
	&=	-\mc A_\nu\fn{x}\Theta^\nu{}_\mu\fn{x}
	=	-\p_\mu\xi^\nu\fn{x}\mc A_\nu\fn{x}.
		\label{inifinitesimal GC on A}
}
The 1-forms $e^\ba\fn{x}=e^\ba{}_\mu\fn{x}\df x^\mu$, $\omega^\ba{}_\bb\fn{x}=\mc \omega^\ba{}_{\bb\mu}\fn{x}\df x^\mu$, and $\mc A\fn{x}=\mc A_\mu\fn{x}\df x^\mu$ are trivially invariant under the GC transformation:
\al{
\delta_\tx{GC}\,e^\ba\fn{x}
	&=	-\p_\mu\xi^\nu\fn{x}e^\ba{}_\nu\fn{x}\df x^\mu
		+e^\ba{}_\mu\fn{x}\df x^\nu\,\p_\nu\xi^\mu\fn{x}
	=	0,\\
\delta_\tx{GC}\,\omega^\ba{}_\bb\fn{x}
	&=	-\p_\mu\xi^\nu\fn{x}\omega^\ba{}_{\bb\nu}\fn{x}\df x^\mu
		+\omega^\ba{}_{\bb\mu}\fn{x}\df x^\nu\,\p_\nu\xi^\mu\fn{x}
	=	0,\\
\delta_\tx{GC}\,\mc A\fn{x}
	&=	-\p_\mu\xi^\nu\fn{x}\mc A_\nu\fn{x}\df x^\mu
		+\mc A_\mu\fn{x}\df x^\nu\,\p_\nu\xi^\mu\fn{x}
	=	0.
}

For the infinitesimal GC-gauge transformation $M^\alpha{}_\beta\fn{x}=\delta^\alpha_\beta+\Theta^\alpha{}_\beta\fn{x}$, the GC-gauge field (as well as the Levi-Civita spin connection, both if they exist) transforms as
\al{
\delta_\tx{GC}\,\Upsilon^\alpha{}_{\beta\mu}\fn{x}
	&=	\Theta^\alpha{}_\gamma\fn{x}\Upsilon^\gamma{}_{\beta\mu}\fn{x}
		-\Upsilon^\alpha{}_{\delta\mu}\fn{x}\Theta^\delta{}_\beta\fn{x}
		-\Upsilon^\alpha{}_{\beta\nu}\fn{x}\Theta^\nu{}_\mu\fn{x}
		-\p_\mu\Theta^\alpha{}_\beta\fn{x}.
}
The next-to-last term comes from the rotation of the spacetime index and is peculiar to the GC-gauge field, compared to the ordinary and LL-gauge transformations in Eqs.~\eqref{infinitesimal ordinary gauge} and \eqref{infinitesimal LL in matrix} (or \eqref{infinitesimal LL more explicit}), respectively.
The last term is the inhomogeneous transformation that characterizes the gauge-field transformation.

It is important that the antisymmetric part has a vanishing inhomogeneous term under the GC transformation:
\al{
\delta_\tx{GC}\,\Upsilon^\alpha{}_{[\beta\mu]}\fn{x}
	&=	\Theta^\alpha{}_\gamma\fn{x}\Upsilon^\gamma{}_{[\beta\mu]}\fn{x}
		-\Upsilon^\alpha{}_{\delta[\mu}\fn{x}\Theta^\delta{}_{\beta]}\fn{x}
		-\Theta^\nu{}_{[\mu}\fn{x}\Upsilon^\alpha{}_{\beta]\nu}\fn{x},
}
due to the condition~\eqref{diff condition} for the GC transformation.
That is, the antisymmetric part is in vain for covariantizing the GC-covariant derivative.
This is the infinitesimal version of the discussion in the last paragraph in Sec.~\ref{GC gauge field section}.

\begin{figure}\centering
\includegraphics[width=0.8\textwidth]{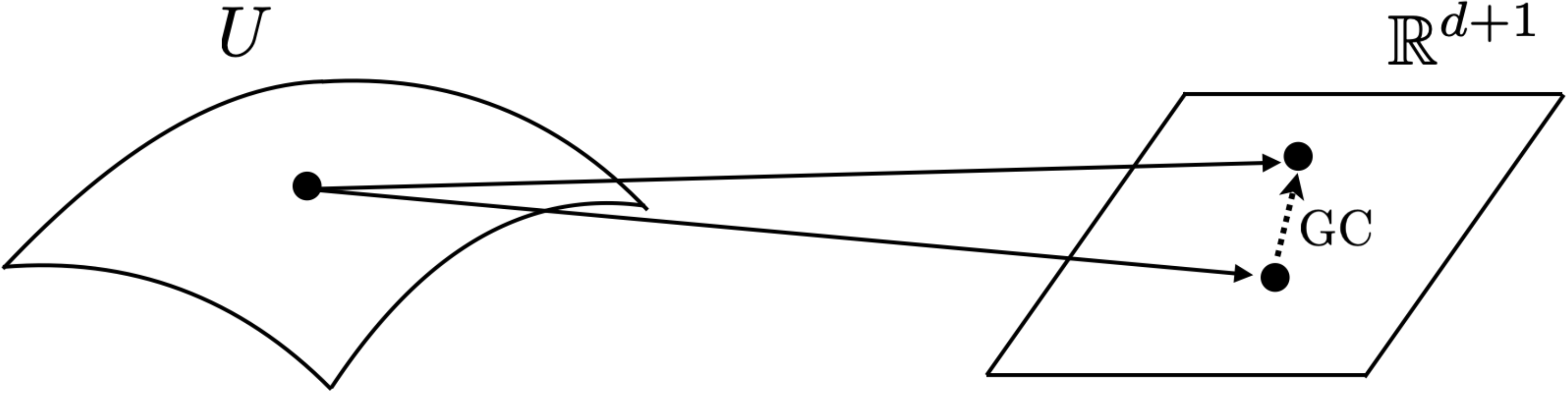}\\
\includegraphics[width=0.8\textwidth]{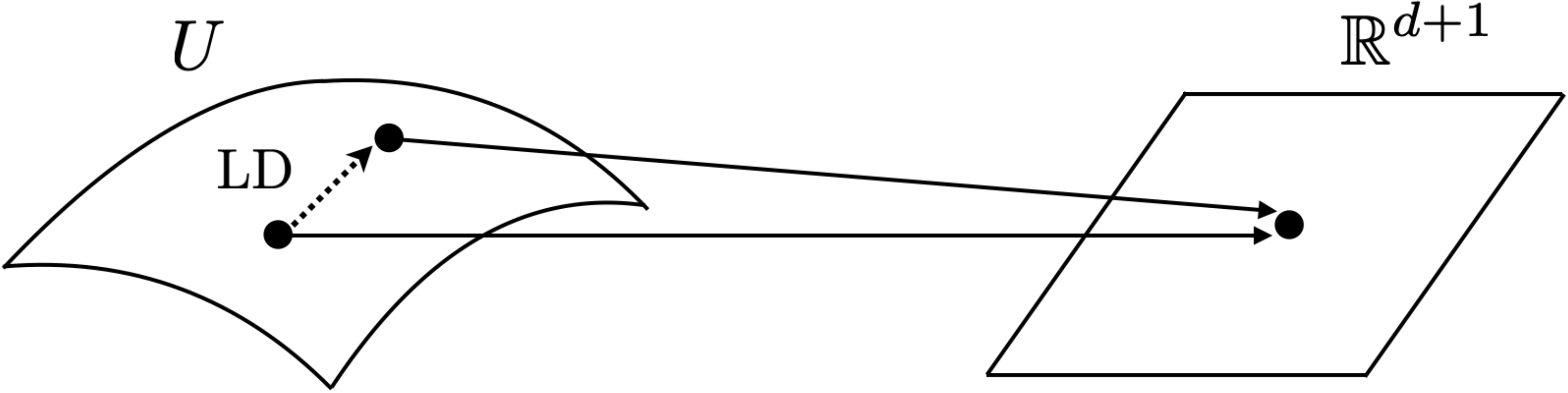}
\caption{Schematic figure for the GC (upper) and LD (lower) transformations.
}\label{GC and LD}
\end{figure}

\subsection{Lie-derivative transformation}\label{Lie derivative section}
Choosing the coordinate system (chart) $x$ for each open subset $U$ of a manifold $\mc M$ can be regarded as a diffeomorphism from $U$ to $\mathbb R^{d+1}$.
The GC transformation between two different coordinate systems is a map between two different diffeomorphisms $U\to\mathbb R^{d+1}$ from the same $U$; see the upper panel in Fig.~\ref{GC and LD}.

Instead, one may consider two different maps from $U$ to the same coordinate values in $\mathbb R^{d+1}$. The transformation between these two maps can also be regarded as a diffeomorphism. We can define a derivative of such a diffeomorphism, namely, the Lie derivative, which we will call the LD transformation below; see the lower panel in Fig.~\ref{GC and LD}.
In the literature~\cite{Daum:2009dn,Daum:2010bc,Daum:2010qt,Daum:2013fu}, the self-diffeomorphism group on $\mc M$ is called the diffeomorphism, or \diff\ in short, and this LD transformation can be regarded as its infinitesimal.\footnote{
Such a transformation can be interpreted as either ``passive'' or ``active,'' and it is claimed that they are distinct concepts; see, e.g.,\ Fig.~10 and Sec.~4 in Ref.~\cite{Gaul:1999ys}. The GC and LD transformations in our language may correspond to the passive and active interpretations, respectively.
}

A Lie derivative along an infinitesimal vector field $\xi^\mu\fn{x}$ is defined as the difference between fields on two distinct spacetime points that happen to have the same coordinate value $x'$ before and after the infinitesimal GC transformation~\eqref{infinitesimal GC}:
\al{
\mc L_\xi\Phi\fn{x}
	:=	\Phi\fn{x'}-\Phi'\fn{x'}
	&=	\Phi\Fn{x+\xi\fn{x}}-\Phi'\fn{x'}\nn
	&=	\Pn{\Phi\fn{x}+\xi\fn{x}\Phi\fn{x}}-\Phi'\fn{x'}
	=	\xi\fn{x}\Phi\fn{x}+\pn{\Phi\fn{x}-\Phi'\fn{x'}}\nn
	&=	\xi\fn{x}\Phi\fn{x}-\delta_\tx{GC}\Phi\fn{x},
	\label{Lie derivative given}
}
where $\xi\fn{x}$ in an argument (of $\Phi$ in the first line, in this case) denotes the $\pn{d+1}$ variables $\pn{\xi^0\fn{x},\dots,\xi^d\fn{x}}$, whereas those in other places denote the differential operator $\xi\fn{x}:=\xi^\mu\fn{x}\p_\mu$.
In other words, the Lie derivative~\eqref{Lie derivative given} is the difference between the original field at the GC-transformed point and the GC-transformed field.
In terms of the Lie derivative, the GC transformation can be written as
\al{
\delta_\tx{GC}\,\Phi\fn{x}
	&=	\xi\fn{x}\Phi\fn{x}-\mc L_\xi\Phi\fn{x}.
 \label{app:eq: GC trans}
}
By construction, the Lie derivative does not change the basis, unlike the GC transformation~\eqref{GC on dx} and \eqref{GC on p}.

We spell out the concrete forms using Eqs.~\eqref{inifinitesimal GC on e}--\eqref{inifinitesimal GC on A}:
\al{
\mc L_\xi\,e^\ba{}_\mu\fn{x}
	&=	\xi\fn{x}e^\ba{}_\mu\fn{x}+\p_\mu\xi^\nu\fn{x}e^\ba{}_\nu\fn{x},\\
\mc L_\xi\,\omega^\ba{}_{\bb\mu}\fn{x}
	&=	\xi\fn{x}\omega^\ba{}_{\bb\mu}\fn{x}+\p_\mu\xi^\nu\fn{x}\omega^\ba{}_{\bb\nu}\fn{x},
}
and
\al{
\mc L_\xi\,\phi\fn{x}
	&=	\xi\fn{x}\phi\fn{x},\\
\mc L_\xi\,\psi\fn{x}
	&=	\xi\fn{x}\psi\fn{x},\\
\mc L_\xi\,\mc A_\mu\fn{x}
	&=	\xi\fn{x}\mc A_\mu\fn{x}+\p_\mu\xi^\nu\fn{x}\mc A_\nu\fn{x}.
}
(In the language of differential geometry,
\al{
\mc L_\xi\,e^\ba\fn{x}
	&=	\xi\fn{x}e^\ba\fn{x}+\Braket{e^\ba\fn{x},\,\df\xi\fn{x}},\\
\mc L_\xi\,\omega^\ba{}_\bb\fn{x}
	&=	\xi\fn{x}\omega^\ba{}_\bb\fn{x}+\Braket{\omega^\ba{}_\bb\fn{x},\,\df\xi\fn{x}},\\
\mc L_\xi\,\mc A\fn{x}
	&=	\xi\fn{x}\mc A\fn{x}+\Braket{\mc A\fn{x},\,\df\xi\fn{x}},
}
where the exterior derivative on the vector field $\xi\fn{x}=\xi^\mu\fn{x}\p_\mu$ is given by $\df\xi\fn{x}:=\Pn{\p_\mu\xi^\nu\fn{x}\df x^\mu}\p_\nu$ and the inner product defined through that of the bases $\Braket{\df x^\lambda,\p_\nu}=\delta^\lambda_\nu$, reads
\al{
\Braket{\mc A\fn{x},\,\df\xi\fn{x}}
	&=	\Braket{\mc A_\lambda\fn{x}\df x^\lambda,\,\Pn{\p_\mu\xi^\nu\fn{x}\df x^\mu}\p_\nu}
	=	\mc A_\lambda\fn{x}\Pn{\p_\mu\xi^\nu\fn{x}\df x^\mu}\Braket{\df x^\lambda,\,\p_\nu}\nn
	&=	\p_\mu\xi^\nu\fn{x}\mc A_\nu\fn{x}\df x^\mu,
}
etc.

One might find it uneasy to transform the coordinate bases as in Eqs.~\eqref{GC on dx} and \eqref{GC on p}. Then it might be tempting to define another transformation, which we call the LD transformation:
\al{
\delta_\tx{LD}\,\Phi\fn{x}
	&:=	\delta_\tx{GC}\,\Phi\fn{x}-\xi\fn{x}\Phi\fn{x}
	=	-\mc L_\xi\,\Phi\fn{x}.
 \label{app: eq: LD trans}
}

An advantage of the GC transformation over the LD one is that the former commutes with gauge symmetries whereas the latter does not~\cite{Daum:2009dn,Daum:2010bc,Daum:2010qt,Daum:2013fu} because the latter is the difference between two distinct spacetime points, which are gauge transformed differently.

\subsection{(Semi)direct product of general-coordinate and gauge transformations}
\label{app: (Semi-)direct product}

Since the GC transformation acts on spacetime indices, it acts on those of the gauge field. Therefore,  when we apply both GC and gauge transformations for a system, one may worry about the order of transformations, that is, a GC transformation after a gauge transformation or the other way around. To clarify this, we show explicit computations.

We first perform a gauge transformation and then a GC transformation:
\al{
\Psi\fn{x}
	\sr{\tx{gauge}}\lra
	\wc\Psi\fn{x}
	&=	U\fn{x}\Psi\fn{x}\nn
	\sr{\tx{GC}}\lra
	\wc\Psi'\fn{x'}
	&=	U'\fn{x'}\Psi'\fn{x'}
	=	U\fn{x}\Psi\fn{x},\\
\mc A_\mu\fn{x}
	\sr{\tx{gauge}}\lra
	\wc A_\mu\fn{x}
	&=	U\fn{x}\mc A_\mu\fn{x}U^{-1}\fn{x}-\p_\mu U\fn{x}U^{-1}\fn{x}\nn
	\sr{\tx{GC}}\lra
	\wc{\mc A}'_\mu\fn{x'}
	&=	U'\fn{x'}\mc A'_\mu\fn{x'}U^{\pr-1}\fn{x'}-\p'_\mu U'\fn{x'}U^{\pr-1}\fn{x'}\nn
	&=	\sqbr{
			U\fn{x}\mc A_\nu\fn{x}U^{-1}\fn{x}
			-\p_\nu U\fn{x}U^{-1}\fn{x}}\mt{M^{-1}\fn{x}}^\nu{}_\mu,
}
where $U'\fn{x'}$ is the pullback defined by $U'\Fn{x'\fn{x}}=U\fn{x}$. In the opposite order, we obtain
\al{
\Psi\fn{x}
	\sr{\tx{GC}}\lra
	\Psi'\fn{x'}
	&=	\Psi\fn{x}\nn
	\sr{\tx{gauge}}\lra
	\wc{\Psi'}\fn{x'}
	&=	U\fn{x}\Psi\fn{x},\\
\mc A_\mu\fn{x}
	\sr{\tx{GC}}\lra
	\mc A'_\mu\fn{x'}
	&=	\mc A_\nu\fn{x}\mt{M^{-1}\fn{x}}^\nu{}_\mu\nn
	\sr{\tx{gauge}}\lra
	\wc{\mc A'}_\mu\fn{x'}
	&=	\sqbr{U\fn{x}\mc A_\nu\fn{x}U^{-1}\fn{x}-\p_\mu U\fn{x}U^{-1}\fn{x}}\mt{M^{-1}\fn{x}}^\nu{}_\mu.
}
Obviously, the GC and gauge groups commute each other: The generators of the gauge and GC transformations $\delta_\theta^\tx{gauge}$ and $\delta_\xi^\tx{GC}$ satisfy
\al{
&\commutator{\delta_\theta^\tx{gauge}}{\delta_\xi^\tx{GC}}\Psi\fn{x}=0,\\
&\commutator{\delta_\theta^\tx{gauge}}{\delta_\xi^\tx{GC}}\mc A_\mu\fn{x}=0.
}
where $\theta(x)$ and $\xi_\mu(x)$ are their transformation parameters, respectively.
Thus, they form a direct product: 
\al{
\tx{GC}\times\tx{gauge}.
}For instance, our action is invariant under $\tx{GC}\times SO(1,d)$.
Indeed, the commutativity between the GC and gauge groups is because of the definition of the GC transformations \eqref{eq: GC for vielbein}--\eqref{eq: GC for gauge field}.

In the literature (see e.g.\ Refs.~\cite{Daum:2009dn,Daum:2010bc,Daum:2010qt,Daum:2013fu}), instead of GC, one has imposed the symmetry~\eqref{app: eq: LD trans} that acts only on path-integrated quantum fields.
We write it $\diff^\tx{LD}$.
In $\diff^\tx{LD}$, the first term in Eq.~\eqref{app:eq: GC trans} is absent, which results in the nonvanishing commutator of $\diff^\tx{LD}$ and elements of a gauge transformation $\mf g$.
In particular, their commutator becomes the generator of the gauge transformation $\mf g$ with gauge parameters $-\xi^\nu\fn{x}\p_\nu\theta\fn{x}$.
In this case, the group becomes a semidirect product:
\al{
\diff^\tx{LD}\ltimes \tx{gauge}.
    \label{DHR}
}
This can be seen explicitly. To this end, let us deal with infinitesimal transformations.
First, $\diff^\tx{LD}$ and subsequent gauge transformations yield
\al{
\Psi\fn{x}
    &\sr{\diff^\tx{LD}}\lra    \Psi\fn{x}-\xi^\mu\fn{x}\p_\mu\Psi\fn{x}\nn
    &\sr{\tx{gauge}}\lra    \Psi\fn{x}+\theta\fn{x}\Psi\fn{x}-\xi^\mu\fn{x}\p_\mu \theta\fn{x}\Psi\fn{x}
    -\xi^\mu\fn{x}\p_\mu\Psi\fn{x}
    -\xi^\mu\fn{x}\theta\fn{x}\p_\mu\Psi\fn{x},\\
\mc A_\mu\fn{x}
    &\sr{\diff^\tx{LD}}\lra    \mc A_\mu\fn{x}-\xi^\nu\fn{x}\p_\nu \mc A_\mu\fn{x}-\mc A_\nu\fn{x}\p_\mu\xi^\nu\fn{x}\\
    &\sr{\tx{gauge}}\lra    \mc A_\mu\fn{x}+\theta\fn{x}\mc A_\mu\fn{x}-\mc A_\mu\fn{x}\theta\fn{x}
            -\p_\mu\theta\fn{x}\nn
    &\qquad\quad
            -\xi^\nu\fn{x}\p_\nu\Pn{
                \mc A_\mu\fn{x}+\theta\fn{x}\mc A_\mu\fn{x}-\mc A_\mu\fn{x}\theta\fn{x}-\p_\mu\theta\fn{x}
                }\nn
    &\qquad\quad
            -\Pn{
                \mc A_\nu\fn{x}
                +\theta\fn{x}\mc A_\nu\fn{x}
                -\mc A_\nu\fn{x}\theta\fn{x}
                -\p_\nu\theta\fn{x}
                }\p_\mu\xi^\nu\fn{x}\nn
    &\quad =  \mc A_\mu+\theta\mc A_\mu-\mc A_\mu\theta-\p_\mu\theta\nn
    &\qquad\quad
        -\xi\mc A_\mu
        -\xi\theta\,\mc A_\mu-\theta\,\xi\mc A_\mu
        +\xi\mc A_\mu\,\theta+\mc A_\mu\,\xi\theta
        +\xi\p_\mu\theta\nn
    &\qquad\quad
        -\Pn{
                \mc A_\nu
                +\theta\mc A_\nu
                -\mc A_\nu\theta
                -\p_\nu\theta
                }\p_\mu\xi^\nu.
}
Note that $\mc A_\mu=\mc A_\mu^aT^a$ and $\theta=\theta^aT^a$ do not commute here.
On the other hand, a gauge transformation and a subsequent $\diff^\tx{LD}$ yield
\al{
\Psi\fn{x}
    &\sr{\tx{gauge}}\lra    \Psi\fn{x}+\theta\fn{x}\Psi\fn{x}\nn
    &\sr{\diff^\tx{LD}}\lra    \Psi\fn{x}-\xi^\mu\fn{x}\p_\mu\Psi\fn{x}
            +\theta\fn{x}\Psi\fn{x}
            -\theta\fn{x}\xi^\mu\fn{x}\p_\mu\Psi\fn{x},\\
\mc A_\mu\fn{x}
    &\sr{\tx{gauge}}\lra    \mc A_\mu\fn{x}+\theta\fn{x}\mc A_\mu\fn{x}-\mc A_\mu\fn{x}\theta\fn{x}-\p_\mu\theta\fn{x}\nn
    &\sr{\diff^\tx{LD}}\lra    \mc A_\mu\fn{x}-\xi^\nu\fn{x}\p_\nu \mc A_\mu\fn{x}-\mc A_\nu\fn{x}\p_\mu\xi^\nu\fn{x}\nn
    &\qquad\quad
            +\theta\fn{x}\Pn{
                \mc A_\mu\fn{x}-\xi^\nu\fn{x}\p_\nu \mc A_\mu\fn{x}-\mc A_\nu\fn{x}\p_\mu\xi^\nu\fn{x}
                }\nn
    &\qquad\quad
            -\Pn{
                \mc A_\mu\fn{x}-\xi^\nu\fn{x}\p_\nu \mc A_\mu\fn{x}-\mc A_\nu\fn{x}\p_\mu\xi^\nu\fn{x}
                }\theta\fn{x}\nn
    &\qquad\quad
            -\p_\mu\theta\fn{x}\nn
    &\quad=   \mc A_\mu-\xi\mc A_\mu-\mc A_\nu\p_\mu\xi^\nu\nn
    &\qquad\quad
            +\theta\Pn{
                \mc A_\mu-\xi\mc A_\mu-\mc A_\nu\p_\mu\xi^\nu
                }\nn
    &\qquad\quad
            -\Pn{
                \mc A_\mu-\xi\mc A_\mu-\mc A_\nu\p_\mu\xi^\nu
                }\theta
            -\p_\mu\theta.
}
Subtracting these two, we obtain
\al{
&\commutator{\delta_\theta^\tx{gauge}}{\delta_\xi^\tx{LD}}\Psi\fn{x}
    =  -\xi^\mu\fn{x}\p_\mu\theta\fn{x}\Psi\fn{x},\\
&\commutator{\delta_\theta^\tx{gauge}}{\delta_\xi^\tx{LD}}\mc A_\mu\fn{x}
    =   -\xi^\nu\fn{x}\p_\nu\theta\fn{x}\mc A_\mu\fn{x}
        +\mc A_\mu\fn{x}\xi^\nu\fn{x}\p_\nu\theta\fn{x}
        +\xi^\nu\fn{x}\p_\nu\p_\mu\theta\fn{x}
        +\p_\nu\theta\fn{x}\p_\mu\xi^\nu\fn{x}.
}
The commutator becomes the extra gauge transformation with the gauge parameter $\mc L_\xi\theta\fn{x}=-\xi\theta=-\xi^\nu\fn{x}\p_\nu\theta\fn{x}$.
For this noncommutativity, it is important that $\diff^\tx{LD}$ does \emph{not} transform $\theta\fn{x}$ by assumption.
To avoid noncommutativity of LD and \diff, one may further introduce a modified $\diff^\tx{LD}$ denoted by $\wt\delta^\tx{LD}_\xi$ to make it commute with $\mf g$~\cite{Daum:2010qt,Daum:2013fu,Daum:2009dn,Daum:2010bc}.

We comment on the global Poincar\'e transformation $ISO(1,d)$ in the Minkowski space $\mathbf M^{1,d}$. The global Poincar\'e transformation $ISO\fn{1,d}$ contains the translation in $\mathbf M^{1,d}$ as a normal subgroup in the sense that $x\to \Lambda x\to \Lambda x+a\to \Lambda^{-1}\fn{\Lambda x+a}=x+\Lambda^{-1}a$. Since $SO(1,d)\simeq ISO\fn{1,d}/T\fn{1,d}$, we write\footnote{
In the case of $SU(2)\to U(1)$ breaking, i.e. $T\simeq SU(2)/U(1)$,
we write
\al{
SU(2)&=T\ltimes U(1).
}
For the SM, $SU(2)\times U(1)_Y\to U(1)_Q$, we have $T \simeq \pn{SU(2)\times U(1)_Y}/U(1)_Q$ and then write
\al{
SU(2)\times U(1)_Y
    &=  T\ltimes U(1)_Q.
}
}
\al{
ISO\fn{1,d}
    =   T\fn{1,d}\rtimes  SO\fn{1,d}.
}
The local version of $T\fn{1,d}\rtimes  SO\fn{1,d}$ is given by 
\al{
\diff \rtimes SO\fn{1,d},
}
which is opposite Eq.~\eqref{DHR}.
In a gravitational theory based on the global Poincar\'e transformation, we infer to realize the symmetry breaking
\al{
ISO(1,d)=T\fn{1,d} \rtimes  SO\fn{1,d}
    \to \diff.
}

\section{Comment on Lie derivative on spinor}
\label{Lie derivative on spinor}

In this appendix, several definitions of the Lie derivative acting on the spinor are argued.

Once the background-covariant derivative $\bar\D$ is defined for the matter field $\Psi$, we may consider a parallel transport with respect to $\bar\D$:
\al{
\Psi'\fn{x+\xi}
    &=  \Psi\fn{x}-\xi^\mu\bar\D_\mu\Psi\fn{x}.
        \label{parallel transport on matter}
}
Here, we stress that this parallel transport differs from the GC transformation in Eqs.~\eqref{eq: GC for scalar}--\eqref{eq: GC for gauge field} in the sense that the transport~\eqref{parallel transport on matter} compares fields on physically distinct points that happen to have the same coordinate values before and after the GC transformation.

The parallel transport~\eqref{parallel transport on matter} induces another Lie derivative of $\Psi$:
\al{
L_\xi\Psi\fn{x}
    &=  \xi^\mu\fn{x}\bar\D_\mu\Psi\fn{x}.
}
On spinors, this is nothing but the Lie derivative introduced by Weyl~\cite{Weyl:1929fm},
\al{
L_\xi\psi\fn{x}
    &=  \xi^\mu\fn{x}\pn{\p_\mu\psi\fn{x}+{\bar\omega_{\ba\bb\mu}\fn{x}\ov2}\Sigma^{\ba\bb}\psi\fn{x}},
}
if $\bar\omega^\ba{}_{\bb\mu}$ is identified to the Levi-Civita spin connection~\eqref{eq: Omegaabmu}.\footnote{
As said above, whether or not $\bar\omega^\ba{}_{\bb\mu}$ coincides with $\bar\Omega^\ba{}_{\bb\mu}$ is to be determined dynamically in our formalism.
}

One may further extend the above definition to the following form~\cite{Kosmann:1971} (see also Ref.~\cite{Ortin:2002qb}\footnote{
In Ref.~\cite{Ortin:2002qb}, the original Lie derivative by Weyl is said to be $\mc L_U\psi=U^\mu\p_\mu\psi$ without the background LL connection, corresponding to the transformation~\eqref{eq: GC for spinor}.
})
\al{
\mathbb L_\xi\psi\fn{x}
    &:=  \xi^\mu\fn{x}\bar\D_\mu\psi\fn{x}
        +{\bar\D_{[\mu} \xi_{\nu]}\fn{x}\ov4}\bar e^\mu{}_\ba\fn{x}\bar e^\nu{}_\bb\fn{x}\Sigma^{\ba\bb}\psi\fn{x}.
            \label{detour notation}
}
This extension is motivated by the fact that on the flat Minkowski background $\bar e^\ba{}_\mu=\delta^\ba_\mu$ and $\bar\omega^\ba{}_{\bb\mu}=0$, there remains the global $SO(1,d)$ invariance under
\al{
x^\mu\to x^{\pr\mu}=\Lambda^\mu{}_\nu x^\nu=\pn{\delta^\mu_\nu+\theta^\mu{}_\nu+\cdots} x^\nu
= x^\mu + \xi^\mu,
\label{eq: global SO1d transformation}
}
namely
$\xi^\mu\fn{x}=\theta^\mu{}_\nu x^\nu+\cdots$. This is the same as Eq.~\eqref{app: eq: Mmunu} with Eq.~\eqref{Theta equals p xi}. Hence, Eq.~\eqref{detour notation} becomes equivalent to the Lie derivative obtained when the background GC transformation reduces to the global $SO(1,d)$ on the flat background, namely when $M^\mu{}_\nu\fn{x}\to\Lambda^\mu{}_\nu$: \al{
\psi\fn{x}\to\psi'\fn{x'}
    =   \pn{1+{\theta_{\mu\nu}\ov2}\Sigma^{\mu\nu}+\cdots}\psi\fn{x}.
        \label{global Lorentz on spinor}
}
The transformation~\eqref{global Lorentz on spinor} corresponds just to the global $SO(1,d)$ transformation for the spinor.
The definition~\eqref{detour notation} may however be a detour notation in our case. As discussed in Section~\ref{sec: Global background Lorentz invariance after spontaneous symmetry breaking}, in our formulation, the global Lorentz $SO(1,d)$ transformation is accidentally realized as a diagonal subgroup of $SO(1,d)\times \tx{GC}$, so that the detour notation is not necessary.

\section{Degenerate limit of vierbein}
\label{app: Degenerate limit of vierbein}

In this appendix, we explain the detailed definition of the degenerate limit and show its application for explicit several examples.

\subsection{General definition of degenerate limit}
As discussed in the Introduction, we assume that the action at $\Lambda_\tx{G}$ admits the weak-field limit $e^\ba{}_\mu\fn{x}\to0$ just as the SM action does for the limit $H\fn{x}\to0$, etc.
In particular, we postulate that the action admits the degenerate limit for any combination of the $\pn{d+1}^2=16$ components of the vierbein~\cite{Matsuzaki:2020qzf}. This requirement puts a more severe constraint on the action than just requiring a simultaneous limit for all the components, as we will see.  The degenerate configuration of the vierbein appears in the topology change of the background spacetime and is expected to play an important role in quantum gravity~\cite{Tseytlin:1981ks,Horowitz:1990qb}.

In general, a vierbein $e^\ba{}_\mu$ has four eigenvalues in four-dimensional spacetime. In the degenerate limit, at least one eigenvalue goes to zero, resulting in the determinant of the vierbein to be zero: $\ab{e}\to0$.
Note that the degenerate limit does not necessarily mean the null limit $e^\ba{}_\mu\to0$ for all 16 components.
Let us see this fact using a specific example.
By a rescaling $e^\ba{}_\mu\mapsto Ce^\ba{}_\mu$, we obtain $\ab{e}\mapsto C^4\ab{e}$ and $g^{\mu\nu}\mapsto C^{-2}g^{\mu\nu}$, and hence the term 
\al{
\ab{e}g^{\mu\nu}\p_\mu\phi\p_\nu\phi
	&\mapsto	C^2\ab{e}g^{\mu\nu}\p_\mu\phi\p_\nu\phi
}
does not diverge in the null limit $C\to0$. However, this term is in general divergent in a degenerate limit as we will see below.

Now, we define the degenerate limit. The power for the degenerate limit can be counted by $\ab{e}$ rather than by the overall normalization factor $C$:
The inverse of the vielbein and metric, $e_\ba{}^\mu$ and $g^{\mu\nu}$, contains one and two factors of $\ab{e}^{-1}$, respectively,
\al{
e_\ba{}^\mu
	&=	{C_\ba{}^\mu\ov\ab{e}},&
g^{\mu\nu}
	&=	{\eta^{\ba\bb}C_\ba{}^\mu C_\bb{}^\nu\ov\ab{e}^2},
}
where the transpose of the cofactor matrix of $e^\ba{}_\mu$ is denoted by $C_\ba{}^\mu$, which remains finite in the degenerate limit.
Hereafter, we write the power of $\ab{e}$ as
\al{
\powe{e_\ba{}^\mu}&=-1,&
\powe{g^{\mu\nu}}&=-2,&
\tx{etc.}
}
Each upper greek index of the vielbein or metric serves an extra $-1$ power of $\ab{e}$, and its power $-1$ cancels the power $+1$ from a lower index of the metric or vielbein.

This can be explicitly seen as follows. The $D^2=\pn{d+1}^2$ degrees of freedom of the vielbein can be parametrized as
\al{
\bmat{e^\ba{}_\mu}_{\ba,\mu=0,\dots,d}
	&=	\Lambda\diag\fn{\lambda_0,\dots,\lambda_d}M^\t,
	\label{eq: diagonalized vierbein}
}
where $\pn{\lambda_0,\dots,\lambda_d}\in\mathbb R^D$ with $\lambda_0<0$ and $\lambda_i>0$  ($i=1,\dots,d$) and each of $\Lambda,M\in SO_+(1,d)$ has ${D\pn{D-1}\ov2}$ degrees of freedom.\footnote{
Percacci has generalized $M$ to be $M\in GL(D)$ having $D^2$ degrees of freedom.
}
In the degenerate limit, an eigenvalue $\lambda_a$ goes to zero: $\lambda_a\to0$.
The determinant reads
\al{
\ab{e}
	&=	\lambda_0\cdots\lambda_d,
	\label{eq: product of eigenvalues of vielbein}
}
while the metric and its inverse are
\al{
\bmat{g_{\mu\nu}}_{\mu,\nu=0,\dots,d}
	&=	\pn{\Lambda\diag\fn{\lambda_0,\dots,\lambda_d}M^\t}^\t\eta\pn{\Lambda\diag\fn{
\lambda_0,\dots,\lambda_d}M^\t}\nn
	&=	M\diag\fn{-\lambda_0^2,\lambda_1^2,\dots,\lambda_d^2}M^\t,\\
\bmat{g^{\mu\nu}}_{\mu,\nu=0,\dots,d}
	&=	M\diag\fn{-\lambda_0^{-2},\lambda_1^{-2},\dots,\lambda_d^{-2}}M^\t.
 \label{eq: inverse metric in appendix}
}
We see that a contraction cancels a power: For example,
\al{
\bmat{e^\ba{}_\mu g^{\mu\nu}}_{\ba,\nu=0,\dots,d}
	&=	\Lambda\diag\fn{-\lambda_0^{-1},\lambda_1^{-1},\dots,\lambda_d^{-1}}M^\t
}
gives the power $\powe{e^\ba{}_\mu g^{\mu\nu}}=-1$. In one more example,
for $\nabla_\mu e^\ba{}_\nu=\p_\mu e^\ba{}_\nu-e^\ba{}_\lambda\Gamma^\lambda{}_{\nu\mu}$ with the Levi-Civita connection \eqref{Levi-Civita given}, we have
\al{
e^\ba{}_\lambda\Gamma^\lambda{}_{\nu\mu}
    &=  e^\ba{}_\lambda{g^{\lambda\lambda'}\ov2}\pn{-\p_{\lambda'}g_{\nu\mu}+\p_\nu g_{\mu\lambda'}+\p_\mu g_{\lambda'\nu}},
}
which gives $\powe{e^\ba{}_\lambda\Gamma^\lambda{}_{\nu\mu}}=-1$ because the first term does not contain a vierbein whose $\lambda'$ leg is to be contracted.

Other examples are in order: The Levi-Civita connection $\Gamma^\lambda{}_{\nu\mu}$ contains two extra inverse powers of $\ab{e}$ coming from $g^{\mu\nu}$, $\powe{\Gamma^\mu{}_{\rho\sigma}}=-2$, while the Levi-Civita spin connection has $\powe{\Omega^\ba{}_{\bb\mu}}=-2$ because of $\powe{e^\ba{}_\lambda g^{\lambda\rho}}=-1$ and hence $\powe{e^\ba{}_\lambda\Gamma^\lambda{}_{\sigma\mu}}=-1$ (contraction of $e_\bb{}^\sigma$ with $\Gamma^\lambda{}_{\sigma\mu}$ does give the additional power $-1$ because the latter contains the index from $\p_\sigma$, which does not come from the vielbein or from metric).
Note that the contraction of $\Gamma^\lambda{}_{\nu\mu}$ with $e_\ba{}^\nu$ does raise the power by 1 because the former contains the second term in the parentheses in Eq.~\eqref{Levi-Civita given}, $\p_\nu g_{\mu\lambda}$, whose lower $\rho$ index comes from the derivative.

\subsection{Concrete examples}
Here, we more explicitly show the degenerate limit on various terms and list the terms prohibited by having a negative power of eigenvalues of $e^\ba{}_\mu$.
\begin{itemize}
\item The kinetic term of a scalar
\al{
S_\phi=\int\df^4x\ab{e}\sqbr{-{1\ov2}g^{\mu\nu}\paren{\p_\mu\phi}\paren{\p_\nu\phi}}
}
is prohibited because from Eqs.~\eqref{eq: product of eigenvalues of vielbein} and \eqref{eq: inverse metric in appendix}
\al{
\ab{e}g^{\mu\nu}&\propto (\lambda_0\cdots\lambda_3)\diag\fn{-\lambda_0^{-2},\lambda_1^{-2},\lambda_2^{-2},\lambda_3^{-2}}\nn
&\quad=\diag\fn{-{\lambda_1\lambda_2\lambda_3\ov\lambda_0},{\lambda_0\lambda_2\lambda_3\ov\lambda_1},{\lambda_0\lambda_1\lambda_3\ov\lambda_2},{\lambda_0\lambda_1 \lambda_2\ov\lambda_3}}.
    \label{inverse metric counting}
}
When some eigenvalues become zero $\lambda_a\to0$, the matrix contains a divergent component $\propto\lambda_a^{-1}$.
We note that Eq.~\eqref{eq: determinant ee} cannot be used in Eq.~\eqref{inverse metric counting} since $\ab{e}g^{\mu\nu}=\ab{e}e_{(\ba}{}^\mu e_{\bb)}{}^\nu\eta^{\ba\bb}$ which is different from $\ab{e}e_{[\ba}{}^\mu e_{\bb]}{}^\nu$ to be vanishing for the contraction with $\eta^{\ba\bb}$.

\item The symmetric kinetic term for the vierbein 
\al{
S_e=\int\df^4x\ab{e}\sqbr{-{Z_e\ov2}g^{\mu\mu'}g^{\nu\nu'}\eta_{\ba\bb}\paren{\nabla_\mu e^\ba{}_\nu}\paren{\nabla_{\mu'} e^\bb{}_{\nu'}}}
}
(with a factor $Z_e$ being mass dimension $2$) is prohibited because
\al{
\ab{e}g^{\nu\nu'}\eta_{\ba\bb} e^\ba{}_\nu  e^\bb{}_{\nu'}&\propto (\lambda_0\cdots\lambda_3)\diag\fn{-\lambda_0^{-2},\lambda_1^{-2},\lambda_2^{-2},\lambda_3^{-2}}\nn
&\qquad\qquad\times \diag\fn{-\lambda_0,\lambda_1,\lambda_2,\lambda_3}\diag\fn{-\lambda_0,\lambda_1,\lambda_2,\lambda_3} \nn
&\quad=\diag\fn{
-{\lambda_1\lambda_2\lambda_3\ov\lambda_0},
{\lambda_0\lambda_2\lambda_3\ov\lambda_1},
{\lambda_0\lambda_1\lambda_3\ov\lambda_2},
{\lambda_0\lambda_1 \lambda_2\ov\lambda_3}
}.
\label{eq: dete gg}
}
The $a$th diagonal element of this matrix has $\lambda_a^{-1}$ which diverges for $\lambda_a\to0$.

\item The antisymmetrized kinetic term
\al{
S_{e,\tx{antisym}}=\int\df^Dx\ab{e}\sqbr{-{1\ov2}g^{\mu\mu'}g^{\nu\nu'}\paren{\p_{[\mu} e_{\nu]}}\paren{\p_{[\mu'} e_{\nu']}}}
}
is prohibited.
Note that this term is GC invariant because $\nabla_{[\mu}e^\ba{}_{\nu]}=\p_{[\mu}e^\ba{}_{\nu]}$ due to the  torsion-free identity of the Levi-Civita connection $\Gamma^\mu{}_{[\rho\sigma]}=0$.
Even though the power $\powe{\nabla_{\mu}e^\ba{}_{\nu}}=-1$ is raised to $\powe{\p_{[\mu}e^\ba{}_{\nu]}}=0$, we still have the total power $-1$.
\item The gauge kinetic term \al{
S_{A}=\int\df^4x\ab{e}\sqbr{+{1\ov2}g^{\mu\mu'}g^{\nu\nu'}F^\ba{}_{\bb\mu\nu}F^\bb{}_{\ba\mu'\nu'}}
}
is proportional to $\lambda_a^{-3}$ because
\al{
\ab{e}g^{\mu\mu'}g^{\nu\nu'}&\propto (\lambda_0\cdots\lambda_3)\diag\fn{-\lambda_0^{-2},\lambda_1^{-2},\lambda_2^{-2},\lambda_3^{-2}}\diag\fn{-\lambda_0^{-2},\lambda_1^{-2},\lambda_2^{-2},\lambda_3^{-2}}\nn
&\quad=\diag\fn{
-{\lambda_1\lambda_2\lambda_3\ov\lambda_0^3},
{\lambda_0\lambda_2\lambda_3\ov\lambda_1^3},
{\lambda_0\lambda_1\lambda_3\ov\lambda_2^3},
{\lambda_0\lambda_1 \lambda_2\ov\lambda_3^3}
}
\label{eq: dete gg 2}
}
and thus diverges for $\lambda_a\to 0$. Therefore, the gauge kinetic term is not compatible with the degenerate limit.

\item The Einstein-Hilbert action solely made of $e$, \al{
S_\tx{EH}=\int\df^Dx\ab{e}R
}
The Riemann and Ricci tensors that are solely made of $e$ are $R^\mu{}_{\nu\rho\sigma}=\p_\rho\Gamma^\mu{}_{\nu\sigma}-\p_\sigma\Gamma^\mu{}_{\nu\rho}+\Gamma^\mu{}_{\lambda\rho}\Gamma^\lambda{}_{\nu\sigma}-\Gamma^\mu{}_{\lambda\sigma}\Gamma^\lambda{}_{\nu\rho}$ and $R_{\nu\sigma}=\p_\mu\Gamma^\mu{}_{\nu\sigma}-\p_\sigma\Gamma^\mu{}_{\nu\mu}+\Gamma^\mu{}_{\lambda\mu}\Gamma^\lambda{}_{\nu\sigma}-\Gamma^\mu{}_{\lambda\sigma}\Gamma^\lambda{}_{\nu\mu}$. Both give the power $-4$. The Ricci scalar $R=g^{\nu\sigma}\paren{\p_\mu\Gamma^\mu{}_{\nu\sigma}-\p_\sigma\Gamma^\mu{}_{\nu\mu}+\Gamma^\mu{}_{\lambda\mu}\Gamma^\lambda{}_{\nu\sigma}-\Gamma^\mu{}_{\lambda\sigma}\Gamma^\lambda{}_{\nu\mu}}$, which is solely made of $e$, gives the same power $-4$.\footnote{
In $\Gamma^\mu{}_{\lambda\sigma}\Gamma^\lambda{}_{\nu\mu}$, the power $-4$ term is one in which both the lower indices $\mu$ and $\lambda$ come from derivatives. In that case, the lower indices $\nu$ and $\sigma$ are from metrics, and hence erase the power from $g^{\nu\sigma}$.
}
Therefore, it has the power $-3$.
\end{itemize}
These terms are forbidden in the action consistent with the degenerate limit.

In contrast, the following operators can admit the degenerate limit:
\begin{itemize}
\item The cosmological constant term
\al{
\int \df^D x\ab{e}
\label{eq: cosmilogical constant term}
}
is obviously not divergent for $\lambda_a\to0$ thanks to Eq.~\eqref{eq: product of eigenvalues of vielbein}.
\item The linear term in $F^{\ba\bb}{}_{\mu\nu}$, namely 
\al{
S=\int \df^D x\ab{e} e_{[\ba}{}^\mu e_{\bb]}{}^\nu F^{\ba\bb}{}_{\mu\nu}
}
does not diverge because the use of Eq.~\eqref{eq: determinant ee} yields
\al{
\ab{e} e_{[\ba}{}^\mu e_{\bb]}{}^\nu&=
{1\ov2}\ep{\ba\bb\bc\bd} e^\bc{}_\rho e^\bd{}_\sigma \ep{\mu\nu\rho\sigma}
\propto \Big(\diag\fn{-\lambda_0,\lambda_1,\lambda_2,\lambda_3}\Big)^2\nn
&\quad 
=\diag\fn{\lambda_0^2,\lambda_1^2,\lambda_2^2 ,\lambda_3^2}.
}
\item The kinetic term of the spinor field \al{S= \int \df^Dx \ab{e} \ol\psi e_\ba{}^\mu \gamma^\ba {\mathcal D}_\ba\psi
\label{eq: kinetic term of spinor}
}
has
\al{
\ab{e} e_\ba{}^\mu
&=	{1\ov3!}\ep{\ba\bb\bc\bd} e^\bb{}_\nu e^\bc{}_\rho e^\bd{}_\sigma \ep{\mu\nu\rho\sigma}
\propto  \Big( \diag\fn{-\lambda_0,\lambda_1,\lambda_2,\lambda_3}\Big)^3\nn
&\quad
=\diag\fn{-\lambda_0^3,\lambda_1^3,\lambda_2^3,\lambda_3^3}.
}
This term does not contain any inverse of eigenvalues of the vierbein field, so that it is free from divergences for $\lambda_a\to0$. Note that the mass term of the spinor field is also accepted.
\end{itemize}
Hence, imposing the degenerate limit on the action accepts only these three terms. In particular only spinor fields have their kinetic terms, i.e., become dynamical.

\subsection{Rarita-Schwinger field}\label{Rarita-Schwinger field}
Here we examine if the spin-$3/2$ Rarita-Schwinger field is compatible with the irreversible vierbein postulate.

When we regard the lower-indexed $\psi_\mu$ as the fundamental field, its free action is
\al{
S   &=  \int\df^4x\ab{e\fn{x}}\sqbr{
        -i\vep^{\mu\nu\rho\sigma}\fn{x}\ol\psi_\mu\fn{x}\gamma_5\gamma_\nu\fn{x}\p_\rho\psi_\sigma\fn{x}
        }\nn
    &=  \int\df^4x\sqbr{
        i\ep{\mu\nu\rho\sigma}\ol\psi_\mu\fn{x}\gamma_5\gamma_\nu\fn{x}\p_\rho\psi_\sigma\fn{x}
        }.
}
When we regard the upper-indexed one $\psi^\mu$ fundamental, its free action is
\al{
S   &=  \int\df^4x\ab{e\fn{x}}\sqbr{
        -i\vep_{\mu\nu\rho\sigma}\fn{x}\ol\psi^\mu\fn{x}\gamma_5\gamma^\nu\fn{x}\p^\rho\psi^\sigma\fn{x}
        }\nn
    &=  \int\df^4x\ab{e\fn{x}}^2\sqbr{
        i\ep{\mu\nu\rho\sigma}\ol\psi^\mu\fn{x}\gamma_5e{}_\ba{}^\nu\fn{x}\gamma^\ba\fn{x}g^{\rho\rho'}\fn{x}\p_{\rho'}\psi^\sigma\fn{x}
        }.
}
The former action is consistent with the irreversible vierbein postulate, whereas the latter is not.
The compatibility of the Rarita-Schwinger field with the irreversible vierbein postulate is contingent upon whether we consider the field with upper or lower indices as the fundamental entity.

\section{Topological terms}\label{topological section}
There are four topological terms: (i) the Immirzi term, (ii) the Nieh-Yan invariant, (iii) the Pontryagin index, and (iv) the Euler number; see e.g.\ Ref.~\cite{Freidel:2005sn} for others.\footnote{
Recently, another topological term $\int\df\pn{e^\ba\wedge\star T_\ba}$ is proposed~\cite{Giacomo:2023ujb}. This term would be interesting to study on its own, though it is incompatible with the irreversible vierbein postulate because it contains the Hodge dual of the 2-form and hence two inverse metrics.
}
More specifically, they are given in the language of differential forms by
\al{
S_\text{Immirzi}
	&=	{1\ov2}\int \os\omega{\mc F}^{\ba\bb}\wedge e_\ba \wedge e_\bb
	=	\int\df^4x\ep{\mu\nu\rho\sigma}{1\ov4}
		\os\omega{\mc F}^{\ba\bb}{}_{\mu\nu}\fn{x}e_\ba{}_\rho\fn{x}e_\bb{}_\sigma\fn{x},\\
S_\text{Nieh-Yan}
	&=	{1\ov2}\int\df\fn{e^\ba\wedge T_\ba}
	=	{1\ov2}\int \pn{T^\ba\wedge T_\ba-\os\omega{\mc F}_{\ba\bb}\wedge e^\ba\wedge e^\bb}\nn
	&=	\int\df^4x\ep{\mu\nu\rho\sigma}\pn{
			{1\ov8}
			T^\ba{}_{\mu\nu}\fn{x}T_{\ba\rho\sigma}\fn{x}
			-{1\ov4}\os\omega{\mc F}_{\ba\bb\mu\nu}\fn{x}e^\ba{}_\rho\fn{x}e^\bb{}_\sigma\fn{x}
			},\\
S_\text{Pontryagin}
	&=	{1\ov2}\int \os\omega{\mc F}^{\ba\bb}\wedge \os\omega{\mc F}_{\ba\bb}
	=	\int\df^4x\ep{\mu\nu\rho\sigma}{1\ov8}F^{\ba\bb}{}_{\mu\nu}\fn{x}F_{\ba\bb\rho\sigma}\fn{x},\\
S_\text{Euler}
	&=	{1\ov8}\int \epsilon_{\ba\bb\bc\bd} \os\omega{\mc F}^{\ba\bb} \wedge \os\omega{\mc F}^{\bc\bd}
	=	{1\ov32}\int\df^4x\ep{\mu\nu\rho\sigma}\epsilon_{\ba\bb\bc\bd} \os\omega{\mc F}^{\ba\bb}{}_{\mu\nu}\fn{x}\os\omega{\mc F}^{\bc\bd}{}_{\rho\sigma}\fn{x},
}
where we have omitted the coupling constants. We define 
\al{
T^\ba
	&:=	\df e^\ba+\omega^\ba{}_\bb\wedge e^\bb
	=	{1\ov2}\pn{\p_\mu e^\ba{}_\nu\fn{x}+\omega^\ba{}_{\bb\mu}\fn{x}e^\bb{}_\nu\fn{x}}\df x^\mu\wedge\df x^\nu,
}
namely,
\al{
T^\ba{}_{\mu\nu}\fn{x}=\p_{[\mu} e^\ba{}_{\nu]}\fn{x}+\omega^\ba{}_{\bb[\mu}\fn{x}e^\bb{}_{\nu]}\fn{x};
}
and we have used $\df x^\mu\wedge\df x^\nu\wedge\df x^\rho\wedge\df x^\sigma=\ep{\mu\nu\rho\sigma}\df^4x$.
More specifically, the Nieh-Yan terms are computed as
\al{
\df\fn{e^\ba\wedge T_\ba}
	&=	\df e^\ba\wedge T_\ba-e^\ba\wedge\df T_\ba
	=	\pn{T^\ba-\omega^\ba{}_\bb\wedge e^\bb}\wedge T_\ba
		-e^\ba\wedge\pn{\df\omega_{\ba\bb}\wedge e^\bb-\omega{}_{\ba\bb}\wedge\df e^\bb}\nn
	&=	T^\ba\wedge T_\ba
		-\omega^\ba{}_\bb\wedge e^\bb\wedge T_\ba
		-e^\ba\wedge\pn{
			\pn{\os\omega{\mc F}_{\ba\bb}-\omega_{[\ba|\bc|}\wedge\omega^\bc{}_{\bb]}}
				\wedge e^\bb-\omega_{\ba\bb}\wedge\pn{T^\bb-\omega^\bb{}_\bc\wedge e^\bc}}\nn
	&=	T^\ba\wedge T_\ba
		-\omega^\ba{}_\bb\wedge e^\bb\wedge T_\ba\nn
	&\quad
		-e^\ba\wedge e^\bb\wedge\os\omega{\mc F}_{\ba\bb}
		+e^\ba\wedge e^\bb\wedge\omega_{[\ba|\bc|}\wedge\omega^\bc{}_{\bb]}
		+e^\ba\wedge\omega_{\ba\bb}\wedge T^\bb
		-e^\ba\wedge\omega_{\ba\bb}\wedge\omega^\bb{}_\bc\wedge e^\bc\nn
	&=	T^\ba\wedge T_\ba-e^\ba\wedge e^\bb\wedge\os\omega{\mc F}_{\ba\bb}.
}

These terms are compatible with the degenerate limit.
However, barring the field-dependent couplings as discussed in the second-to-last paragraph in Section~\ref{sec: action setting}, they do not affect the quantum dynamics of the vierbein and LL-gauge fields since these topological terms give the propagators of neither the vierbein nor the LL-gauge fields.

\bibliographystyle{JHEP} 
\bibliography{refs}
\end{document}